\documentclass[12pt]{article}
\usepackage{fullpage,amsmath,amssymb,mathtools,hyperref}
\usepackage{titling,enumitem}
\usepackage[skip=0pt]{caption} 
\usepackage{pdflscape}
\usepackage{setspace}
\usepackage{datetime,bigints}
\usepackage[dvipsnames]{xcolor}
\usepackage[most]{tcolorbox}
\usdate
\usepackage{multirow,algorithm} 
\usepackage{float}
\usepackage{xcolor}

\usepackage{algorithm,bbm}
\usepackage[]{algpseudocode}
\usepackage{arydshln}
\usepackage[section]{placeins}
\usepackage[titletoc,toc,title]{appendix}
\usepackage{fancyvrb} 
\usepackage{listings}
\usepackage{bm}
\usepackage{xcolor,colortbl} 
\usepackage[backend=biber,natbib=true,style=ext-authoryear]{biblatex}

\DeclareAutoCiteCommand{textcite}{\textcite}{\textcites}
\DeclareDelimFormat[bib]{nametitledelim}{\addcolon\space}

\DeclareFieldFormat*{title}{#1}
\DeclareFieldFormat{journaltitle}{#1}
\DeclareFieldFormat{issuetitle}{#1}
\DeclareFieldFormat{maintitle}{#1}
\DeclareFieldFormat{booktitle}{#1}

\DeclareFieldFormat{pages}{#1}
\DeclareFieldFormat{postnote}{\mknormrange{#1}}
\DeclareFieldFormat{multipostnote}{\mknormrange{#1}}

\newcommand{\notes}[1]{%
    \linespread{0.2}\vspace{0.1em}%
    \captionsetup{justification=justified}%
    \caption*{\footnotesize #1}%
}

\definecolor{lightgray}{gray}{0.9}
\xdefinecolor{gray}{rgb}{0.8,0.8,0.8}
\xdefinecolor{blue}{RGB}{58,95,205}
\DeclareCaptionFormat{listing}{\rule{\dimexpr\textwidth+17pt\relax}{0.4pt}\vskip1pt#1#2#3}
\definecolor{gray}{gray}{0.85}
\definecolor{LightCyan}{rgb}{0.88,1,1}

\newcolumntype{a}{>{\columncolor{gray}}c}
\newcolumntype{b}{>{\columncolor{white}}c}

\usepackage{xr,caption}

\renewenvironment{thebibliography}[1]{
  \begin{oldthebibliography}{#1}
    \setlength{\itemsep}{0em}
    \setlength{\parskip}{0em}
}
{
  \end{oldthebibliography}
}

\newcommand{\vech}{\text{vech}}

\usepackage{thmtools} 
\usepackage{amsthm}
{
      \theoremstyle{plain}
      \newtheorem{definition}{Definition}
      \newtheorem{theorem}{Theorem}
      \newtheorem{example}{Example}
      
      \newtheorem{lemma}{Lemma}
      \newtheorem{corollary}{Corollary}
      \newtheorem{assumption}{Assumption}
      
  }

\renewcommand{\arraystretch}{1.5}

\def\lQ{\scalebox{-1}[1]{''}}

\makeatletter
\renewenvironment{abstract}{%
    \if@twocolumn
      \section*{\abstractname}%
    \else 
      \begin{center}%
        {\bfseries \normalsize\abstractname\vspace{\z@}}
      \end{center} \vspace{-0.5cm}%
      \quotation
    \fi}
    {\if@twocolumn\else\endquotation\fi}
\makeatother

\begin{document}
\begin{refsection}

  \title{Fitting Dynamically Misspecified Models:\\ An Optimal Transportation Approach}
  \author{ Jean-Jacques Forneron\thanks{Department of Economics, Boston University, 270 Bay State Road, Boston, MA 02215 USA.\newline Email: \href{mailto:jjmf@bu.edu}{jjmf@bu.edu}, Website: \href{http://jjforneron.com}{http://jjforneron.com}}. \and Zhongjun Qu\thanks{Department of Economics, Boston University, 270 Bay State Road, Boston, MA 02215 USA.\newline Email: \href{qu@bu.edu}{qu@bu.edu}. \newline
  We would like to thank Alfred Galichon, Marc Henry, Ivana Komunjer, seminar participants at Penn State, VTSS, the SED 2024, IAAE 2025 conferences, the 2025 Manchester Econometrics Workshop, and the BC/BU Greenline workshop for useful comments and suggestions.}} 
  \date{\today}
  \maketitle 

  \begin{abstract}  
    This paper considers filtering, parameter estimation, and testing for potentially dynamically misspecified state-space models. When dynamics are misspecified, filtered values of state variables often do not satisfy model restrictions, making them hard to interpret, and parameter estimates may fail to characterize the dynamics of filtered variables. To address this, a sequential optimal transportation approach is used to generate a model-consistent sample by mapping observations from a flexible reduced-form to the structural conditional distribution iteratively. Filtered series from the generated sample are model-consistent. Specializing to linear processes, a closed-form Optimal Transport Filtering algorithm is derived. Minimizing the discrepancy between generated and actual observations defines an Optimal Transport Estimator. Its large sample properties are derived. A specification test determines if the model can reproduce the sample path, or if the discrepancy is statistically significant. Empirical applications to DSGE models, affine term structure models, and trend-cycle decomposition illustrate the methodology and the results.
  \end{abstract} 
  

  \baselineskip=18.0pt
  \thispagestyle{empty}
  \setcounter{page}{0}
  
\newpage

\section{Introduction}

Structural estimation is routinely used to evaluate Economic theories and conduct counterfactual analyses with non-experimental data. As noted by \citet{domowitz1982}, to make the analysis tractable - either analytically or numerically - some simplification is needed so that the model merely approximates the actual, more complex, data-generating process. Though misspecified, the model still provides tractable insights about causal mechanisms that can be used for policy evaluation and conduct counterfactual experiments. This paper is specifically interested in multivariate models of the form:
\begin{align} 
 y_t &= g(z_t,v_t;\theta), \quad z_t = h(z_{t-1},v_t;\theta), \label{eq:ssmodel}
\end{align} 
where $y_t$ are observed variables such as output or inflation, $z_t$ are unobserved variables such as productivity, and $v_t$ are structural innovations. The functions $g$ and $h$ are known, or solved numerically, up to parameters of interest $\theta$. This is known as a state-space, or hidden Markov model. Examples include DSGE and structural asset pricing models. It is common to fit the model using a filtering algorithm that recovers the latent variables $z_t$ -- the Kalman or particle filter -- and then maximize the likelihood, or sample Bayesian posterior draws. 

This paper shows that several issues can arise when the dynamics in $g$ or $h$ are misspecified. First, for a given value $\theta$, filtered variables may not satisfy the model restrictions given by $g$ and $h$. For instance, the same average of a filtered series can be substantially non-zero even though the model describes a mean-zero process. This can make inferences on policy-relevant variables difficult, e.g., output gap or natural rate of interest, since their interpretation is model-dependent. Similarly, filtered structural shock processes can be cross-correlated even though the model specifies them as independent, a well-documented phenomenon in the DSGE literature. Second, likelihood estimates $\hat{\theta}_n$ might be hard to interpret since these coefficients may not characterize the dynamics of the filtered variables: the serial correlation of a Kalman filtered shock can differ substantially from its model-implied value. These two points are illustrated using a medium-scale DSGE model. Third, the likelihood is not defined when there are fewer structural shocks than observables. This limits the potential for model-based dimension reduction where a few structural shocks are used to summarize the comovement of financial or economic variables. This is also illustrated in the applications.


This paper considers an optimal transportation approach to fitting dynamics models in (\ref{eq:ssmodel}) with a mean-squared criterion. Fitting here refers to filtering latent variables and estimating the parameters of interest. The basic idea is to first flexibly approximate the true dynamics of the data; using a reduced-form model. Then, for a given $\theta$, a new sample is recursively constructed for which (\ref{eq:ssmodel}) holds. At each time iteration, the procedure maps the observations to model-consistent data by transporting from the reduced-form to the model-predicted conditional distribution, i.e. the conditional mean-squared error between the sample and the model-consistent data is minimized. Finding the least-squares difference between the original and the new sample produces optimal transport estimates for the parameters of interest. A by-product is an optimal transport filtered series for $z_t$. The new data, filtered values, and estimates preserve model dynamics by construction, and are thus internally valid. Note that, unlike with i.i.d. data, the dependence structure here requires a different approach to implementing the transport. This is reflected in the use of a flexible reduced-form model and the iterative nature use of a \textit{conditional} transport, where each step builds on the previous ones and is performed as many times as the sample size.

Although there has been much progress in the computation of non-linear filters and numerical optimal transportation, the generic approach described above can be computationally demanding for larger models. Specializing to linear processes, a plugin rule for the optimal transport map is derived leading to closed-form expressions. The true dynamics of the data are approximated using a sieve approach through a vector autoregression of increasing order. The resulting algorithm has closed-form, is easy to implement, and numerically inexpensive. The associated estimator is semiparametric, as only the linear dynamics are specified. The closed-form plugin map extends to a class of non-linear models, though not as general as (\ref{eq:ssmodel}). Because the transport map reduces to the identity map under correct model specification, the framework encompasses correctly specified structural models as a special case.

For stationary linear processes, we derive the large sample frequentist results for the optimal transport estimator $\hat{\theta}_n$ which minimizes the mean-squared difference between original and model-consistent samples. Under standard regularity conditions, the estimates are shown to be consistent and asymptotically normal at a $\sqrt{n}$-rate. An expression for the asymptotic variance is derived under correct specification and misspecification. A specification test based on the mean-squared discrepancy between the sample paths is proposed and studied. The method and results cover a large class of models with an infinite moving average representation which includes linear state-space models. While the results are confined to frequentist estimation, it can also be of interest to extend the framework and consider quasi-Bayesian posterior sampling and inference. This goes beyond the scope of this paper.

In Machine Learning, the sample Wasserstein distance between distributions is a popular tool for data analyses by optimal transportation. It is generally intractable, non-smooth, suffers from a curse of dimensionality, and is computationally demanding for estimation. This can limit its appeal for estimating models with a moderate or large number of observables and parameters. In the scalar case, the minimum Wasserstein distance estimator is fully parametric but has non-standard limiting distribution, which complicates inference. Entropic regularization is a popular way to circumvent some limitations of the Wasserstein distance, but it introduces bias. In contrast, the setting here is semiparametric -- only first and second order moments of the data and model are involved. The auxiliary model provides these moments for the data. We show that this allows for a computationally trivial closed-form solution to the transport problem, even for medium-scale DSGE models. The closed-form map is smooth, making estimation regular with $\sqrt{n}$-asymptotically Gaussian estimates.
  
Three empirical applications illustrate the methodology and the issues discussed above on well-known models. First, a small-scale DSGE model from \citet{lubik2004} is estimated. The fit for inflation is rejected at the 5\% level, which corroborates previous findings. Second, the medium-scale \citet{smets2007} model is estimated. Unlike previous studies, the fit for consumption is rejected as it does not match volatility and persistence. Further, the Kalman filtered series display irregularities consistent with misspecification, exacerbated by persistence in the data. Next, an affine term structure model based on \citet{hamilton2012} illustrates the dimension-reduction aspect: 3 factors explain most of the variation of 6 yields ranging from 1 month to 5 years. Finally, a trend-cycle decomposition shows that the optimal transport filter produces an interpretable cycle, while the Kalman filter produces a cycle estimate that is systematically positive over 1965-2008.


\newcommand{\vardbtilde}[1]{\tilde{\raisebox{0pt}[0.85\height]{$\tilde{#1}$}}}

\section{Filtering under Dynamic Misspecification}

This section motivates the proposed Optimal Transport Filter (OTF). It illustrates the Kalman Filter (KF) under misspecification using a small DSGE model, then discusses the effects of misspecification on filtering in a general setting.

\subsection{Motivating Example: A Small DSGE Model} \label{sec:motivating}
\begin{figure}[h] \caption{Misspecified LS model: Filtered Shocks and Fitted Variables} \label{fig:LS04_ms}
  \centering \includegraphics[scale=0.45]{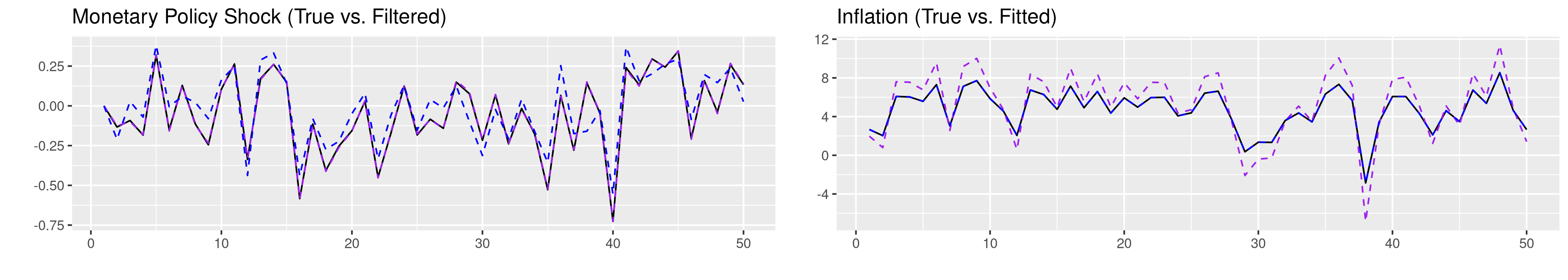}\\\notes{ \textbf{Legend:} First 50 observations of the simulated series from model (\ref{eq:LS}) and fitted series using model (\ref{eq:LS}'). Black solid line: True value, Purple dashed line: Optimal Transport Filter, Blue dashed line: Kalman Filter. }
\end{figure}

The following illustrates two issues with Kalman filtering under misspecification: (i) filtered variables need not satisfy model constraints, (ii) the KF fits observables perfectly. In constrast, the OTF introduced in this paper: (i) recovers latent variables that satisfy model constraints, and (ii) indicates lack of fit in observables, which is informative about the misspecification. The model is taken from \citet[LS]{lubik2004}:
\begin{align} \begin{split}
  &y_{t} =E_{t}y_{t+1}-\tau (r_{t}-E_{t}\pi _{t+1})+g_{t}, \quad \pi _{t} =\beta E_{t}\pi _{t+1}+\kappa (y_{t}-z_{t}), \quad g_{t} =\rho _{g}g_{t-1}+\varepsilon_{gt},  \\ 
  &z_{t} =\rho_{z}z_{t-1}+\varepsilon _{zt}, \quad r_{t} = \rho _{r}r_{t-1}+(1-\rho _{r})\psi _{1}\pi _{t} +(1-\rho _{r})\psi
_{2}(y_{t}-z_{t}) +\varepsilon _{rt}   \end{split}, \label{eq:LS}
\end{align}
where $y_{t},\pi _{t}$, and $r_{t}$ are log-deviations of output, inflation, and the nominal interest rate from their steady states. The shocks $\varepsilon _{rt},$ $\varepsilon _{gt},\varepsilon _{gt}$ are
iid Gaussian with mean zero and variances $\sigma_{r}^{2},\sigma _{g}^{2},\sigma _{z}^{2}$; and $\varepsilon _{gt}$ and $\varepsilon _{zt}$ are correlated with correlation $\rho_{gz}$. Using empirical estimates from Section \ref{sec:emp} below, $n = 10^6$ observations $(\tilde{y}_t,\tilde{\pi}_t,\tilde{r}_t)$ are simulated from model (\ref{eq:LS}). The KF and OTF are applied to the data, without re-estimating the model to focus on filtering. \\
\textbf{Correct specification.} As a benchmark, the KF and OTF are applied to the correct specification (\ref{eq:LS}) using the true parameter values. The $R^2$ between the filtered observables $({y}_t,{\pi}_t,{r}_t)$ and the data $(\tilde{y}_t,\tilde{\pi}_t,\tilde{r}_t)$ equals one for the KF and $0.9999$, or higher, for the OTF.\footnote{$R^2 = 1-\text{ESS}/\text{TSS}=1- \sum_{t=1}^n (\tilde{y}_t-y_t)^2/\sum_{t=1}^n (\tilde{y}_t-\tilde{y})^2$ with $\tilde{y}$ the sample mean of $\tilde{y}_t$. } The three $R^2$ are similar when comparing the true $(\tilde{\varepsilon}_{rt},\tilde{g}_{t},\tilde{z}_{t})$ with the filtered $(\varepsilon_{rt},g_{t},z_{t})$ shocks. As expected, when the model is correctly specified, both filters recover the latent variable accurately and show a perfect fit for the observables.\\
\textbf{Misspecification.} Now suppose a researcher misspecifies the model by assuming the central bank targets expected, rather than current, inflation:
\begin{align} r_{t} = \rho _{r}r_{t-1}+(1-\rho _{r})\psi _{1}\mathbb{E}_t(\pi _{t+1}) +(1-\rho _{r})\psi
_{2}(y_{t}-z_{t}) +\varepsilon _{rt}. \tag{\ref{eq:LS}'}
\end{align}
Here, only the monetary policy equation is misspecified; other equations remain correct. \\
\textbf{\textit{Issue (i)}} is illustrated by Table \ref{tab:LS04_ms}: when the model is misspecified, Kalman filtered shocks can be cross-correlated even for those specified to be independent. In comparison, the Optimal Transport filtered shocks match the model-implied covariance structure. The left panel of Figure \ref{fig:LS04_ms} provides further information. Comparing filtered monetary policy shocks $\varepsilon_{rt}$ with the true  values, the OTF estimates recover the true series accurately with an $R^2$ of $0.9997$.  In contrast, the KF estimates track it less accurately, with an $R^2$ of $0.8035$. \\
\textbf{\textit{Issue (ii)}} is illustrated by the right panel of Figure \ref{fig:LS04_ms}: the KF fits inflation perfectly with an $R^2$ of $1.0$, whereas the OTF indicates a notable lack of fit with an $R^2$ of $0.5955$. The KF distorts the shocks to achieve this perfect fit. The $R^2$ between true and fitted interest rates are $1.0$ and $0.9966$ for KF and OTF, respectively. The monetary equation is misspecified but the lack of fit mainly appears in the inflation variable. Because this is a simultaneous equation system, misspecification in one equation can affect all variables. In this case the OTF indicates which variables are most affected by the misspecification.\\
Going beyond this simulation illustration, a comparison of latent processes recovered by the KF and OTF using real data for the \citet[SW]{smets2007} model can be found in Figure \ref{fig:SWfilter} and Supplemental Figure \ref{tab:SWshocks}. Moreover, Section \ref{sec:emp} compares actual and OT filtered observables for the LS and SW models and conducts formal specification tests to determine whether the discrepancy between the observed and fitted series is statistically significant.

\begin{table}[h] \caption{Misspecified LS model: Covariance Matrices of True and Filtered Shocks} \label{tab:LS04_ms} \centering
  \small
  \setlength\tabcolsep{4.0pt}
    \renewcommand{\arraystretch}{0.85} 
  { \small
    \begin{tabular}{c|ccc|ccc|ccc} \hline \hline
        & \multicolumn{3}{c|}{True (unobserved)} & \multicolumn{3}{c|}{Optimal Transport Filter} & \multicolumn{3}{c}{Kalman Filter}\\ \hline
        & $\varepsilon_r$ & $\varepsilon_g$ & $\varepsilon_z$ & $\varepsilon_r$ & $\varepsilon_g$ & $\varepsilon_z$ & $\varepsilon_r$ & $\varepsilon_g$ & $\varepsilon_z$\\ \hline
        $\varepsilon_r$ & 0.0802 & 0 & 0 & 0.0802 & 0 & 0 & 0.0582 & 0.0149 & 0.0017\\
        $\varepsilon_g$ & 0 & 0.0296 & 0.0674 & 0 & 0.0296 & 0.0674 & 0.0149 & 0.0227 & 0.0688\\
        $\varepsilon_z$ & 0 & 0.0674 & 0.5398 & 0 & 0.0674 & 0.5397 & 0.0017 & 0.0688 & 0.5413\\
            \hline\hline
    \end{tabular} } \\\notes{ \textbf{Legend:} True: model implied covariance matrix. Optimal Transport and Kalman Filters: sample covariance matrix of filtered shocks. Simulated sample size $n=10^6$.}
\end{table}

\subsection{Discussion: Standard Filtering under Misspecification}

Consider a general state-space model as in (\ref{eq:ssmodel}), associated with a state transition density
\(p(z_t|z_{t-1})\) and measurement density \(p(y_t|z_{t})\) (omitting $\theta$ for brevity).
Standard filtering such as KF and particle filter proceeds recursively as follows: starting at time $t=0$ with initial distribution $p(z_0)$:
(a) form beliefs \( p(z_1)=\int p(z_1|z_0)p(z_0)\,dz_0, \)
(b) observe $y_1$ and update beliefs using Bayes' rule to obtain the filtered distribution
\( p(z_1|y_1)\propto p(y_1|z_1)p(z_1), \) (c) predict
\( p(z_2|y_1)=\int p(z_2|z_1)p(z_1|y_1)dz_1, \) (d) continue for $t=2,3,\dots$

However, under misspecification, in step (b) the researcher observes $\tilde{y}_1$ instead of $y_1$, with a predictive density that differs from
\( p(y_1|z_1). \)
The resulting filtered distribution for $z_1$ is biased because $p({y}_1|z_1)$ is replaced by $p(\tilde{y}_1|z_1)$. The prediction in step (c) is also distorted because it uses this filtered distribution as input. The mismatch occurs at each step when a new observation enters the filter. As a result, the inferred latent variables become difficult to interpret within the structural model. This issue does not improve with a larger samples.

To address this mismatch, we propose constructing a model-consistent sample $y_1,\dots,y_n$ using the observed data $\tilde{y}_1,\dots,\tilde{y}_n$. The main idea is that, at each $t$, an optimal transport step is used to map the observation $\tilde{y}_t \sim \tilde{p}(\tilde{y}_t|\tilde{y}_{t-1},\dots)$ into a model-consistent value $y_t \sim p(y_t|y_{t-1},\dots)$ which is closest to the data. Filtering in steps (b) and (c) then proceeds using $y_t$ rather than $\tilde{y}_t$, eliminating the model-data mismatch. The filtered state variables $z_t$ satisfy the model restrictions by construction. 

Next we focus on linear state-space models to provide a complete analysis of filtering, estimation and inference, forecasting, and model specification testing. Section \ref{sec:extension} then presents a filtering algorithm for nonlinear state-space models based on the same idea outlined above. 

\section{\large{Optimal Transport Filtering, Estimation, and Forecasting}} \label{sec:main}


For the remainder of the paper, $\tilde{y}_t$ denotes the observations collected by the researcher and $y_t$ denotes data generated from (\ref{eq:ssmodel}). The data generating process for $\tilde{y}_t$ is given by the conditional distribution $\tilde{p}(\tilde{y}_t|\tilde{y}_{t-1},\tilde{y}_{t-2},\dots)$, which needs to be estimated. For a parameter value $\theta$, filtering methods can be used to evaluate the model's predictive distribution $p(y_t|y_{t-1},\dots;\theta)$, this is not estimated. To reduce notation, the same $p$ refers to the joint and marginal distributions of $y$ and $z$. 

\subsection{Optimal Transport Filtering}

Consider linear state-space models of the form:
\begin{align}
  y_t = \mu(\theta) + A(\theta) z_t + B(\theta) v_t, \quad z_{t} = C(\theta) z_{t-1} + D(\theta) v_t,  \label{eq:Lssmodel}
\end{align}
where $v_t \sim (0,I)$ are white noise, with its higher order moments left unspecified. The dependence on the parameters $\theta$ will be omitted in the Algorithm below to simplify notation. Specification (\ref{eq:Lssmodel}) sets a particular linear structure in (\ref{eq:ssmodel}). The number of structural shocks $v_t$ can be greater, equal, or less than the number of observed outcomes $y_t$. Model (\ref{eq:Lssmodel}) includes linearized DSGE models and affine term structure models as special cases. 

The model is characterized by the conditional mean and covariance of $y_t$ and its prediction error at each $t$. To construct the transport map, the proposed filter requires computing their data-implied counterparts, which requires a flexible approximation to the DGP. Under mild conditions (given below), $\tilde{y}_t$ admits an infinite-order vector autoregressive (VAR) representation. A natural auxiliary model is a VAR(k), a sieve approximation of the VAR($\infty$):
\begin{align}
  \tilde{y}_t = \tilde{\mu} + \sum_{j=1}^k \Psi_j [\tilde{y}_{t-j} - \tilde{\mu}] + e_t.
\end{align}
This is a reduced-form VAR, and $e_t \sim (0,\tilde{\Sigma}_k)$ is a prediction error that may differ in dimension from the structural shocks. As $k \to \infty$, $\tilde{\Sigma}_k$ converges to the innovation covariance matrix of the VAR($\infty$).  In practice, the number of lags $k$ should be sufficiently large so that no significant residual autocorrelation remains. See \citet{kuersteiner2005}, and references therein, for automated lag-length selection procedures. This is a semi-parametric model because $e_t$ is specified up to the first two unconditional moments and is distributionally unrestricted.

Algorithm \ref{algo:KOTF} below presents the OTF, using as inputs: 1) the residuals $\hat{e}_t$  from a vector autoregression of $\tilde{y}_t$ on its $k$ lags, the covariance matrix of these residuals, and 2) for a given parameter value $\theta$, the model-implied state-space coefficients $\mu,A,B,C,D$.  Algorithm \ref{algo:KOTF}  only involves matrix operations and can be readily applied to models where the KF is used.

\begin{algorithm}[H]
  \caption{Optimal Transport Filter: Linear State-Space Models}\label{algo:KOTF} { \small
  \begin{algorithmic}[1]
    \Procedure{\textsc{otf}}{}\newline
    \textbf{Inputs:} 1) Sample: data $\tilde{y}_1,\dots,\tilde{y}_n$, residuals $\hat{e}_1,\dots,\hat{e}_n$ from a VAR($k$) regression with $k \geq 1$\newline\hphantom{\textbf{Inputs:}} 2) Model: coefficients $\mu,A,B,C,D$. Initial beliefs $z_0 \sim (m_{0|0},V)$ \newline
    \textbf{Outputs:} 1) Mapped data $y_1,\dots,y_n$,  2) Filtered states $z_{t|t} \sim (m_{t|t},V)$
    \For{$t \in \{1,\dots,n\}$} 
    \State{\textbf{Predict:} $m_{t|t-1} = C m_{t-1|t-1}$, $\mu_{t|t-1} = \mu + A m_{t|t-1}$}  \Comment{(KF)}
    \State{\textbf{Transport:} $y_t = \mu_{t|t-1} + P \hat{e}_t$} \Comment{(OT)}\newline
    \hphantom{\textbf{Compute:}} where $P = \tilde{\Sigma}_{nk}^{-1/2}[\tilde{\Sigma}_{nk}^{1/2} \Sigma \tilde{\Sigma}_{nk}^{1/2}]^{1/2}\tilde{\Sigma}_{nk}^{-1/2}$ \,\, (Transport Map)\newline
    \hphantom{\textbf{Compute:}} and $\Sigma = \text{var}_{t|t-1}(y_t)$, $\tilde{\Sigma}_{nk} = \widehat{\text{var}}_{t|t-1}(\tilde{y}_t)$ (Innovation Variance)
    \State{\textbf{Update:} $m_{t|t} = m_{t|t-1} + K P \hat{e}_t$} \Comment{(KF)}\newline
    \hphantom{\textbf{Compute:}} where $K = \overline{V} A^\prime \Sigma^\dagger$ (Kalman gain)\newline
    \hphantom{\textbf{Compute:}} and $\overline{V} = \text{var}_{t|t-1}(z_t)$
    \EndFor
    \EndProcedure
  \end{algorithmic}}
  \label{alg_2}
\end{algorithm}

Algorithm \ref{algo:KOTF} combines time-invariant KF iterations with an optimal transport (OT) map $P$. It adjusts the innovations $\hat{e}_t$, whose sample variance is $\tilde{\Sigma}_{nk} = \hat{\text{var}}(\tilde{y}_t|\tilde{y}_{t-1},\dots)$, to match the variance $\Sigma(\theta) = \text{var}(y_t|y_{t-1},\dots)$ implied by model (\ref{eq:Lssmodel}). The predictive distributions are summarized by $m_{t|t} = \mathbb{E}(z_t|y_t,\dots,y_1)$, $m_{t|t-1} = \mathbb{E}(z_t|y_{t-1},\dots,y_1)$, $\mu_{t|t-1} = \mathbb{E}(y_t|y_{t-1},\dots,y_1)$, and $V = \text{var}(z_t|y_t,\dots)$; $\tilde{\Sigma}_{nk}^{1/2}$ and $\tilde{\Sigma}_{nk}^{-1/2}$ are the matrix square root of $\tilde{\Sigma}_{nk}$ and its inverse. There are three main steps in the Algorithm: \textbf{Filtering}, \textbf{Transport}, and \textbf{Update}. The following discusses each step in more detail. After that, a simple example will illustrate how the series $y_t$ is constructed and how estimation is performed.

\noindent \textbf{Predict.} The prediction step is a standard KF operation. The matrices $V,\Sigma$, and $K$ solve the system of equations \citep[see][Ch4, for further details]{anderson1979}: $\bar{V} = C V C^\prime + DD^\prime,K = \bar{V} A^\prime \Sigma^{\dagger},V = (I - KA)\bar{V},\Sigma = A C V C^\prime A^\prime + (B + AD)(B + AD)^\prime$, 
where $\Sigma^{\dagger}$ denotes the Moore-Penrose inverse of $\Sigma$ if it is singular, and otherwise its inverse. 
Besides $V$ and $\Sigma$ defined above, the matrix $\overline{V} = \text{var}(z_t|y_{t-1},\dots)$ measures the one-step-ahead prediction error for $z_t$. These matrices are standard KF quantities. 

\noindent \textbf{Transport.} The transport step maps an observation $\tilde{y}_t$ with conditional distribution $\tilde{p}_{t|t-1}$ to a $y_t$ with conditional distribution $p_{t|t-1}$. This is done by solving the minimization problem: \[ \min_{\pi_{t|t-1}} \mathbb{E}_{\pi_{t|t-1}}(\|\tilde{y}_t - y_t\|^2), \] where ${\pi_{t|t-1}}$ represents any joint distribution with marginal distributions $\tilde{p}_{t|t-1}$ and $p_{t|t-1}$: \[ y_t|\{y_{t-1},y_{t-2},\dots\} \sim ( \mu_{t|t-1}, \Sigma ), \quad  \tilde{y}_t|\{\tilde{y}_{t-1},\tilde{y}_{t-2},\dots\} \sim ( \tilde{\mu}_{t|t-1}, \tilde{\Sigma} ),\]
where $(\tilde{\mu}_{t|t-1},\mu_{t|t-1})$ are the conditional means of $(\tilde{y}_t,y_t)$ and $(\tilde{\Sigma},\Sigma)$ the forecast error covariance matrices; $(\mu_{t|t-1},\Sigma)$ are produced by the Kalman recursions in the Predict step; and $(\tilde{\mu}_{t|t-1},\tilde{\Sigma})$ are estimated from the data using the vector autoregression. 

This setup ensures that the generated $y_t$ represents a draw from the model distribution $p_{t|t-1}$. The solution is unique and in closed form, given by $y_t=\mu_{t|t-1}+P\hat{e}_t$, as shown in Step 4 of the Algorithm. All subsequent belief updating and filtering operations are based on the new $y_t$ rather than the original $\tilde{y}_t$, ensuring model consistency. In the absence of misspecification, $P$ is the identity matrix and $y_t$ equals $\tilde{y}_t$.

The following provides details on how the transport map is derived and why it is semiparametrically valid. Here, the marginal distributions are only specified up to second moments. As a result, $\pi_{t|t-1}$ are also defined up to second moments:
\[ \left( \begin{array}{c} y_t \\ \tilde{y}_t \end{array} \right) \Bigg| \left( \begin{array}{c} y_{t-1},y_{t-2},\dots \\ \tilde{y}_{t-1},\tilde{y}_{t-2},\dots  \end{array} \right)  \sim \left( \left( \begin{array}{c} \mu_{t|t-1} \\ \tilde{\mu}_{t|t-1} \end{array} \right), \left( \begin{array}{cc} \Sigma & C_{t|t-1} \\ C_{t|t-1}^\prime & \tilde{\Sigma} \end{array} \right) \right), \]
where $C_{t|t-1}$ is the conditional covariance between $y_t$ and $\tilde{y}_t$, and $\tilde{\mu}_{t|t-1},{\mu}_{t|t-1},\tilde{\Sigma},$ and $\Sigma$ are the same as above.
Recall that the Optimal Transport problem is to minimize $\mathbb{E}_{\pi_{t|t-1}}(\|\tilde{y}_t-y_t\|^2) = \|\tilde{\mu}_{t|t-1} - \mu_{t|t-1}\|^2 + \text{trace}(\tilde{\Sigma}+\Sigma) - 2 \text{trace}(C_{t|t-1})$ in this setup, with the additional
constraint that $\pi_{t|t-1}$ is a proper distribution, i.e. $C_{t|t-1}$ cannot be arbitrary. This implies that the optimal transportation problem can be written as a semidefinite program:   
\begin{align} \min_{C} \Big( -2 \text{trace} ( C ) \Big) \text{ subject to } \left( \begin{array}{cc} \Sigma & C \\ C^\prime & \tilde{\Sigma} \end{array} \right) \geq 0. \label{eq:OT_L} \end{align}Any transport map between $\tilde{y}_t$ and $y_t$ with covariance $C$ that solves $(\ref{eq:OT_L})$ is optimal. Since the distributions are not fully specified, the map is not uniquely defined. In particular, the linear map which solves the Gaussian case:
\begin{align*} T : \tilde{y}_t \to \mu_{t|t-1} + P (\tilde{y}_t - \tilde{\mu}_{t|t-1}), \text{ where } P = \tilde{\Sigma}^{-1/2} \left( \tilde{\Sigma}^{1/2} \Sigma \tilde{\Sigma}^{1/2}\right)^{1/2} \tilde{\Sigma}^{-1/2}, \label{eq:map_L} 
\end{align*}
is optimal and preserves the linearity of the process. These derivations follow from \citet{dowson1982}, \citet{olkin1982}, and \citet{givens1984}. See \citet{peyre2019}, Remark 2.31, for additional discussion of the Gaussian case. The Transport step in Algorithm \ref{algo:KOTF} uses the plugin estimate $\tilde{\Sigma}_{nk}$ for $P$, and $\hat{e}_t$ for $\tilde{y}_t-\tilde{\mu}_{t|t-1}$. 

\noindent \textbf{Update.} A key difference with KF is in the update step. The standard KF update is $m_{t|t} = m_{t|t-1} + K \tilde{e}_t$ where $\tilde{e}_t$ are prediction errors for $\tilde{y}_t$ computed using model (\ref{eq:Lssmodel}) and the $K$ matrix is the Kalman gain. Here, the prediction errors $\hat{e}_t = \tilde{y}_t - \tilde{\mu}_{t|t-1}$ are based on the auxiliary VAR model and are transported using the matrix $P$ to have variance $\Sigma(\theta)$. Enforcing the model-based covariance structure ensures the new data is model-consistent. If the model is misspecified such that $\tilde{e}_t$ are serially correlated and/or do not have covariance matrix $\Sigma$, then the Kalman filtered variable will not satisfy model constraints.

\noindent \textbf{Accommodating some non-linearities:} The plugin transport map extends to non-linear models of the form:
\[ y_t = \mu(x_t;\theta) + \Sigma^{1/2}(x_t;\theta) v_t, \]
where $v_t \sim (0,I)$ and $x_t$ is observed, or can be perfectly inferred, at time $t-1$. The solution to (\ref{eq:OT_L}) now changes with $t$: $P_{t|t-1} = \tilde{\Sigma}_{t|t-1}^{-1/2}( \tilde{\Sigma}_{t|t-1}^{1/2} \Sigma(x_t;\theta) \tilde{\Sigma}_{t|t-1}^{1/2} )^{1/2} \tilde{\Sigma}_{t|t-1}^{-1/2}$; the map becomes $T_{t|t-1} : \tilde{y}_t \to \mu(x_t;\theta) + P_{t|t-1} (\tilde{y}_t - \tilde{\mu}_{t|t-1})$. The requirement that $x_t$ is observable accommodates (G)ARCH but not stochastic volatility models, for instance. Choices of auxiliary models $\tilde{p}_{t|t-1}$ used for these models in simulation-based estimation are referenced below.
\subsection{Optimal Transport Estimation (OTE)}

An output of the OTF Algorithm \ref{algo:KOTF} is the model-consistent series $y_t$, which will be referred to as \textit{coupled series}, or \textit{coupling}. The following considers estimating the parameters $\theta$ by minimizing the discrepancy between the original sample $\tilde{y}_t$ and its coupling $y_t$.

The coupling $y_t$ depends on two sets of parameters: the structural coefficients $\theta$, and reduced-form auxiliary parameters $\psi_k$. For stationary linear processes approximated by a finite-order VAR(k), $\psi_k = (\tilde{\mu}^\prime,\text{vech}(\tilde{\Sigma})^\prime,\text{vec}(\Psi_1)^\prime,\dots,\text{vec}(\Psi_k)^\prime)^\prime$, where $\text{vec}$, $\text{vech}$ denote the vectorization and half vectorization \citep[Ch2.4]{magnus2019}. In practice, OTF relies on OLS estimates $\hat{\psi}_{nk} = (\tilde{\mu}_{n}^\prime,\text{vech}(\tilde{\Sigma}_{nk})^\prime,\text{vec}(\hat{\Psi}_1)^\prime,\dots,\text{vec}(\hat{\Psi}_k)^\prime)^\prime$.

The estimation is conducted as follows: given parameters $(\theta,\hat{\psi}_{nk})$, use Algorithm \ref{algo:KOTF} to generate a series $y_t(\theta;\hat{\psi}_{nk})$ and compute the loss function:
\begin{align*}
  Q_n(\theta;\hat{\psi}_{nk}) = \frac{1}{n} \sum_{t=1}^n \|y_t(\theta;\hat{\psi}_{nk}) - \tilde{y}_t\|^2_{W_n},
\end{align*}
for some symmetric positive definite weighting matrix $W_n$. The optimal transport estimator (OTE) is the minimizer $\hat{\theta}_n$ of $Q_n$. 
For a d-dimensional vector, i.e., $y_t = (y_{t,1},\dots,y_{t,d})$, setting $W_n = \text{diag}(\text{var}(\tilde{y}_{t,1}),\dots,\text{var}(\tilde{y}_{t,d}))^{-1}$ gives the qualitative interpretation that $\hat{\theta}_n$ maximizes the average R-squared between $\tilde{y}_t$ and its coupling $y_t$, i.e. $R^2_j = 1- [\sum_t (y_{t,j} - \tilde{y}_{t,j})^2 ]/[\sum_t (\tilde{y}_{n,j} - \tilde{y}_{t,j})^2 ]$ for $j \in \{1,\dots,d\}$. This choice of $W_n$ was used in all simulated and empirical examples. For DSGE models, it is common to incorporate prior information. This can be accommodated here by penalization: $Q_n(\theta;\hat{\psi}_{nk}) - \frac{1}{n} \log(\pi(\theta))$, where $\pi$ is the prior density. Under suitable regularity conditions, the first-order asymptotic properties of $\hat{\theta}_n$ are unchanged.

Although the OTE might appear abstract, Theorem \ref{th:cons} in Section \ref{sec:Asym} implies  that under regularity conditions, given below, the sample loss $Q_n$ converges in probability to:
\begin{align*}
Q(\theta;\psi_0) = \|\tilde{\mu} - \mu(\theta)\|^2_W + \sum_{j=0}^\infty \text{trace} \left( \tilde{\Sigma}^{1/2} \{ \tilde{\Lambda}_j - \Lambda_j(\theta) P(\theta;\tilde{\Sigma}) \}^\prime W \{ \tilde{\Lambda}_j - \Lambda_j(\theta)P(\theta;\tilde{\Sigma}) \} \tilde{\Sigma}^{1/2} \right), 
\end{align*}
where $W_n$ converges to $W$; $\tilde{\Lambda}_0=\Lambda(\theta)_0=I_d$; $(\tilde{\Lambda}_j,\Lambda(\theta)_j)_{j\geq 1}$ are the vector moving average coefficients; $(\tilde{\mu},\mu(\theta))$ the unconditional means; $(\tilde{\Sigma},\Sigma(\theta))$ the forecast error covariances of $(\tilde{y}_t,y_t(\theta))$; and $P(\theta;\tilde{\Sigma})$ is the transportation matrix introduced above. From this expression, the OTE can be interpreted as minimizing a distance between the sample and model's impulse responses at all horizons. This is illustrated with a simple example below.

The OTE is closely related to simulation-based estimators: the simulated method of moments, indirect inference, and the efficient method of moments. Both rely on the original sample $\tilde{y}_t$ and model-implied data $y_t(\theta)$. An important difference is that OT constructs $y_t$ from reduced-form innovations $\hat{e}_t$ evaluated from the sample $\tilde{y}_t$, whereas simulation-based estimator draw structural innovations from a known parametric distribution. Similar to these methods, OT estimation could minimize a distance between moments computed from $\tilde{y}_t$ and $y_t(\theta)$, respectively. A common criticism about fitting a particular set of moments is that the corresponding coefficients may not provide a good fit on non-fitted moments. The following Lemma addresses this particular concern.

\begin{lemma} \label{lem:Qbound} Suppose $\tilde{y}_t$ and $y_t(\theta;\psi_0)$ are covariance stationary. Let ${\bf{\tilde{y}}}_t = (\tilde{y}_t,\tilde{y}_{t-1},\dots)$, ${\bf{y}}_t(\theta;\psi_0) = (y_t(\theta;\psi_0),y_{t-1}(\theta;\psi_0),\dots)$ denote the infinite history of $\tilde{y}_t$ and ${y}_t(\theta;\psi_0)$. Let: 
\begin{align*} \mathcal{M} = \Big\{ m:{\bf{\tilde{y}}}_t \to  \sum_{j \geq 0} A_{j}^\prime \tilde{y}_{t-j} + \sum_{j,j^\prime \geq 0}  \tilde{y}_{t-j}^\prime B_{j,j^\prime} \tilde{y}_{t-j^\prime},\text{ s.t. } \sum_{j=0}^\infty \|A_j\|^2 \leq M_1^2, \sum_{j,j^\prime \geq 0} \|B_{j,j^\prime}\|_{\infty} \leq M_2 \Big\}, \end{align*}
denote the set of quadratic moments over the infinite histories. Then:
\[ \sup_{m \in \mathcal{M}}\Big\|\mathbb{E}\left( m({\bf{\tilde{y}}}_t)\right) - \mathbb{E} \left( m({\bf{{y}}}_t(\theta;\psi_0)) \right) \Big\| \leq \sqrt{C_W} (M_1+M_2 \sqrt{\mathbb{E}(\|\tilde{y}_t\|^2)}) \sqrt{Q(\theta;\psi_0) } + C_W M_2 Q(\theta;\psi_0), \]
for some $C_W$ which depends on $W$. If $\tilde{\mu} = \mu(\theta)$, $M_1$ can be removed from the upper bound.
\end{lemma}

Lemma \ref{lem:Qbound} shows that the OT loss, $Q$, bounds the worst-case fit over a large class of quadratic moments which includes means, covariances, and autocovariances. These are routinely used in (simulated) minimum distance (SMD) estimations of (\ref{eq:Lssmodel}), where the fit is not guaranteed for moments that do not enter the estimation. Similarly, the Kullback-Leibler divergence, used in MLE, does not provide this type of guarantee for moments.

\noindent \textbf{Accommodating some non-linearities (cont'd):} For non-linear models discussed in the previous subsection, the choice of $\tilde{p}$ is related to the choice of auxiliary models in simulation-based estimation, and must provide an adequate approximation to the DGP. \citet{gallant1996} suggest several models, including the SNP estimator of \citet{gallant1987} and a nonparametric ARCH model.\footnote{\citet{gallant1987} call $\tilde{p}$ the score generator for the Efficient Method of Moments.} \citet{altissimo2009}, \citet{kristensen2012} consider kernel-density estimates when the model is Markovian in the observables. Once chosen, their parameter estimates represent $\hat{\psi}_{nk}$, and the estimation of structural parameters proceeds in the same way, using the same objective function as in the linear case.

\subsection{A Pen \& Pencil Example} \label{sec:pen}

The following example illustrates how the OTF constructs a model-consistent series, how the parameters are estimated, and how the estimates compare with quasi-maximum likelihood. Consider a DGP and a structural model given, respectively, by
\begin{align*}
\tilde{y}_t &= \tilde{\mu} + \rho \tilde{y}_{t-1} + \tilde{\sigma} e_t, \qquad
y_t = \mu + \sigma \left( v_t + \beta_1 v_{t-1} + \dots + \beta_q v_{t-q} \right),
\end{align*}
where $e_t \sim (0,1)$ and $v_t \sim (0,1)$. An MA($q$) model is thus fitted to data generated by an AR(1). Assume the lag order $q$ is finite, so that the model is misspecified. Also assume the MA($q$) parameters specify a stationary and invertible process so that $v_t$ are the prediction errors, that is, the shocks in the Wold decomposition.

Use an AR($k$) model with $k \ge 1$ as the auxiliary model, and let $\hat{e}_t$ denote the standardized residuals from this regression. By Algorithm \ref{alg_2}, for a given value of $\theta = (\mu,\beta_1,\dots,\beta_q,\sigma)$, the OTF maps the prediction errors from the AR($k$) into the prediction errors of the MA($q$). This yields the following coupled series:
\(
y_t = \mu + \sigma \left[\hat{e}_t + \beta_1 \hat{e}_{t-1} + \dots + \beta_q \hat{e}_{t-q}\right].
\)
Accordingly, the OTE for $\theta$ minimizes the loss function
\[
Q_n(\theta;\hat{\psi}_{nk}) 
= \frac{1}{n}\sum_{t=1}^n (\tilde{y}_t-y_t)^2
= \frac{1}{n}\sum_{t=1}^n (\tilde{y}_t - \mu - \sigma \hat{e}_t - \sigma \beta_1 \hat{e}_{t-1} - \dots - \sigma \beta_q \hat{e}_{t-q})^2,
\]
which amounts to regressing $\tilde{y}_t$ on $(\hat{e}_t,\hat{e}_{t-1},\dots,\hat{e}_{t-q})$ and an intercept.

Since the difference between $(\hat{e}_t,\dots,\hat{e}_{t-q})$ and $(e_t,\dots,e_{t-q})$ is asymptotically negligible, $\hat{\theta}_n$ converges in probability to $\theta_0 = (\tilde{\mu},\rho,\rho^2,\dots,\rho^q,\tilde{\sigma})$. The OTE recovers the impulse response function of $\tilde{y}_t$ to $e_t$ up to the specified order $q$, even though the estimated model is misspecified. The maximum likelihood estimator of $\theta$ does not have this property. For example, when $q=1$, conditional on $v_{0} = 0$ the maximum likelihood estimator minimizes:
\[ L_n(\theta) = \frac{1}{n} \sum_{t=1}^n ( \tilde{y}_t - \mu + \sum_{k=1}^{t-1} (-\beta_1)^k \tilde{y}_{t-k} )^2/\sigma^2 + n \log(\sigma^2), \]
which does not have closed-form expressions for $\beta_1$, $\sigma^2$ when $\rho \neq 0$.

\begin{figure}[h]
  \caption{Misspecified MA(1) model: MLE and OTE estimates ($n=10^4$)} \label{fig:Toy} \centering
  \includegraphics[scale=0.45]{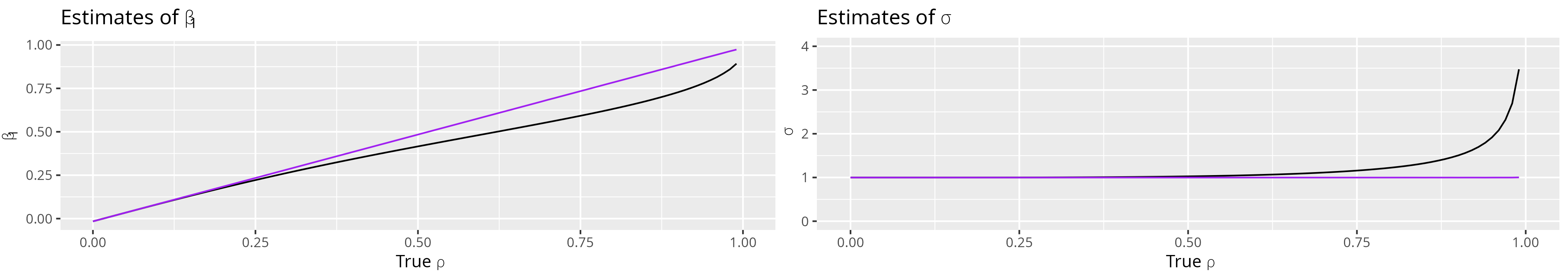}\\
  \notes{ \textbf{Legend:} Estimates from a simulated sample of $n=10^4$ observations, with $q=1$. Black solid line: MLE estimates, Purple solid line: Optimal Transport Estimates (OTE). }
\end{figure}

Figure \ref{fig:Toy} compares MLE and OTE for an MA(1) model as a function of the true AR(1) coefficient $\rho \in [0,0.95]$, fixing $(\tilde{\mu},\tilde{\sigma})=(0,1)$. As expected, the OTE of the MA(1) coefficient $\beta_1$ varies linearly with $\rho$, since it converges to $\rho$ asymptotically. It recovers the impulse response at horizon $h=1$. The MLE of $\beta_1$ depends nonlinearly on $\rho$ and systematically underestimates the impulse response at horizon $h=1$. For $\sigma$, the OT estimate matches the true value $\tilde{\sigma}=1$, whereas the ML estimate diverges nonlinearly as $\rho$ increases.

\subsection{Optimal Transport Forecasting}
The standard practice to forecast from a sample ($\tilde{y}_1,\dots,\tilde{y}_n$) using model (\ref{eq:Lssmodel}) is to apply a standard filter to recover latent variables $(z_1,\dots,z_n)$, and use the model dynamics to predict $z_{n+h|n}(\theta)$ and then $y_{n+h|n}(\theta)$ at $h \geq 1$. As shown above, when the true data generating process does not coincide with (\ref{eq:Lssmodel}), the filtered variables are not model-consistent and, by extension, neither are the resulting forecasts. Under misspecification, mismatch between model and data implies that the resulting forecast cannot be interpreted through the structural model.

Using Algorithm 1, there are \textit{a priori} two approaches to constructing a model-consistent forecast conditional on the parameter estimates: \textbf{(i) direct forecast}: predict $y_{n+h|n}^{D}(\theta;\hat{\psi}_{nk})$ using the model-implied conditional mean of the observable in model (\ref{eq:Lssmodel}), conditional on the generated sample $y_1(\theta;\hat{\psi}_{nk}),\dots,y_n(\theta;\hat{\psi}_{nk})$; \textbf{(ii) indirect forecast}: predict $\tilde{y}_{n+h|n}$ using the conditional mean given by the auxiliary model evaluated at $\hat{\psi}_{nk}$, and then apply Algorithm 1 to transport the extended sample $\tilde{y}_1,\dots,\tilde{y}_n,\tilde{y}_{n+1|n},\tilde{y}_{n+2|n},\dots$ to produce the forecast $y_{n+h|n}^{D}(\theta;\hat{\psi}_{nk})$. The following lemma shows that the two forecasts are identical.

\begin{lemma} \label{lem:DI}
  For any $\theta \in \Theta$ and $h\geq 1$, the following holds: $y_{n+h|n}^{I}(\theta;\hat{\psi}_{nk}) = y_{n+h|n}^{D}(\theta;\hat{\psi}_{nk}).$
\end{lemma}

An implication of Lemma \ref{lem:DI} is that the reduced form forecast can be decomposed into $\tilde{y}_{n+h|n} = y_{n+h|n}^{D} + u_{n+h|n}$ where $y_{n+h|n}^{D}$ decomposes the forecast in terms of strutural quantities and $u_{n+h|n}$ is the variation that the structural model cannot explain. Among competiting structural models, those with a larger $R^2$ can explain a greater share of the reduced-form forecast.

\section{Related Literatures} \label{sec:lit} 
Textbook references on optimal transport (OT) include \citet{villani2003} for theory, \citet{peyre2019} for computation, and \citet{galichon2018} for Economics. In statistics, much of the methodology and theory considers OT between iid samples. \citet{dudley1969} showed that the empirical Wasserstein distance suffers from a curse of dimensionality, unlike the plug-in approach used here.  The literature is much more limited for dependent data. \citet[pp8-9]{o2022} construct couplings between finite state Markov Chains using dynamic programming methods. They do not consider parameter estimation. 

Several papers consider parameter estimation using the Wasserstein distance. \citet{bassetti2006a} and \citet{bassetti2006b} study the estimation of location and scale for univariate distributions. They derive consistency and a non-standard limiting distribution for the estimator; \citet{bernton2019a} extend their results. As a minimum-distance estimator, alternatives to OTE include the Simulated Method of Moments, Indirect Inference \citep{gourieroux1996}, and adversarial estimation using GANs \citep{kaji2023}. \citet{genevay2018} discuss the advantages of using OT over classifiers found in GANs. \citet{forneron2023} considers semi-nonparametric simulation-based estimation, but assumes correctly specified dynamics. These methods do not recover the latent variables. Algorithm \ref{alg_2} is closely related to a goodness-of-fit plot in the Real Business Cycle literature. \citet[Figures 2-6]{plosser1989} and \citet[Figures 7, 13]{king1999} compute historical productivity shocks outside the model and use them to simulate a one-shock RBC economy. They plot simulated against real data to show the fit of calibrated models. 

Misspecification robust filtering also considers model misspecification but aims to recover the true latent variable. The main goal is to reduce sensitivity to local misspecification over a pre-specified neighborhood, see e.g. \citet{sayed2001} and \citet{shafieezadeh2018}. This relates to \citet{hansen2008}'s approach to robustness in Economics. Here, the model can be globally misspecified; the filtered values are computed under model constraints. 

There is a rich literature on quasi-maximum likelihood estimation under misspecification starting with \citet{white1982}. Result on quasi-ML estimation of hidden Markov models are fewer: \citet{mevel2004} consider discrete state-space models, \citet{douc2012} give conditions for parameter consistency for more general state-space models. These estimators rely on Kalman or particle filters which, as illustrated above, recover latent variables that typicall do not satisfy moment contraints under misspecification. There a also a number of misspecification-robust moment-based estimators, e.g. \citet{schennach2007} or \citet{antoine2021} which fit to model to a selected set of moments.  

Several papers consider estimation and policy analysis with misspecified DGSE models. \citet{del2007} and \citet{del2009} use a DSGE-VAR framework where a hyperparameter penalizes between a reduced form and structural model. Here, the flexible VAR is used to enforce the model structure with the coupling. This ensures the parameters are internally valid, i.e. characterize the dynamics of $y_t$. \citet{watson1993} proposed a goodness of fit measure based on the magnitude (in the frequency domain) of the measurement error required for the model to match the second-order moments of the data. He does not consider filtering, estimation, inference or specification testing.   

\section{Large Sample Properties of OTE} \label{sec:Asym}
The following derives consistency and asymptotic normality results for a class of linear processes, which includes linear state-space models described by (\ref{eq:Lssmodel}):
\begin{align}
  y_t = \mu(\theta) + A(\theta) z_t + B(\theta) v_t, \quad z_{t} = C(\theta) z_{t-1} + D(\theta) v_t.  \tag{\ref{eq:Lssmodel}}
\end{align}

\paragraph{Notation:} The parameters $\theta \in \Theta \subseteq \mathbb{R}^{d_\theta}$. Norms: for a matrix $A = (a_{ij})$ of size $n \times m$, the baseline norm is $\|A\| = \sqrt{\text{trace}(A'A)}$, the operator norm is $\|A\|_{op} = \sqrt{\lambda_{\max}(A'A)}$, the sup norm is $\|A\|_{\infty} = \max_{i,j} |a_{ij}|$. Eigenvalues: for a symmetric matrix $A$ of size $n \times n$, $\lambda_j(A)$ denotes the $j$-th eigenvalue, $1 \leq j \leq n$, in increasing order, $\lambda_{\max}(A) = \lambda_n(A)$ and $\lambda_{\min}(A) = \lambda_1(A)$; $\underline{\lambda} \preceq A$ implies $\lambda_{\min}(A) \geq \underline{\lambda}$ and $A \preceq \overline{\lambda}$ implies $\lambda_{\max}(A) \leq \overline{\lambda}$. For a matrix $A$ of size $n \times m$, the singular values are given by $\sigma_j(A) = \sqrt{\lambda_{j}(A'A)}$ if $m < n$, or $\sigma_j(A) = \sqrt{\lambda_{j}(AA')}$ if $n < m$, for $j = 1,\dots,\min(n,m)$; $\sigma_{\min}(A) = \sigma_1(A)$ and $\sigma_{\max}(A) = \sigma_{\min(m,n)}(A) = \|A\|_{op}$.

\subsection{Consistency and Asymptotic Normality} \label{sec:cons_asy}

The true data-generating process (DGP) and the model are assumed to be left-invertible; i.e. they admit a one-sided infinite vector moving-average (VMA) representation.\footnote{Some models can feature non-invertibility, this is the case with permanent income \citep{fernandez2007}. Algorithm \ref{algo:KOTF} only relies on second-order moments whereas identification and estimation of non-invertible models rely on higher-order cumulants, which is beyond the scope of this paper.}

\begin{assumption}
  \label{ass:Causal} $\tilde{y}_t$ and $y_t$ admit causal VMA$(\infty)$ representations:
  \begin{align*}
    \tilde{y}_t = \tilde{\mu} + e_t + \sum_{j=1}^\infty \tilde{\Lambda}_j e_{t-j}, \quad
    y_t(\theta) = \mu(\theta) + \xi_t + \sum_{j=1}^\infty \Lambda_j(\theta) \xi_{t-j},
  \end{align*}
  for any $\theta \in \Theta$,   where $e_t$ and $\xi_t$ are white noise with variance $\tilde{\Sigma}$ and $\Sigma(\theta)$.
\end{assumption}

The VMA innovations $\xi_t$ need not coincide with, or span, the structural innovations $v_t$ \citep{fernandez2007}. The number of structural innovations can be greater than, equal to, or less than the number of observables. In the latter case, the models are stochastically singular.\footnote{For instance: multivariate RBC models with a single shock to productivity are stochastically singular. See \citet{komunjer2011}, \citet{qu2018} for identification and estimation with stochastic singularity.} Our empirical applications cover all three situations.

Using the VMA representation, Algorithm \ref{algo:KOTF} involves the following quantities: $\tilde{\Sigma} = \text{var}(e_t)$, $\Sigma(\theta) = \text{var}(\xi_t)$, $\tilde{\mu}_{t|t-1} = \tilde{\mu} + \sum_{j=1}^\infty \tilde{\Lambda}_j e_{t-j}$, and $\mu_{t|t-1} = \mu(\theta) + \sum_{j=1}^\infty \Lambda_j(\theta) \xi_{t-j}$. Note that $\Sigma(\theta)$ is the same as that in Algorithm \ref{algo:KOTF}, where $y_t$ admits a state-space representation. Take the transport map $P(\theta;\tilde{\Sigma})$, computed using $\tilde{\Sigma}$ and $\Sigma(\theta)$, the coupled series $y_t$ is:
\[ y_t(\theta;\psi_0) = \mu(\theta) + P(\theta;\tilde{\Sigma}) e_t + \sum_{j=1}^\infty \Lambda_j(\theta) P(\theta;\tilde{\Sigma}) e_{t-j}. \]
The index $\psi_0$ refers to the innovations and the variance used to compute the coupling. For $\psi = \psi_0$, the true errors $e_t$ and $\tilde{\Sigma}$ are used, as above. For $\psi = \hat{\psi}_{nk}$, the residuals $\hat{e}_t$ and sample variance $\tilde{\Sigma}_{nk}$ are used, with the convention that $\hat{e}_t = 0$ for $t \leq 0$. For $\psi = \psi_{k}$, the error $e_{t,k}$ and $\tilde{\Sigma}$ are used, where $e_{t,k} = \tilde{y}_t - \tilde{\mu} - \sum_{j=1}^k \tilde{\Psi}_j[\tilde{y}_{t-j}-\tilde{\mu}]$ are the VAR($k$) errors.
The KF steps in Algorithm \ref{algo:KOTF} compute the VMA coefficients $\Lambda_j(\theta)$ using the state-space representation (\ref{eq:Lssmodel}). Suppose $\|C(\theta)\|_{op} < 1$, iterate on the KF and OT steps to find:
\[ y_t(\theta;\psi_0) = \mu(\theta) + P(\theta;\tilde{\Sigma}) e_t + \sum_{j=1}^\infty A(\theta)C^j(\theta)  K(\theta) P(\theta;\tilde{\Sigma}) e_{t-j}, \]
where $K(\theta)$ is the Kalman gain. Thus, $\Lambda_0 = I_d$ and $\Lambda_j(\theta) = A(\theta)C^j(\theta)  K(\theta)$ for each $j \geq 1$.

\begin{assumption}
  \label{ass:DGPdata} (i). $\sum_{j=1}^\infty j^{1/2}\|\tilde{\Lambda}_j\|<\infty$ and $\text{det}\left( \sum_{j=0}^\infty \tilde{\Lambda}_j z^j \right) \neq 0$ for all $|z| \leq 1$ with $z \in \mathbb{C}$; (ii). $e_t$ is strictly stationary, $\mathbb{E}_{t-1}(e_t)=0$, $\mathbb{E}(e_te_t^\prime) = \tilde{\Sigma}$, and $0 < \underline{\lambda} \preceq \tilde{\Sigma} \preceq \overline{\lambda} < \infty$; (iii). for some $r>4$, $\mathbb{E}(\|e_t\|^{2r}) < \infty$, and $e_t$ is $\alpha$-mixing with size $-a$, where $a > r/(r-2)$.
\end{assumption}

Assumption \ref{ass:DGPdata} provides several sufficient conditions for $\tilde{y}_t$ to admit a VAR($\infty$) representation and to study the OLS estimates \citep[Ch7]{hannan2012}. The mixing conditions are needed to derive near-epoch dependence (NED) properties for $y_t(\theta;\hat{\psi}_{nk})$, its derivatives, and asymptotic results for $\tilde{\Sigma}_{nk}$.
 Assumption \ref{ass:DGPdata} allows for unmodelled dependence in higher-order moments, such as conditional heteroskedasticity (ARCH, GARCH) or stochastic volatility that satisfy a strong-mixing condition.

\begin{assumption}
  \label{ass:InnovModel} 
  $\Theta$ is convex and compact and $\theta \to (\mu(\theta),\Sigma(\theta),\Lambda_1(\theta),\dots)$ is three times continuously differentiable, such that: (i). $\text{rank}[\Sigma(\theta)] = r_\Sigma$ for all $\theta \in \Theta$, and $0 \preceq \Sigma(\theta) \preceq \overline{\lambda} < \infty$; (ii). $\sup_{\theta \in \Theta} \|\mu(\theta)\| < \infty$ and $\sum_{j=0}^\infty \sup_{\theta \in \Theta} \|\Lambda_j(\theta)\|_{op} < \infty$; (iii). for $s=1,\dots,3$ and any $i_1,\dots,i_s \in \{1,\dots,d_\theta\}$, $\sup_{\theta \in \Theta} \|\partial^s_{\theta_{i_1},\dots,\theta_{i_s}} \mu(\theta)\|< \infty$, $\sup_{\theta \in \Theta} \| \partial^s_{\theta_{i_1},\dots,\theta_{i_s}} \text{vec}[\Sigma(\theta)]\|_{\infty}< \infty$, and $ \sum_{j=0}^\infty \sup_{\theta \in \Theta} \| \partial^s_{\theta_{i_1},\dots,\theta_{i_s}} \text{vec}[\Lambda_j(\theta)]\|_{\infty} < \infty$.
\end{assumption}

\begin{assumption}
  \label{ass:NED}
  There exists $C \geq 0$, $b \geq 2$, and $\varepsilon > 0$ such that for $s=1,\dots,3$ and any $i_1,\dots,i_s \in \{1,\dots,d_\theta\}$:
  $\sum_{j = m + 1}^\infty \sup_{\theta \in \Theta} \|\Lambda_j(\theta)\|_{op} \leq C m^{-(b+\varepsilon)}$, $\sum_{j = m + 1}^\infty \|\tilde{\Lambda}_j\|_{op} \leq C m^{-(b+\varepsilon)}$, and $\sum_{j = m + 1}^\infty \sup_{\theta \in \Theta} \|\partial^s_{\theta_{i_1},\dots,\theta_{i_s}} \text{vec}[\Lambda_j(\theta)]\|_{\infty} \leq C m^{-(b+\varepsilon)}$.
\end{assumption}

Assumptions \ref{ass:InnovModel} and \ref{ass:NED} restrict the dependence of $\tilde{y}_t$, $y_t$, and its derivatives. The constant rank condition is discussed below. The following Lemma gives conditions on the state-space representation (\ref{eq:Lssmodel}) for which Assumption \ref{ass:InnovModel} holds. Lemma \ref{lem:NED} in Appendix \ref{apx:prelim} further shows that Assumption \ref{ass:NED} also holds for any $b \geq 2$ and $\varepsilon > 0$, with an appropriate constant $C > 0$.

\begin{lemma}[State-Space Model - VMA representation] 
  \label{lem:SStoVMA} 
  If $\Theta$ is convex and compact, and the following conditions hold: (i). $\text{rank}[\Sigma(\theta)] = r_{\Sigma}$ for all $\theta \in \Theta$; (ii). $\Sigma(\cdot)$, $\mu(\cdot)$, $A(\cdot)$, $B(\cdot)$, $C(\cdot)$, and $D(\cdot)$ are three times continuously differentiable with bounded derivatives; (iii). $\inf_{\theta \in \Theta} \inf_{|z| \leq 1, z \in \mathbb{C}}|\text{det}(I - C(\theta)z )| > 0$, then Assumption \ref{ass:InnovModel} holds.
\end{lemma}

The constant rank condition $\text{rank}[\Sigma(\theta)]=r_{\Sigma}$, which appears in Assumption \ref{ass:NED} and Lemma \ref{lem:NED}, and the full rank condition $0 < \underline{\lambda} \preceq \tilde{\Sigma}$ (Assumption \ref{ass:DGPdata}) are particularly important for the transport map to be well behaved. Much like the square root of a scalar, $x \to \sqrt{x}$, the matrix square root $A \to A^{1/2}$, used in the transport map, is not continuously differentiable at a singular $A$. The following Lemma derives a new result for the differentiability of $\theta \to A(\theta)^{1/2}$ when $A(\cdot)$ is singular with a constant rank.  The proof involves a constructive local block decomposition which can be used to compute differentials analytically. Unlike an eigenvalue decomposition, the block decomposition is smooth under multiplicity of eigenvalues. This should be of independent interest as the matrix square root appears in a variety of settings.



\begin{lemma}[Matrix Square Root, Constant Rank] \label{lem:ConstantRank}Suppose $\Theta \subset \mathbb{R}^{d_\theta}$ is convex and compact and $\theta \to A(\theta) \geq 0$ is $s$-times continuously differentiable for some $s \geq 1$. Assume that $A(\theta)$ has constant rank $r$, where $1 \leq r \leq d = \text{dim}(A)$ and $0 < \underline{\lambda} \leq \inf_{\theta} \lambda_{r}[A(\theta)] \leq \sup_{\theta} \lambda_{\max}[A(\theta)] \leq \overline{\lambda} < \infty$. Then:   
  (i). There exists $\delta > 0$, such that for any $\theta_0 \in \Theta$, there exists $M(\theta)$ and $B(\theta)$ that are $s$-times continuously differentiable on $\mathcal{B}_\delta(\theta_0) = \{\theta \in \Theta, \|\theta-\theta_0\| \leq \delta\}$, such that $0 < \underline{\lambda}_B \preceq B(\theta) \preceq \overline{\lambda}_B < \infty$, $M(\theta)M(\theta)^\prime = I_d$, and $A(\theta)  = M(\theta) \text{blockdiag}[B(\theta), 0_{m,m}] M(\theta)^\prime$ where $m = d-r$.  
  (ii). For all $\theta_0 \in \Theta$, the square root $A(\theta)^{1/2}  = M(\theta) \text{blockdiag}[B(\theta)^{1/2}, 0_{m,m}] M(\theta)^\prime$ is $s$-times continuously differentiable on $\mathcal{B}_\delta(\theta_0)$.   
  (iii). The square root $\theta \to A(\theta)^{1/2}$ is $s$-times continuously differentiable on $\Theta$.
\end{lemma}

The KF recursions are well defined under stochastic singularity \citep[p39]{anderson1979}; however, the likelihood is not defined.
If the constant rank condition fails, the transport map becomes non-smooth and the KF steps in Algorithm \ref{algo:KOTF} become sensitive to numerical accuracy and can be unstable \citep[Ch6.5]{anderson1979}. 

\begin{lemma}[Data: VAR($\infty$) representation, VAR($k$) approximation] \label{lem:VARinf} Suppose Assumptions \ref{ass:Causal}, \ref{ass:DGPdata}, and \ref{ass:NED} hold. Then $\tilde{y}_t$ admits a VAR($\infty$) representation:
  $\tilde{y}_t = \tilde{\mu} + \sum_{j=1}^\infty \Psi_j (\tilde{y}_{t-j} - \tilde{\mu}) + e_t$,
  where $\sum_{j=1}^\infty j^{1/2}\|\Psi_j\|<\infty$, $\Psi_0 = I_d$, and $\text{det}\left( \sum_{j=0}^\infty \Psi_j z^j \right) \neq 0$ for any $|z| \leq 1$. Further, suppose $k \to \infty$ such that $k^3/n \to 0$ and $\sqrt{n} \sum_{j = k+ 1}^\infty \|\Psi_j\| \to 0$, and let $\tilde{\Sigma}_{nk} = \frac{1}{n} \sum_{t=1}^n \hat{e}_t \hat{e}_t^\prime$ and $\tilde{\Sigma}_{n} = \frac{1}{n} \sum_{t=1}^n e_t e_t^\prime$. Then: (i). $\max_{j=1,\dots,k} \|\hat{\Psi}_j - \Psi_j\| = O_p( \sqrt{\log(n)/n})$; 
  (ii). $\tilde{\Sigma}_{nk} - \tilde{\Sigma}_{n}  = o_p(1/\sqrt{n})$; 
  (iii). $\tilde{y}_n - \tilde{\mu} = O_p(n^{-1/2})$ and $\tilde{\Sigma}_n - \tilde{\Sigma} = O_p(n^{-1/2})$. 
\end{lemma}

Lemma \ref{lem:VARinf} combines several existing results for the auxiliary parameters $\hat{\psi}_{nk}$ from the literature, mainly \citet{lewis1985} and \citet{hannan2012}. The conditions on the order of the VAR order, $k$, depend on the decay of the VAR coefficients. If model (\ref{eq:Lssmodel}) is correctly specified and the conditions for Lemma \ref{lem:SStoVMA} hold - or if the true model is a finite order stationary VARMA - then $\|\Psi_j\|_{op} = O(\bar{\rho}^j)$ for some $\bar{\rho} \in [0,1)$ and $\sqrt{n} \sum_{j=k+1}^\infty \|\Psi_j\| = o(1)$ as long as $\log(n)/k \to 0$. In these cases, the order $k$ can increase very slowly.
  
\begin{theorem}[Consistency] \label{th:cons}
  Suppose Assumptions \ref{ass:Causal}-\ref{ass:NED} hold, $k$ satisfies the conditions of Lemma \ref{lem:VARinf}, $W_n \overset{p}{\to} W > 0$, and
  $Q(\theta;\psi_0) = \lim_{n \to \infty} \frac{1}{n} \sum_{t=1}^n \mathbb{E}\left( \|y_t(\theta;\psi_{0}) - \tilde{y}_t \|_{W}^2 \right)$ is uniquely minimized at $\theta = \theta_0$. If $k$ is such that $\sqrt{n}\sum_{j=k+1}^\infty \sup_{\theta \in \Theta} \|\Lambda_j(\theta)\|_{op} = o(1)$,
then $\hat{\theta}_n \overset{p}{\to} \theta_0$.
\end{theorem}

Theorem \ref{th:cons} shows that the estimator $\hat{\theta}_n$ is consistent for the minimizer $\theta_0$ of $Q(\cdot;\psi_0)$. 
\begin{theorem}[Asymptotic Normality] \label{th:asym_normal}  Suppose the conditions for Theorem \ref{th:cons} hold with $\theta_0 \in \text{interior}(\Theta)$. Let $u_{t,k} = y_t(\theta_0;\psi_k) - \tilde{y}_t$, $u_t = y_t(\theta_0;\psi_0) - \tilde{y}_t$, $G_t(\theta_0;\psi_k) = \text{vec}[ \partial_\theta y_t(\theta_0;\psi_k)^\prime ]$ and
  \begin{align*} 
    M &= \mathbb{E}\left( \partial_\theta y_t(\theta_0;\psi_0)^\prime W \partial_\theta y_t(\theta_0;\psi_0) \right) + \mathbb{E}\left( \left[ u_{t}^\prime W \otimes I \right]\partial_\theta G_t(\theta_0;\psi_0) \right),\\
    D_{\theta,\psi}(k) &= \mathbb{E}\left[ \partial_\theta y_t(\theta_0;\psi_k)^\prime W \partial_\psi y_t(\theta_0;\psi_{k0}) +  \left[ u_{t,k}^\prime W \otimes I \right]\partial_\psi G_t(\theta_0;\psi_{k0}) \right]. \end{align*}  
  Suppose $M$ is invertible and there exists $\underline{k} \geq 1$ and $c_1 > 0$ such that for all $k \geq \underline{k}$: $0 < c_1  \leq \sigma_{\min}[D_{\theta,\psi}(k)] < \infty$.
  Define $Z_{k,t} = ((\tilde{y}_t-\tilde{\mu})^\prime,\text{vec}[ e_t \tilde{Y}_{t-1,k}^\prime\Gamma_k^{-1} ]^\prime,\text{vech}[e_te_t^\prime - \tilde{\Sigma}]^\prime)^\prime$, with $\tilde{Y}_{t-1,k} = ((\tilde{y}_{t-1}-\tilde{\mu})^\prime,\dots,(\tilde{y}_{t-k}-\tilde{\mu})^\prime)^\prime$ and $\Gamma_k = \mathbb{E}( \tilde{Y}_{t-1,k}\tilde{Y}_{t-1,k}^\prime)$. Then, the sequence of covariance matrices
  $V_{n,k} = M^{-1} \text{var} \left[ \frac{1}{\sqrt{n}} \sum_{t=1}^n \{\partial_\theta y_t(\theta_0;\psi_k)^\prime W u_{t,k} + D_{\theta,\psi}(k)Z_{k,t} \}\right] M^{-1},$ 
  is bounded from above. If, in addition, $V_{n,k}^{-1} = O(1)$, then: $\sqrt{n}V_{n,k}^{-1/2}(\hat{\theta}_n - \theta_0) \overset{d}{\to} \mathcal{N}(0,I).$
\end{theorem}

Theorem \ref{th:asym_normal} establishes the asymptotic normality of the estimates $\hat{\theta}_n$. The invertibility of $M$ and the lower bound $c_1$ are local identification conditions. The requirement $V_{n,k}^{-1} = O(1)$ is standard for central limit theorems \citep[e.g.][Th5.20]{white2014}. The boundedness of $V_{n,k}$ implies a $\sqrt{n}$-rate of convergence for $\hat{\theta}_n - \theta_0$. Note that this rate does not apply to all functionals of $\hat{\psi}_{nk}$; some may converge more slowly \citep[e.g.][Th6]{lewis1985}.

The residual $u_t = y_t(\theta_0;\psi_0) - \tilde{y}_t$ measures the model-data discrepancy. Like in OLS, it reflects a simple decomposition of $\tilde{y}_t$ into fitted values $y_t(\theta_0;\psi_0)$ and residuals $u_t$. The $R^2$ introduced earlier measure their relative magnitudes. With $n = \infty$, $R^2 = 1$ indicates correct specification. Formal specification testing is considered in the next subsection. 

\paragraph{Computing standard errors.} The following describes how to compute standard errors assuming correct specification and allowing for misspecification. For models considered in Section \ref{sec:emp}, bootstrap inference could be rather computationally cumbersome. The plugin estimates for $M,D_{\theta,\psi}(k)$, etc, are shown to be consistent in the proof of Theorem \ref{th:asym_normal}. Consistency of standard errors can be deduced from these results, the proof is omitted for brievety.

Under correct specification, $u_t = 0$ and $u_{t,k} = o(n^{-1/2})$ don't contribute to the asymptotic standard errors. First, evaluate $\partial_{\theta} y_t(\hat{\theta}_n;\hat{\psi}_{nk})$ and $\partial_{\psi} y_t(\hat{\theta}_n;\hat{\psi}_{nk})$ with finite-differences or by automatic differentiation. Then, compute $\hat{M}_{n} = \frac{1}{n} \sum_{t=1}^n \partial_{\theta} y_t(\hat{\theta}_n;\hat{\psi}_{nk})^\prime W_n \partial_{\theta} y_t(\hat{\theta}_n;\hat{\psi}_{nk})$ and $\hat{D}_{n,\theta,\psi}(k) = \frac{1}{n} \sum_{t=1}^n \partial_{\theta} y_t(\hat{\theta}_n;\hat{\psi}_{nk})^\prime W_n \partial_{\psi} y_t(\hat{\theta}_n;\hat{\psi}_{nk})$. Let $L_n(\hat{\psi}_{nk})$ denote the Gaussian quasi-likelihood for the auxiliary VAR($k$) model, $H_n(\hat{\psi}_{nk})$ its Hessian, and $\partial_{\psi} L_t(\hat{\psi}_{nk})$ the score for $\tilde{y}_t$. Take $\hat{S}_{t,k} = \hat{M}_{n}^{-1} \hat{D}_{n,\theta,\psi}(k) H_n(\hat{\psi}_{nk})^{-1} \partial_{\psi} L_t(\hat{\psi}_{nk})$. Then, $\hat{V}_{n,k}$ is the HAC estimator for the long-run variance of $\hat{S}_{t,k}$. Standard errors are computed from $\hat{V}_{n,k}/n$ in a standard fashion.

Allowing for misspecification requires estimating several additional terms. Compute $\hat{u}_t = \hat{u}_{t,k} = y_t(\hat{\theta}_n;\hat{\psi}_{nk}) - \tilde{y}_t$, $\partial_\theta \hat{G}_t(\hat{\theta}_n;\hat{\psi}_{nk})$, and $\partial_\psi \hat{G}_t(\hat{\theta}_n;\hat{\psi}_{nk})$ with $\hat{G}_t(\hat{\theta}_n;\hat{\psi}_{nk}) = \text{vec}[\partial_\theta y_t(\hat{\theta}_n;\hat{\psi}_{nk})]$. Then, compute $\hat{M}_{n} = \frac{1}{n} \sum_{t=1}^n \partial_{\theta} y_t(\hat{\theta}_n;\hat{\psi}_{nk})^\prime W_n \partial_{\theta} y_t(\hat{\theta}_n;\hat{\psi}_{nk}) + \frac{1}{n} \sum_{t=1}^n [\hat{u}_t^\prime W_n] \otimes \partial_\theta \hat{G}_t(\hat{\theta}_n;\hat{\psi}_{nk})$ and $\hat{D}_{n,\theta,\psi}(k) = \frac{1}{n} \sum_{t=1}^n \partial_{\theta} y_t(\hat{\theta}_n;\hat{\psi}_{nk})^\prime W_n \partial_{\psi} y_t(\hat{\theta}_n;\hat{\psi}_{nk}) + \frac{1}{n} \sum_{t=1}^n [\hat{u}_t^\prime W_n] \otimes \partial_\psi \hat{G}_t(\hat{\theta}_n;\hat{\psi}_{nk})$. Using the same Gaussian quasi-Likelihood terms as above, evaluate $\hat{S}_{t,k} = \hat{M}_{n}^{-1} \{ \partial_\theta y_t(\hat{\theta}_n;\hat{\psi}_{nk})^\prime W_n \hat{u}_t + \hat{D}_{n,\theta,\psi}(k) H_n(\hat{\psi}_{nk})^{-1} \partial_{\psi} L_t(\hat{\psi}_{nk})\}$. Finally, $\hat{V}_{n,k}$ is the estimator for the long-run variance of $\hat{S}_{t,k}$.

\subsection{Specification Testing} \label{sec:spec}

The population loss $Q(\theta_0;\psi_0)$ defines a distance between the VMA($\infty$) representations of $\tilde{y}_t$ and $y_t(\theta;\psi_0)$. When the model is correctly specified in terms of second-order moments, the minimizer $\theta_0$ yields $Q(\theta_0;\psi_0)= 0$. When the model is misspecified, however, the minimum is strictly positive: $Q(\theta_0;\psi_0) > 0$. The following considers a specification test based on the sample analog $Q_n(\hat{\theta}_n;\hat{\psi}_{nk})$ of the optimal transport distance $Q(\theta_0;\psi_0)$. 

\begin{assumption} \label{ass:stronger} Suppose that: (i). $[ k \log(n)]^8/n = o(1)$; (ii). $[\log(n)]^{4}/k = o(1)$; (iii). $\mathbb{E}(\|e_t\|^{16}) < \infty$; (iv). $\alpha(j) \leq C (1+j)^{-(a+\varepsilon)}$ for $a \geq 6$, $\varepsilon > 0$ and all $j \geq 1$; (v). Assumption \ref{ass:NED} holds with $b \geq 6$; (vi). $\|W_n - W\| = O_p( n^{-1/2} )$, (vii) $\sqrt{nk} \sum_{j=k+1}^\infty \|\Psi_j\|_{op} = o([\log(n)]^{-2})$.
\end{assumption}

Assumption \ref{ass:stronger} is more restrictive than those needed for Theorems \ref{th:cons} and \ref{th:asym_normal}. When the model is correctly specified, the loss $Q_n(\hat{\theta}_n;\hat{\psi}_{nk})$ is asymptotically determined by the distance between $\hat{\psi}_{nk}$ and $\psi_k$. 
 The distributional results below build on a strong approximation result for NED processes with dependence changing with the lag structure, indexed by $k$.

\begin{theorem}[Specification Test] \label{th:spec} Suppose the conditions for Theorems \ref{th:cons} and \ref{th:asym_normal} and Assumption \ref{ass:stronger} hold. If  the model is correctly specified, then:
  \begin{align*} nQ_n(\hat{\theta}_n;\hat{\psi}_{nk}) 
    &= n \mathcal{Z}_{n,k}^\prime M_k \mathcal{Z}_{n,k} + o_p(k^{1/2}[\log(n)]^{-2}),
   \end{align*}
  where $M_k = \mathbb{E} \left[ (\partial_\psi y_t(\theta_0;\psi_{k0}) + \partial_\theta y_t(\theta_0;\psi_k) M^{-1}E_k )^\prime W (\partial_\psi y_t(\theta_0;\psi_{k0}) + \partial_\theta y_t(\theta_0;\psi_k) M^{-1}E_k ) \right]$ , 
  with $E_k = - \mathbb{E}[ \partial_\theta y_t(\theta_0;\psi_{k0})^\prime W \partial_\psi y_t(\theta_0;\psi_{k0}) ]$, $\mathcal{Z}_{n,k} \sim \mathcal{N}( 0, S_{n,k}/n )$, $S_{n,k} = n\text{var}[\overline{Z}_{n,k}]$, and $M$, $\overline{Z}_{n,k}$ are defined in Theorem \ref{th:asym_normal}. 
      If $S_{n,k}$ and $M_k$ are such that $\text{trace}\left( S_{n,k}M_k \right) \geq O(k)$ and $\text{trace}\left( [S_{n,k}M_k]^2 \right) \geq O(k)$,   then for any $\alpha \in (0,1)$: 
     \[ \mathbb{P}\left( nQ_n(\hat{\theta}_n;\hat{\psi}_{nk}) > c_{n,k}(1-\alpha) \right) = \alpha + o(1), \]
     where $c_{n,k}(1-\alpha)$ is the $1-\alpha$ quantile of $n \mathcal{Z}_{n,k}^\prime M_k \mathcal{Z}_{n,k}$.
\end{theorem} 

Theorem \ref{th:spec} shows that $nQ_n(\hat{\theta}_n;\hat{\psi}_{nk})$ can be approximated by a weighted sum of independent $\chi^2_1$ random variables. The derivatives $\partial_\psi y_t(\theta;\psi_{k0})$ are given in Lemma \ref{lem:NED}. \citet[Ch15]{lutkepohl2005} provides formulas for $S_{n,k}$ in the homoskedastic case. The conditions $\text{trace}(S_{n,k}M_k) \geq O(k)$ and $\text{trace}([S_{n,k}M_k]^2) \geq O(k)$ are analogous to the rank conditions needed to ensure the J-test for GMM has a $\chi^2_{k-d}$ distribution, where $k$ is the number of moments and $d$ the number of parameters. 
The Theorem states that under the null hypothesis of correct specification, the asymptotic size of the test is $\alpha$. The test is also consistent against distant alternatives, see Lemma \ref{lem:cons_test} in Appendix \ref{apx:Spec}. Power against local alternatives depends on the ratio $k/n$. A detailed analysis of local power is left to future research. 

The test in Theorem \ref{th:spec} involves all variables $\tilde{y}_t$ used in the estimation. In certain settings, the researcher might inquire how well the model fits a specific variable $\tilde{y}_{t,j}$, e.g. consumption if the object of interest is welfare. The following Corollary specializes to a single variable, using a selection matrix $D_j$. The proof is the same as Theorem \ref{th:spec}, and it is omitted.
\begin{corollary}[Specification Test on a Single Variable] \label{cor:spec} Suppose the conditions for Theorems \ref{th:cons} and \ref{th:asym_normal} and Assumption \ref{ass:stronger} hold. Let $Q_{n,j}(\hat{\theta}_n;\hat{\psi}_{nk}) = \frac{1}{n} \sum_{t=1}^n \| y_{t} - \tilde{y}_{t}\|^2_{D_j W_n D_j}$ for $j \in \{1,\dots,d\}$, where $D_j = \text{diag}(\mathbbm{1}_{j=1},\dots,\mathbbm{1}_{j=d})$.   
  If the model is correctly specified, then:
  \[ nQ_{n,j}(\hat{\theta}_n;\hat{\psi}_{nk}) = n \mathcal{Z}_{n,k}^\prime M_{k,j} \mathcal{Z}_{n,k} + o_p(k^{1/2}[\log(n)]^{-2}),\]
  where $M_{k,j} = \mathbb{E} \left[ (\partial_\psi y_t(\theta_0;\psi_{k0}) + \partial_\theta y_t(\theta_0;\psi_k) M^{-1}E_{k,j} )^\prime D_j W D_j (\partial_\psi y_t(\theta_0;\psi_{k0}) + \partial_\theta y_t(\theta_0;\psi_k) M^{-1}E_k ) \right],$  
  with $E_{k,j} = - \mathbb{E}[ \partial_\theta y_t(\theta_0;\psi_{k0})^\prime D_j W D_j \partial_\psi y_t(\theta_0;\psi_{k0}) ]$, and $M$ and $\mathcal{Z}_{n,k}$ are as defined in Theorem \ref{th:spec}.
  If $S_{n,k}$ and $M_{k,j}$ are such that $\text{trace}\left( S_{n,k}M_{k,j} \right) \geq O(k)$ and $\text{trace}\left( [S_{n,k}M_{k,j}]^2 \right) \geq O(k)$,   then for any $\alpha \in (0,1)$: 
  $\mathbb{P}\left( nQ_{n,j}(\hat{\theta}_n;\hat{\psi}_{nk}) > c_{n,k,j}(1-\alpha) \right) = \alpha + o(1)$,
     where $c_{n,k,j}(1-\alpha)$ is the $1-\alpha$ quantile of $n \mathcal{Z}_{n,k}^\prime M_{k,j} \mathcal{Z}_{n,k}$.
\end{corollary} 

\section{Monte Carlo Simulations} \label{sec:MC}
The Monte Carlo simulations are based on the \citet[LS]{lubik2004} model, given by (\ref{eq:LS}). The observables are log levels of output, inflation, and interest rate (both annualized), which satisfy $%
Y_{t}=(0,$ $\pi ^{\ast },$ $\pi ^{\ast }+r^{\ast })^{\prime }+(y_{t},$ $4\pi
_{t},$ $4r_{t})^{\prime }$, where output is detrended, and $\pi ^{\ast }$
and $r^{\ast }$ are annualized steady-state rates of inflation and real
interest rate with $\beta =(1+r^{\ast }/100)^{-1/4}$. The data are generated using the posterior means from Bayesian inference on the full sample with LS's prior.\footnote{Table \ref{tab:LSpar}, Appendix \ref{apx:emp}, includes a description of the parameters, the bounds imposed on the parameters, and the prior $\pi$ used to regulate the estimates.} The VAR includes a constant and $4$ lags as regressors. 
\begin{table}[ht] \caption{ LS Model: Average Estimate, Standard Deviation, Rejection Rates  } \label{tab:LSsim}
  \centering
  \setlength\tabcolsep{4.0pt}
    \renewcommand{\arraystretch}{0.85} 
  { \small
  \begin{tabular}{l|ccccccccccccc} \hline \hline
    & ${\tau }^{-1}$ & ${r}^{\ast }$ & ${\kappa }$ & ${\psi }_{1}$ & ${\psi }_{2}$ & ${\rho }_{r}$ & ${\rho }_{g}$ & ${\rho }_{z}$ & ${\sigma }_{r}$ & ${\sigma }_{g}$ & ${\sigma }_{z}$ & ${\rho }_{gz}$ & ${\pi }^{\ast }$ \\ \hline
    \textsc{true} & 3.18 & 1.87 & 0.50 & 1.33 & 0.21 & 0.76 & 0.89 & 0.86 & 0.26 & 0.13 & 0.97 & 0.80 & 4.01\\ \hline 
    & \multicolumn{13}{c}{$n = 192$} \\ \hline 
    \textsc{mean} & 1.93 & 1.84 & 0.39 & 1.29 & 0.18 & 0.73 & 0.86 & 0.82 & 0.26 & 0.20 & 1.04 & 0.60 & 3.94\\
    \textsc{std} & 0.42 & 0.33 & 0.13 & 0.16 & 0.02 & 0.05 & 0.04 & 0.05 & 0.04 & 0.03 & 0.17 & 0.21 & 0.63 \\ \hline
    \textsc{rej}$_c$ & 0.14 & 0.07 & 0.05 & 0.00 & 0.00 & 0.01 & 0.03 & 0.04 & 0.01 & 0.05 & 0.03 & 0.04 & 0.11 \\
    \textsc{rej}$_r$ & 0.01 & 0.07 & 0.05 & 0.09 & 0.00 & 0.00 & 0.07 & 0.08 & 0.00 & 0.07 & 0.03 & 0.02 & 0.10 \\ \hline
    \textsc{len}$_c$ & 5.16 & 1.20 & 0.99 & 2.51 & 4.14 & 0.30 & 0.16 & 0.19 & 0.20 & 0.22 & 0.68 & 1.22 & 2.13 \\
    \textsc{len}$_r$ & 7.66 & 1.25 & 1.99 & 6.29 & 11.21 & 0.78 & 0.18 & 0.24 & 0.35 & 0.40 & 0.95 & 2.25 & 2.17 \\  \hline \hline
    & \multicolumn{13}{c}{$n = 500$} \\ \hline \hline
    \textsc{mean} & 2.08 & 1.87 & 0.38 & 1.30 & 0.18 & 0.75 & 0.89 & 0.84 & 0.26 & 0.16 & 1.03 & 0.63 & 4.08\\
    \textsc{std} & 0.45 & 0.23 & 0.11 & 0.11 & 0.02 & 0.03 & 0.02 & 0.03 & 0.03 & 0.02 & 0.11 & 0.15 & 0.44 \\ \hline
    \textsc{rej}$_c$ & 0.25 & 0.07 & 0.14 & 0.00 & 0.00 & 0.00 & 0.01 & 0.07 & 0.04 & 0.04 & 0.04 & 0.01 & 0.10 \\
    \textsc{rej}$_r$ & 0.04 & 0.07 & 0.03 & 0.00 & 0.00 & 0.00 & 0.02 & 0.06 & 0.01 & 0.06 & 0.04 & 0.01 & 0.10 \\ \hline
    \textsc{len}$_c$ & 3.59 & 0.81 & 0.65 & 1.44 & 2.67 & 0.18 & 0.10 & 0.10 & 0.14 & 0.12 & 0.43 & 0.84 & 1.57\\
    \textsc{len}$_r$ & 4.50 & 0.82 & 0.91 & 2.84 & 5.96 & 0.37 & 0.10 & 0.12 & 0.18 & 0.16 & 0.50 & 1.17 & 1.58\\ \hline \hline
  \end{tabular}\notes{ \textbf{Legend:} 200 Monte Carlo replications. \textsc{mean}/\textsc{std}: average and empirical standard error of estimates. \textsc{rej}$_c$, \textsc{rej}$_r$: rejection rates for $5\%$ level t-test. \textsc{len}: median length of $95\%$ confidence intervals. }
  }
\end{table}

The baseline sample size corresponds to the full sample estimation below with $n=192$. A larger sample size of $n = 500$ is also considered. The prior from LS is used to regularize the estimates. Table \ref{tab:LSsim} reports the averages and standard deviations of the estimates, rejection rates using standard errors that assume correct specification and those that allow for misspecification, and the length of resulting $95\%$ confidence intervals. The standard error estimates used to compute the tests and confidence intervals do not account for the prior regularization, which is assumed to be \textit{asymptotically} negligible. 

Most estimates are centered at the true value when $n=192$. A few estimates are somewhat biased towards the prior mode, most notably the risk aversion $\tau^{-1}$, which lies between the true value $3.18$ and the prior mode $1.88$. The rejection rates are generally close to the 5\% level or conservative; significant overrejection is observed only for $\tau^{-1}$ when using the non-robust standard errors, driven by prior-induced bias. The robust standard errors tend to be larger, producing lower rejection rates and wider confidence intervals. The estimation precision improves when $n$ is increased to $500$, and the other conclusions remain similar. The specification test for all variables has a rejection rate of 0.05 and 0.04 for $n=192$ and $n = 500$, respectively. For consumption only, the rejection rates are $0.05$ and $0.04$. Both are close to the nominal level. Additional Monte Carlo simulation for the medium-scale \citet{smets2007} model can be found in Table \ref{tab:SW_MC}, Appendix \ref{apx:emp}.

\section{Empirical Illustrations}  \label{sec:emp}
Three macroeconomic and one financial applications illustrate different aspects of the OT filter and estimation. The first two revisit a small and medium-scale DSGE model using the same sample period 1960Q1-2007Q4 for both. 
\subsection{Small New-Keynesian Model} \label{sec:LS}

This empirical application further considers the LS model. Following LS, there are two specifications: determinacy, with a unique equilibrium, and indeterminacy, where sunspot equilibria exist. The parameters are described in Table \ref{tab:LSest}. 
The sample is constructed and divided into subsamples as in \citet{clarida2000}: the full sample (1960Q1-2007Q4), the pre-Volcker period (1960Q1-1979Q2), and the post-Volcker period (1979Q3-2007Q4). They are associated with determinate, indeterminate, and determinate policy regimes, respectively. To remain consistent with LS and subsequent analyses, their log-prior density $\pi$ is used to penalize the OT loss so that the estimates minimize $n Q_n(\theta;\hat{\psi_n}) -  \log \pi(\theta)$. The baseline auxiliary model is a VAR($4$), results with $k=2$ are reported in Appendix \ref{apx:emp}. The weighting matrix $W_n$ is diagonal with the inverse of the variances of the three observables.

\begin{figure}[ht] \caption{LS Model: Actual and Fitted Values} \label{fig:LS} \centering
  \includegraphics[scale = 0.5]{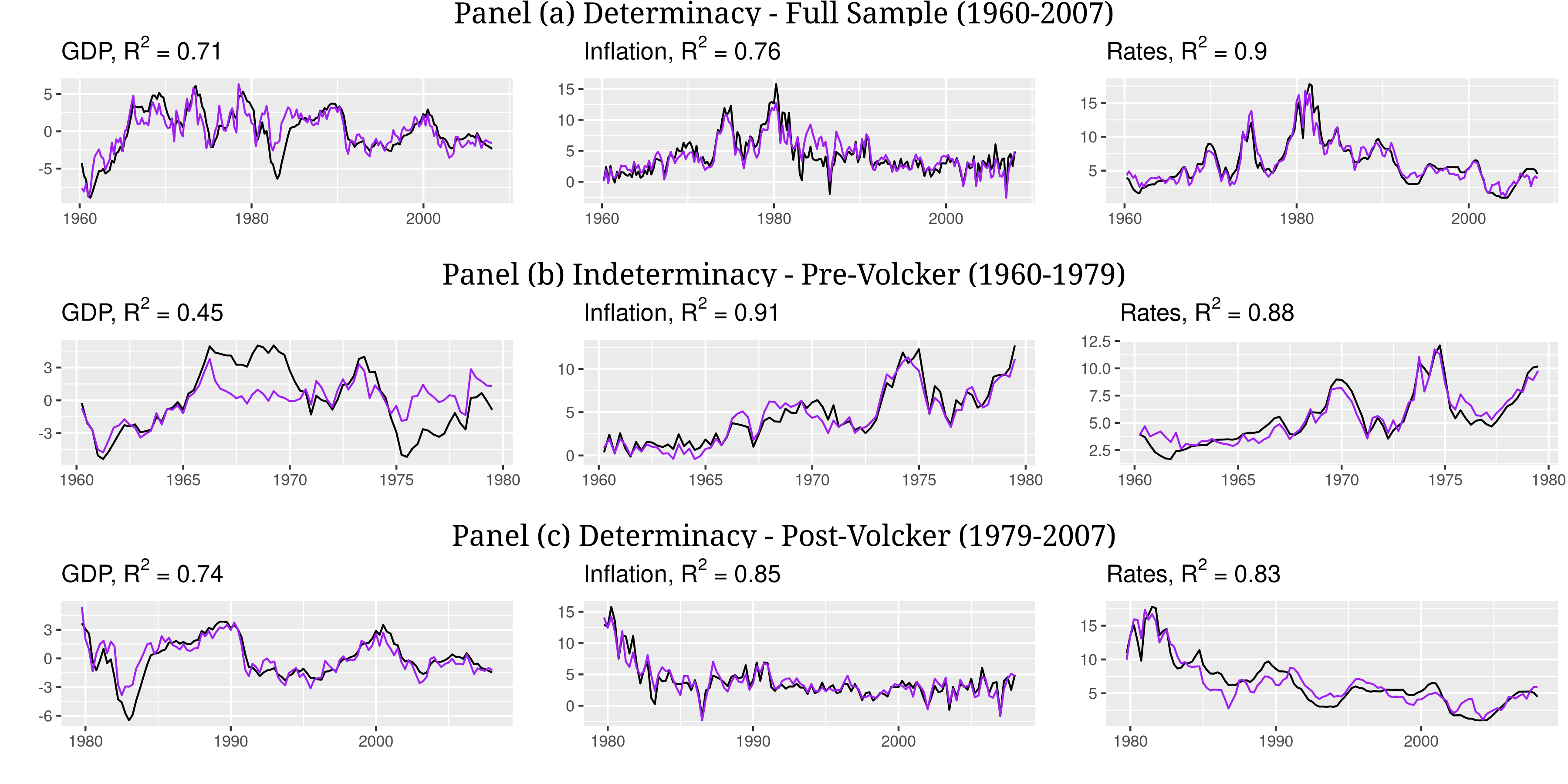}  
  \notes{ \textbf{Legend:} Black solid line = data, Purple solid line = coupling. $R^2$ is computed for each variable. }
\end{figure}

Point estimates and standard errors are reported in Table \ref{tab:LSest2}. The estimates for the pre-Volcker and post-Volcker periods are in line with those in LS, computed using Bayesian likelihood inference. In constrast to likelihood-based estimation, the methods developped here enable us to contrast the actual data with their model-implied values (i.e. the coupling) to obtain an intuitive understanding of their discrepancies. Figure \ref{fig:LS} contrasts actual and model-consistent data for GDP, inflation, and interest rate series, respectively, for the full sample and the two subsamples. For GDP, the actual data exhibit a deeper recession and lower inflation rates in the 1980s than those implied by the model. In other words, the model overpredicts the levels of GDP and inflation compared to the data. This finding confirms that this model, with time-invariant parameters, is unable to capture the rich GDP and inflation dynamics present in the data for the full sample period. 

The formal specification test, in Table \ref{tab:LSest}, investigates tensions between data and model. In the full sample, the test rejects inflation at the 5\% significance level. With $k=2$ the model is rejected overall and for each variable individually (Table \ref{tab:LSest_k2}, Appendix \ref{apx:emp}). The pre and post-Volcker samples do not reject the model. This finding is consistent with LS. None of the variables are individually rejected on the two subsamples.

\begin{table}[ht]
  \caption{LS Model: Specification Test ($k = 4$ lags)} 
 \centering
 \setlength\tabcolsep{4.0pt}
  \renewcommand{\arraystretch}{0.85} 
 {\small \
 \begin{tabular}{ll|ccc|ccc|ccc}
 \hline\hline
 \multicolumn{2}{c|}{\multirow{2}{*}{Specification Test}} & 
 \multicolumn{3}{c|}{Determinacy} & \multicolumn{3}{c|}{Indeterminacy} & 
 \multicolumn{3}{c}{Determinacy} \\ 
 &  & \multicolumn{3}{c|}{(Full Sample)} & \multicolumn{3}{c|}{(Pre-Volcker)}
 & \multicolumn{3}{c}{(Post-Volcker)} \\ \hline
 \multicolumn{2}{c|}{} & \textsc{stat} & $10\%$ & $5\%$ & 
 \textsc{stat} & $10\%$ & $5\%$ & \textsc{stat} & $10\%$ & $5\%$ \\ \hline
 \multicolumn{2}{c|}{All} & 121.2 & 140.0 & 189.6 & 59.4 & 106.2 & 169.6 & 
 65.9 & 164.8 & 235.3 \\ \hline
 \multicolumn{2}{c|}{Output} & 56.0 & 90.6 & 123.8 & 42.8 & 51.1 & 86.0 & 30.1
 & 131.8 & 188.6 \\ 
 \multicolumn{2}{c|}{Inflation} & 45.2 & 27.1 & 37.8 & 7.0 & 33.9 & 55.0 & 
 16.6 & 22.2 & 29.0 \\ 
 \multicolumn{2}{c|}{Interest Rate} & 20.0 & 31.1 & 42.1 & 9.6 & 22.7 & 36.6
 & 19.2 & 22.0 & 30.8 \\ \hline\hline
 \end{tabular}%
 \notes{ \textbf{Legend:} \textsc{stat}:
 $nQ_n(\hat{\theta}_n;\hat{\psi}_{nk})$. $10\%$, $5\%$: critical values for
 specification test at corresponding significance levels. All: joint test on all variables. Output, Inflation, Interest Rate: test
 on individual variables. $n = 192,78,114$ for the full, pre and post-Volcker
 samples. } } \label{tab:LSest}
 \end{table}
\subsection{Medium-Scale DSGE Model} \label{sec:SW}

The second empirical application considers the \citet[SW]{smets2007} model. Table \ref{tab:SWpar}, Appendix \ref{apx:emp}, includes parameter interpretations and prior distributions from SW used in our estimation. This model includes 36 free parameters and is estimated on 7 observables: consumption, investment, output and wage growth, hours worked, inflation, and interest rate. There are as many shocks: productivity, exogenous spending, monetary policy, investment-specific technology, price markup, wage markup, and risk premium shocks. 

\begin{table}[htb!]
  \caption{SW Model: Estimates and Standard Errors (selected coefficients)} \label{tab:SWest}
  \centering
  \setlength\tabcolsep{4.5pt}
   \renewcommand{\arraystretch}{0.85} 
   { \small
  \begin{tabular}{ll|ccc||cc}
  \hline\hline
  \multirow{2}{*}{${\theta }$} & \multirow{2}{*}{Parameter Interpretation} & 
  \multicolumn{3}{c||}{OT Estimate} & \multicolumn{2}{c}{Posterior} \\ 
  &  & {\ \textsc{est}} & {\ \textsc{sd}$_{c}$} & {\ \textsc{sd}$_{r}$} & {\ 
  \textsc{mean}} & \textsc{std} \\ \hline
  ${\varphi }$ & {\ Investment adjustment cost} & 3.00 & 1.50 & 3.42 & {\ 6.12}
  & 0.99 \\ 
  ${\sigma }_{c}$ & {\ Elast. of Intertemporal substitution} & 1.01 & 0.12 & 
  0.52 & {\ 1.50} & 0.14 \\ 
  ${\lambda }$ & {\ Habit persistence} & 0.74 & 0.12 & 0.38 & {\ 0.71} & 0.04 \\ 
  ${\sigma }_{l}$ & {\ Labor supply elasticity} & 1.00 & 2.38 & 2.92 & {\ 2.25}
  & 0.55 \\ 
  ${r}_{\pi }$ & {\ Taylor rule: inflation weight} & 1.80 & 1.06 & 0.77 & {\
  2.03} & 0.16\\ 
  ${r}_{\Delta y}$ & {\ Taylor rule: output gap change weight} & 0.16 & 0.05 & 
  0.10 & {\ 0.21} & 0.02 \\ 
  ${r}_{y}$ & {\ Taylor rule: output gap weight} & 0.16 & 0.16 & 0.08 & {\ 0.10%
  } & 0.02 \\ 
  ${\rho }$ & {\ Taylor rule: interest rate smoothing} & 0.90 & 0.08 & 0.07 & {%
  \ 0.82} & 0.02 \\ 
  ${\rho }_{a}$ & {\ Productivity shock AR} & 0.94 & 0.03 & 0.22 & {\ 0.97} & 0.01 \\ 
  ${\rho }_{b}$ & {\ Risk premium shock AR} & 0.68 & 0.10 & 0.19 & {\ 0.28} & 0.12 \\ 
  ${\rho }_{g}$ & {\ Exogenous spending shock AR } & 0.86 & 0.09 & 1.22 & {\
  0.97} & 0.01 \\ 
  ${\rho }_{i}$ & {\ Investment shock AR} & 0.47 & 0.15 & 0.13 & {\ 0.70} & 0.06 \\ 
  ${\rho }_{r}$ & {\ Monetary policy shock AR} & 0.47 & 0.28 & 0.71 & {\ 0.17}
  & 0.07 \\  \hline\hline
  \end{tabular}%
  \notes{ \textbf{Legend:}  Full table of estimates, prior distribution, and estimation bounds can be found in Tables \ref{tab:SWpar}, \ref{tab:SWest_full}, Appendix \ref{apx:emp}. \textsc{mean}, \textsc{std} = posterior mean and standard deviation. } }  
   \end{table}
   
\noindent\textbf{Estimation.} To illustrate the scope of OTE, the full model and $3$ singular versions are estimated using the same method. Estimates are reported in Tables \ref{tab:SWest} (for the full model) and \ref{tab:SWsing} (for singular models), Appendix \ref{apx:emp}. Likelihood-based posterior estimates computed with SW's prior are reported in Table \ref{tab:SWest} as a reference. All specifications rely on a VAR(4) auxiliary model; the inverse of the variances of the observables is used as $W_n$.

The $3$ singular models remove, in order, the risk premium, wage markup, and price markup shocks reducing to $6$, $5$, and $4$ shocks for $7$ observables. The choice of shocks to remove follows \citet{qu2018}, to reflect a view that they have a weaker structural interpretation than the remaining 4. It is interesting to examine their impact on the model fit. The standard likelihood approach cannot estimate singular DSGE models because the covariance matrix of the one-step-ahead forecasting errors $\Sigma(\theta)$ is singular. \citet{qu2018} used a composite likelihood framework and did not formally test the resulting models. OTE handles both singular and nonsingular models within the same framework. 

Additionally, note that although SW chose to fit their model to seven variables, it has implications for additional macro variables including the price of capital and capital utilization rate. Adding any of these variables to the set of observables will immediately make the model singular. In fact, for all medium-scale DSGE models, nonsingularity arises only because we restrict the estimation to a limited set of macro variables.

For the full model, the OT estimates are similar to the posterior means. In all but two cases, the posterior means fall within the 95\% confidence intervals obtained from OT estimates and robust standard errors. For the remaining two cases, OT produces a more persistent risk premium shock process with a lower residual standard deviation. The robust standard errors are almost always greater than those assuming correct model specification.
 
For singular models, removing the risk premium shock has little effect on the estimates --- they are close to the nonsingular case; all confidence intervals overlap with their nonsingular counterparts. When the wage markup shock is also removed, ${\small \xi }_{w}$ (wage stickiness) and $\iota _{w}$ (wage indexation) decrease, while $\rho _{r}$ (monetary policy shock persistence) increases, though their confidence interval still overlaps due to substantial estimation uncertainty. When further removing the price markup shock, ${\small \xi }_{p}$ (price stickiness) and ${\small \iota }_{p}$ (wage stickiness) both drop noticeably. Although not reported here, the effects of these parameter differences on the model can be further assessed by plotting the impulse response functions.  

\paragraph{Specification Testing.} The specification test does not reject the original model at the 5\% significance level, as shown in Table \ref{tab:SWspec}. However, the test rejects the model's fit for consumption, even with 7 shocks. Further investigation reveals that the model under-predicts contractions, e.g. in 1974Q4 consumption fell by 2.38\% vs. 1.34\% for the fitted values. Fitted consumption is less volatile (standard deviation of 0.53 vs. 0.68), and more persistent (autocorrelation of 0.46 vs. 0.18). For singular models, when the risk premium shock is removed, the specification test rejects the full model at the 5\% significance level; for individual tests, the results for consumption, wage, and interest rate reject the null hypothesis. When the wage markup shock is removed, the tests on output and labor also reject, implying that 5 out of 7 variables are now rejected. Finally, when the price markup shock is removed, the conclusions remain the same as in the five-shock case, with only two variables—investment and inflation—not rejected by the test at the 5\% level.

This is the first attempt to formally test singular DSGE models. The results pinpoint model features that remain compatible or become incompatible with data once shocks are removed. They can be useful tools for researchers to determine which latent processes and mechanisms contribute to the fit of a model within a unified framework.

   \begin{table}[ht]
    \centering
   \caption{SW Model: Specification Testing With(out) Stochastic Singularity} \label{tab:SWspec}
  \centering
  \setlength\tabcolsep{4.0pt}
  \renewcommand{\arraystretch}{0.85} 
  { \small
  \begin{tabular}{l|ccc||ccc||ccc||ccc}
  \hline\hline
  & \multicolumn{3}{c||}{7 shocks} & \multicolumn{3}{c||}{6 shocks} & 
  \multicolumn{3}{c||}{5 shocks} & \multicolumn{3}{c}{4 shocks} \\ 
  & \textsc{stat} & 10\% & 5\% & \textsc{stat} & 10\% & 5\% & \textsc{stat} & 
  10\% & 5\% & \textsc{stat} & 10\% & 5\% \\ \hline
  All & 256.6 & 320.7 & 395.0 & 411.4 & 282.1 & 364.5 & 597.7 & 291.4 & 357.0
  & 632.0 & 288.6  & 353.3 \\ \hline
  Cons. & 60.4 & 31.9 & 36.8 & 56.5 & 32.6 & 39.9 & 192.9 & 6.3 & 8.9 & 192.7
  & 6.1 & 8.5 \\ 
  Invest. & 25.6 & 33.2 & 38.6 & 34.1 & 31.9 & 36.6 & 45.6 & 41.0 & 47.3 & 45.0
  & 41.1 & 47.7 \\ 
  Output & 34.3 & 37.0 & 41.8 & 34.9 & 32.4 & 36.7 & 54.4 & 39.8 & 45.1 & 57.9
  & 39.2 & 44.8 \\ 
  Labor & 16.3 & 35.4 & 47.2 & 14.7 & 48.4 & 66.8 & 86.9 & 49.3 & 63.5 & 97.6
  & 47.8 & 60.0 \\ 
  Infl. & 52.8 & 92.5 & 122.7 & 75.1 & 93.5 & 124.2 & 67.6 & 101.1 & 137.8 & 
  69.6 & 104.1 & 146.1 \\ 
  Wage & 22.6 & 64.8 & 80.3 & 108.3 & 26.2 & 33.3 & 76.4 & 30.9 & 37.6 & 98.3 & 
  27.4 & 33.1 \\ 
  Int. Rate & 44.6 & 47.9 & 63.3 & 87.9 & 33.8 & 45.2 & 73.9 & 49.9 & 67.3 & 
  70.9 & 48.8 & 65.6 \\ \hline\hline
  \end{tabular}%
  \notes{ \textbf{Legend:} All: specification test on all $7$ variables
  (consumption, investment, output, labor, inflation, wage, interest rate).
  \textsc{stat}: test statistic for specification test. $5\%$, $10\%$:
  critical values.} }
  \end{table}
\noindent\textbf{Filtering the Latent Shock Processes.} 
The OTF produces model-consistent values of latent variables, including shock processes. Using the original SW model, we compare these filtered values with their counterparts produced by the KF, which does not enforce model consistency. The same parameter values (OT estimates) are used to ensure comparability.
\begin{figure}[h]
  \centering
  \caption{SW Model: Filtered Shock Processes} \label{fig:SWfilter}
  \includegraphics[scale=0.45]{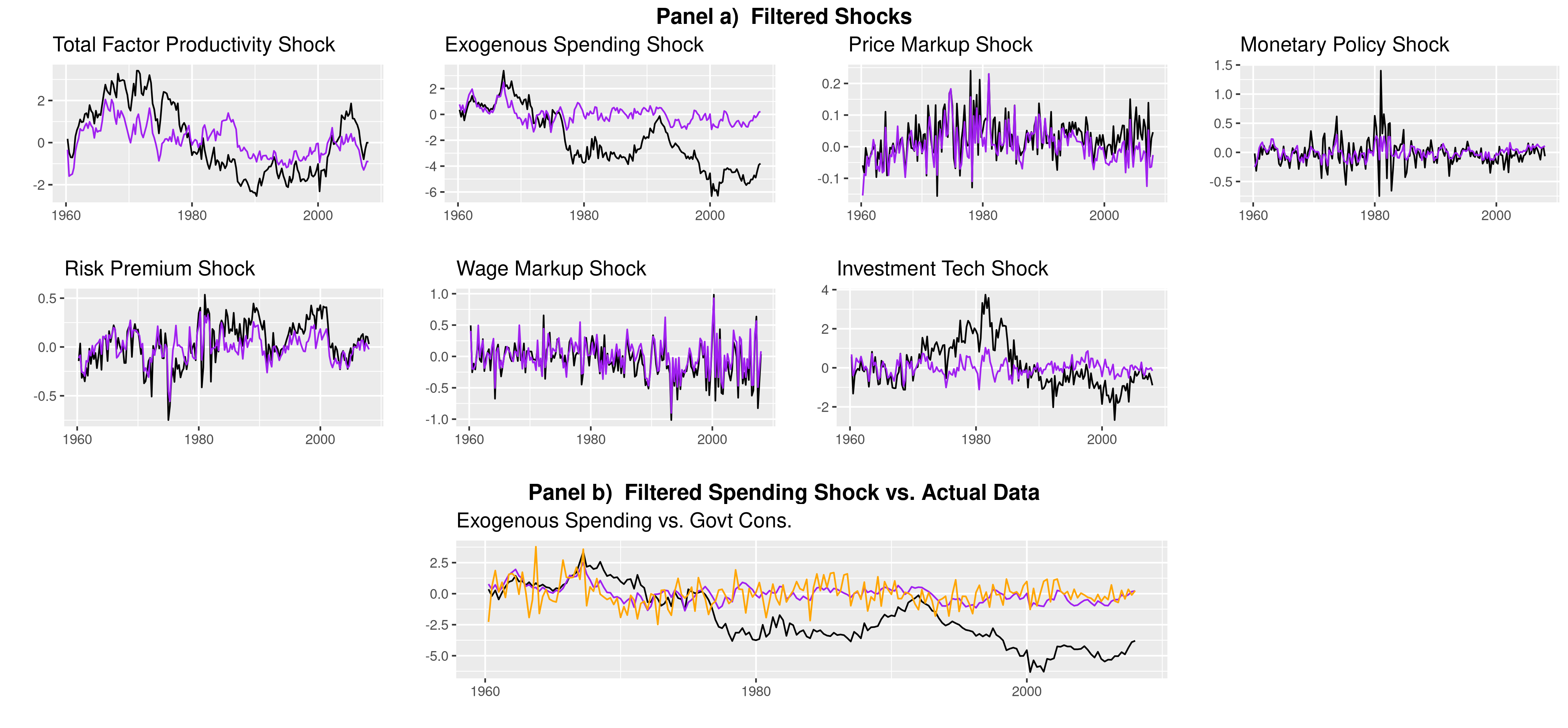}\\
  \notes{ \textbf{Legend:} Black solid line: Kalman Filter (KF), Purple solid line: Optimal Transport Filter (OTF). Both filters are applied using the same OT estimates found in Table \ref{tab:SWest}, Appendix \ref{apx:emp}. Orange line: Real Government Consumption Expenditures and Gross Investment; Source: FRED (GCEC1). }
\end{figure}

Figure \ref{fig:SWfilter}, panel a) displays the results for the 7 shocks separately. For TFP, the KF yields a puzzling conclusion: the economy was under a positive TFP process from 1960 until about 1980, and then affected by a negative TFP process between 1980 and 2000. In contrast, the OT filter produces a process with negative values during the early 1960s, the mid-1970s recessions, and the slowdown leading up to the 2001 recession. These estimates are clearly more interpretable than the KF values. The investment technology shock shows a similar pattern: the KF yields a mostly positive process from the 1970s to the mid-1980s, which then switches to a mostly negative process. The OT estimates do not have this problem.

For exogeneous spending, KF yields negative values most of the time and exhibits a downward trend, in sharp contrast with the zero-mean assumption. The OTF estimates do not have this problem. For the monetary policy shock, KF produces large oscillating values in the early 1980s, which is puzzling given the monetary tightening that characterizes this period. In contrast, the OTF produces mostly positive values for this period, consistent with this characterization. The last $3$ shocks show closer resemblances between KF and OTF.

Table \ref{tab:SWshocks}, Appendix \ref{apx:emp}, presents the cross and serial-correlations of the 7 shock processes. OT consistently gives values close to the true ones, while KF shows significant discrepancies in several cases; for example, the true first-order serial correlations for investment and monetary policy shocks are both 0.47, but KF yields 0.83 and 0.09. This illustrates that, when using KF, the parameter estimates may not capture the dynamics of the filtered values.


As a validation exercise, Figure \ref{fig:SWfilter}, panel b), compares the exogenous spending shocks process with government consumption data. They are closely related since exogeneous spending is the difference between output and the sum of consumption, investment and capital utilization \citep[][p588]{smets2007}. The correlation between the data and filtered values is 0.36 and 0.05 for OTF and KF, respectively. Although this data was \textit{not used} in the estimation, the OTF captures some of its variation using other series and the model. These results demonstrate that the OT filter, by enforcing model consistency, produces filtered values that obey model assumptions and can be more interpretable in practice.

To wrap up, we have considered this medium-scale DSGE model to illustrate that the proposed methods can be used to estimate singular and nonsingular models, testing their specifications, and produce filtered variables, all within the same framework. The methods' applications are not restricted to macroeconomics; we next consider a financial application.

\subsection{Affine Term Structure Model} \label{sec:ATSM}

The third application considers a term structure model where three latent factors explain six yields. Algorithm \ref{alg_2} provides a way to assess the extent to which the unaltered, stochastically singular, structural model fits the empirical data; no measurement errors are introduced.

The specification of the structural model follows \citet{ang2003} and \citet{hamilton2012}. Let $F_{t}$ denote three latent factors, which follow a Gaussian VAR:%
\(
F_{t+1}=c+\rho F_{t}+v_{t+1}, 
\)
with $v_{t+1} \overset{iid}{\sim} \mathcal{N}(0,I)$. 
Affine
term structure models assume that the one-period short rate $r_{t}$ is an
affine function of the state vector. They specify the stochastic
discount factor $M_{t,t+1}$ to be a function of $\lambda_{t}$, the market prices of risk, which is also an affine function of $F_{t}$:
$r_{t}=\delta _{0}+\delta _{1}^{\prime }F_{t}$, $M_{t,t+1}=\exp [-r_{t}-(1/2)\lambda _{t}^{\prime }\lambda _{t}-\lambda_{t}^{\prime }v_{t+1}]$, $\lambda _{t}=\lambda +\Lambda F_{t}$.
As highlighted in \citet{hamilton2012}, under the risk-neutral measure all assets are discounted by the short-term interest rate: $%
P_{t}=\exp (-r_{t})\int P_{t+1}(F_{t+1})\phi (F_{t+1};c^{Q}+\rho
^{Q}F_{t},I)dF_{t+1}$, where $F_{t+1}=c^{Q}+\rho ^{Q}F_{t}+v_{t+1}^{Q}$,
with $v_{t+1}^{Q} \overset{iid}{\sim} \mathcal{N}(0,I)$, $c^{Q}=c-\lambda $, and $\rho ^{Q}=\rho -\Lambda $.

It is well-documented that parameter normalizations are necessary to
identify the model. Following \citet{ang2003} and \citet{hamilton2012}, we set: $c=0,\delta _{1}\geqslant 0$,
and $\rho ^{Q}$ lower triangular. Then, the yield $y_{t}^{n}$ on an $n$-period
pure-discount bond is given by:
\begin{align*}
  y_{t}^{n} &= a_{n}+b_{n}^{\prime }F_{t}, \quad b_{n} =\frac{1}{n}\left( I+\rho ^{Q\prime }+...+\left( \rho ^{Q\prime }\right) ^{n-1} \right) \delta _{1} \\
a_{n} &=\delta _{0}+\frac{1}{n}\left( b_{1}^{\prime }+2b_{2}^{\prime
}+...+(n-1)b_{n-1}^{\prime }\right) c^{Q}-\frac{1}{2n} \left( b_{1}^{\prime
}b_{1}+4b_{2}^{\prime }b_{2}+...+(n-1)^{2}b_{n-1}^{\prime }b_{n-1} \right).
\end{align*}%
The parameters to be estimated are $\delta _{0},\delta _{1},\rho ^{Q},c^{Q}$%
, and $\rho$. The diagonal elements of $\rho ^{Q}$ are required to be in
decreasing order to ensure identification. The auxiliary model is VAR($4$). The yields are weighted diagonally by the inverse of their variance. The OT objective $Q_n$, is penalized by prior distribution which enforces an ordering of the factors.

\begin{table}[ht]
  \caption{Affine Term Structure Model: Parameter Estimates} \label{tab:ATSMest}
  \centering
  \setlength\tabcolsep{4.0pt}
    \renewcommand{\arraystretch}{0.85} 
  { \small
\begin{tabular}{c|ccc|ccc} \hline \hline
&  \multicolumn{3}{c|}{Prior [\textsc{mean}, \textsc{sd}]} & \multicolumn{3}{c}{
OT Estimates} \\ \hline
\multirow{3}{*}{$\rho ^{Q}$} 
& [0.9,0.2] & [0,0.2] & - & 0.999 & - & - \\ 
& [0,0.2] & [0.8,0.2] & - & 0.022 & 0.963 & - \\ 
& [0,0.2] & [0,0.2] & [0.6,0.2] & 0.018 & 0.209 & 0.723 \\  \hline
$\delta _{1}$ 
& [0.01,0.1] & [0.01,0.1] & [0.01,0.1] & 0.003 & 0.023 & 0.037 \\ \hline
$\delta _{0}$
& [0.4,0.2] & - & - & 0.441 & - & - \\ \hline
$c^{Q}$
& [0,1] & [0,1] & [0,1] & 1.254 & -0.026 & 0.536 \\ \hline
\multirow{3}{*}{$\rho$} 
& [0.9,0.2] & [0,0.2] & [0,0.2] & 0.958 & 0.012 & 0.071 \\ 
& [0,0.2] & [0.8,0.2] & [0,0.2] & 0.004 & 0.908 & 0.102 \\ 
& [0,0.2] & [0,0.2] &   [0.6,0.2] & 0.011 & 0.133 & 0.770 \\ \hline \hline
\end{tabular}\notes{ \textbf{Legend:} A weakly informative prior is used to enforce the ordering of the factors. } }  
\end{table}

The estimates in Table \ref{tab:ATSMest} are similar to \citet[Table 5]{hamilton2012}.\footnote{\citet{hamilton2012} use four yields and introduce measurement error in one of the series.}
The first factor is very persistent, and the off-diagonal elements of $\rho$ are
small in magnitude. The estimate of $\delta_{0}$ is close to the mean of the short-term rate. The main difference is that the first element of $\delta _{1}$ in our case is smaller than theirs (their estimate is 0.017).
The parameters $\rho^{Q},\delta _{1}$ and $c^{Q}$ are highly correlated: it is possible to move their values jointly with little effect on $Q_n$, suggesting weak identification.For this reason standard errors are not reported. The specification test is not reported either. The prior helps stabilize the estimates.

\begin{figure}[ht] \caption{Affine Term Structure Model: Actual and Model-Consistent Yields}
  \centering
  \includegraphics[scale=0.45]{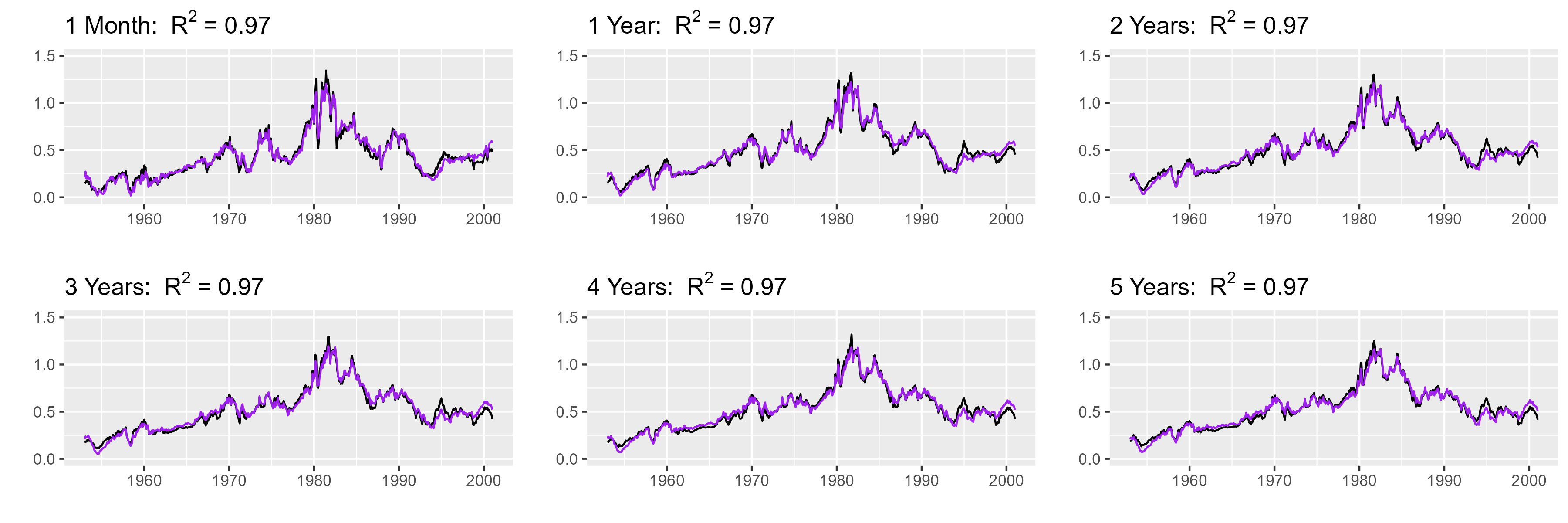}  
  \notes{\textbf{Legend:} Actual yields: Black solid line, Coupling: Purple solid line. Fit shown for prior regularized estimates; the fit is virtually identical without prior regularization.}
  \label{fig:ATSM}
  \end{figure}

What is intriguing is whether, with three shocks, the model can approximate
the dynamics of six observables. Figure \ref{fig:ATSM} compares actual yield data with their couplings. These values track each other closely, with the $R^{2}$ equal to 0.97 in all six cases. 

With three shocks, the model accounts for 97\% of the dynamics in the data for this sample period. The plots suggest no apparent model misspecification for this sample period. Adding measurement errors in this case would be a shortcut to obtaining parameter estimates with a likelihood and cannot reveal additional information regarding the baseline model specification. Our approach allows us to obtain parameter estimates with a graphical method to assess the model fit. Finally, we conjecture that incorporating more information to improve the identification of $\rho^{Q},\delta _{1}$ and $c^{Q}$ can be beneficial for further improve this
model.

\subsection{Trend-Cycle Decomposition} \label{sec:TrendCycle}

The last application illustrates how the methodology can be applied in a setting where the stationarity assumptions used in the theoretical analysis are not satisfied. Trend-cycle decompositions are routinely used to date business cycles for many countries. One approach is to model the log real GDP using an unobserved component model as in \citet{watson1986}:
\[ \underbrace{y_t = \tau_t + c_t}_{\text{ $\log$(GDP) }}, \text{ where } \underbrace{\tau_t = \mu + \tau_{t-1} + \eta_t}_{\text{trend component}} \text{ and } \underbrace{c_t = \rho_1 c_{t-1} + \rho_2 c_{t-2} + e_t}_{\text{cycle component}}, \] 
where $(\eta_t,e_t) \overset{iid}{\sim} \mathcal{N}(0, \text{diag}(\sigma_{\eta}^2,\sigma_{e}^2) )$. 
While this DGP is non-stationary, and thus not covered by the parameter estimation results above, it illustrates that the methodology applies to a broader set of filtering problems.

\begin{figure}[H] \caption{U.S. GDP -- 1947Q1-2023Q4 --  Trend-Cycle Decomposition} \label{fig:watson86}
  \centering
  \includegraphics[scale=0.5]{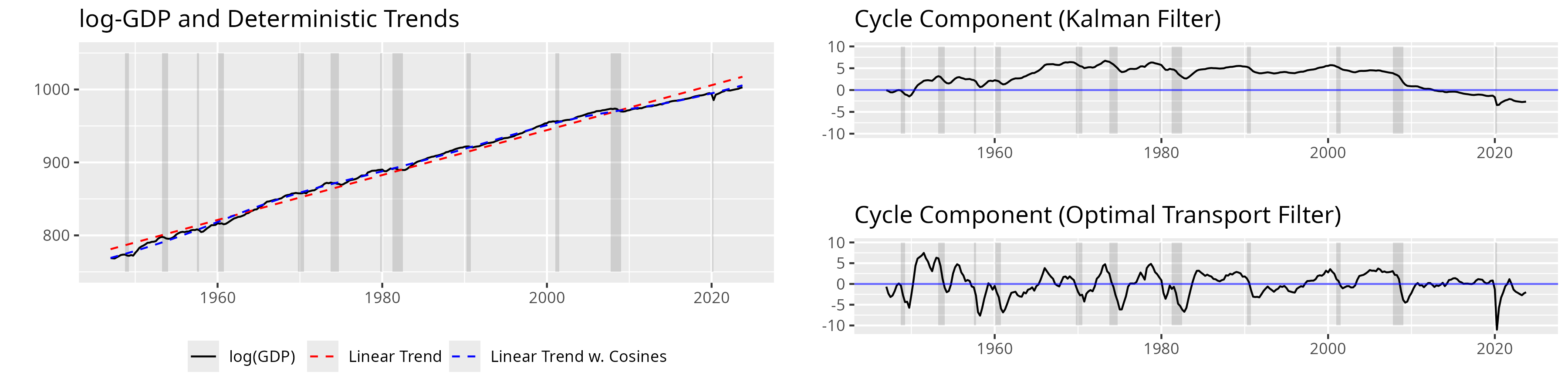}\vspace{-0.3cm}
  \notes{ \textbf{Legend:} Blue dashed line: linear trend w. cosines $\delta_0 + \delta_1 t + \delta_2 \cos( 2 \pi t/n ) + \delta_3 \cos( 2 \pi 2 t/n ) + \delta_4 \cos( 2 \pi 3 t/n ) + \delta_5 \cos( 2 \pi 4 t/n )$. Red dashed line: Kalman Filter estimates of $\tau_t$. Vertical bars: NBER recession dates. }
\end{figure}

Figure \ref{fig:watson86} displays U.S. log-GDP between 1947 and 2023 (left panel) and the extracted cycle components (right panels). The left panel indicates that the KF (red dashed line) does not fully capture changes in the trend growth rate over the three-quarters of a century spanned by the data. In particular, the estimated trend is systematically below the log-GDP between 1965 and 2008. This issue was already raised by \citet{perron2009}. A flexible trend estimate (blue dashed line), described below the figure, better captures the gradual changes in the trend component. The number of cosine terms is chosen to have the shortest periodicity at 20 years, larger than the business cycle frequency, i.e. 1.5 to 8 years.

The top right panel reports the KF estimates for the cycle component. The KF estimated trend is systematically below GDP throughout the period 1965-2008 (left panel). Likewise, the filtered cycle component (top right panel) is systematically positive during this period. This would suggest that the economy is characterized by a continuous expansion over this forty-year period. Clearly, the cycle estimates are contaminated by misspecification in the trend. 
Finally, the bottom right panel shows the OTF values, computed using the flexible trend as a basis for the reduced form model. See Appendix \ref{apx:motivating} for details. No terms were added to handle the Covid recession. Unlike KF estimates, it marks turning points (vertical bars) with good accuracy and visually appears to be stationary with zero mean. 


\section{Extension: Non-Linear State-Space Models} \label{sec:extension}

The following provides a general-purpose Algorithm to perform Optimal Transport Filtering for non-linear non-Gaussian state-space models. It is similar to Algorithm \ref{algo:KOTF} with the difference that the \textbf{Predict}, \textbf{Transport}, and \textbf{Update} steps accomodate more general state-space models (\ref{eq:ssmodel}). It is not closed form and more computationally demanding than Algorithm \ref{algo:KOTF}.

\begin{algorithm}[H]
  \caption{Optimal Transport Filter}\label{algo:OTF} { \small
  \begin{algorithmic}[1]
      \Procedure{\textsc{otf}}{}\newline
      \textbf{Inputs:} 1) Sample: $\tilde{y}_1,\dots,\tilde{y}_n$, predictive distribution $\tilde{p}(\tilde{y}_t|\tilde{y}_{t-1},\dots)$\newline\hphantom{\textbf{Inputs:}} 2) Model: conditional distribution $p(y_t,z_t|z_{t-1};\theta)$, initial beliefs $z_0 \sim p_{0|0}(z_0)$\newline
      \textbf{Outputs:} 1) Mapped data $y_1,\dots,y_n$, 2) Filtered states $z_{t|t} \sim p( z_t|y_t,\dots,y_1 )$.
      \For{$t \in \{1,\dots,n\}$} 
      \State{\textbf{Predict:} Using the model, compute} \Comment{(Filter)}\newline
      \hphantom{\textbf{Inputs:}} $p_{t|t-1}(y_t,z_t) = \int p(y_t,z_t|z_{t-1})p_{t-1|t-1}(z_{t-1})dz_{t-1}$  
      \State{\textbf{Transport:} Find $\pi_{t|t-1}$ which solves $\min_{\pi \in \Pi_{t|t-1}} \mathbb{E}_{\pi}(\|y_t-\tilde{y}_t\|^2)$ }  \Comment{(OT)} \newline
      \hphantom{\textbf{Inputs:}} where $\Pi_{t|t-1} = \{ \pi(y_t,\tilde{y}_t) \text{ s.t. } \int \pi(y_t,\tilde{y}_t)d\tilde{y}_t = p_{t|t-1}(y_t), \int \pi(y_t,\tilde{y}_t)dy_t = \tilde{p}_{t|t-1}(\tilde{y}_t) \}$
      \State{\textbf{Update:} $p_{t|t}(z_t) \propto p_{t|t-1}(z_t,y_t)/\pi_{t|t-1}(y_t|\tilde{y}_t)$ } \Comment{(Filter)}\newline
      \hphantom{\textbf{Inputs:}} using the distributions $p_{t|t-1}$ from step 3 and $\pi_{t|t-1}$ from step 4.
      \EndFor
      \EndProcedure
  \end{algorithmic}}
  \label{alg_1}
\end{algorithm}

The following briefly discusses the main steps of the Algorithm. An estimate of $\tilde{p}_{t|t-1}$ is required as inputs, as for linear state-space models. More details on the filtering (Filter) and optimal transport (OT) steps can be found in Algorithm \ref{alg_3} in Appendix \ref{apx:Extra_Algo}. 

\noindent \textbf{Predict.} This step can be carried out using  simulations as in the particle filter algorithm. For discrete or discretized models, these can be carried out using only matrix operations \citep[see e.g.][]{chopin2020}. This produces draws for Transport step below. 

\noindent \textbf{Transport.} When $\tilde{p}_{t|t-1}$ and $p_{t|t-1}$ are non-Gaussian, the transport still has a closed form when both $\tilde{y}_t$ and $y_t$ are scalar: $y_t = F_{t|t-1}^{-1} \circ \tilde{F}_{t|t-1}(\tilde{y}_{t|t-1})$. For multivariate outcomes, the solution is generally not closed form and needs to be computed numerically. From a finite number of draws $\tilde{y}_t^b \sim \tilde{p}_{t|t-1}$ and $y_t^b \sim p_{t|t-1}$, $b=1,\dots,B$, the optimal transport problem can be solved exactly as a linear program or approximated using entropic regularization and Sinkhorn’s algorithm. The latter is implemented in Algorithm \ref{alg_3}.

\noindent \textbf{Update.} As in Algorithm \ref{algo:KOTF}, the update step is standard but applied to $y_t$ instead of $\tilde{y}_t$ to ensure the filtered variables are model-consistent.

\section{Conclusion} 
This paper has introduced a computationally attractive method for filtering and estimation of potentially misspecified dynamic models using dynamic optimal transportation. Empirical applications illustrate how this can be used to visually assess the fit of a model, by comparing the actual and the coupled time-series, and to formally test the model specification over all or some specific variables. Several extensions could be of interest in future research. Deriving a plugin map for general non-linear state-space models in (\ref{eq:ssmodel}) could be useful to circumvent the curse of dimensionality. A useful Corollary to Theorem \ref{th:spec} would be to consider an Anderson-Rubin type statistic for models that are potentially weakly identified.

\printbibliography[heading=subbibliography]

\end{refsection}
\begin{appendices}
  \renewcommand\thetable{\thesection\arabic{table}}
  \renewcommand\thefigure{\thesection\arabic{figure}}
  \renewcommand{\theequation}{\thesection.\arabic{equation}}
  \renewcommand\thelemma{\thesection\arabic{lemma}}
  \renewcommand\thetheorem{\thesection\arabic{theorem}}
  \renewcommand\thedefinition{\thesection\arabic{definition}}
    \renewcommand\theassumption{\thesection\arabic{assumption}}
  \renewcommand\theproposition{\thesection\arabic{proposition}}
    \renewcommand\theremark{\thesection\arabic{remark}}
    \renewcommand\thecorollary{\thesection\arabic{corollary}}
\setcounter{equation}{0}
\setcounter{lemma}{0}
\clearpage \baselineskip=18.0pt
\appendix

\makeatletter
\@addtoreset{assumption}{section}
\makeatother

\begin{refsection}

\section{Definitions} \label{apx:definitions}
The following recalls two notions of dependence, further details can be found in \citet[Ch14,15,18]{Davidson2021}.
Take the sequence $e_t$ found in Assumptions \ref{ass:Causal} and \ref{ass:DGPdata}. Let $\mathcal{F}_{t} = \sigma(e_t,e_{t-1},\dots)$ be the sigma-algebra constructed on $e_t,e_{t-1},\dots$. Similarly, let $\mathcal{F}_{t-m}^{t+m} = \sigma(e_{t+m},e_{t+m-1},\dots,e_{t-m+1},e_{t-m})$. For two sub sigma-algebras $\mathcal{G},\mathcal{H}$, their strong-mixing coefficient is defined as $\alpha(\mathcal{G},\mathcal{H}) = \sup_{ G \in \mathcal{G}, H \in \mathcal{H} }  | \mathbb{P}(G \cap H) - \mathbb{P}(G)\mathbb{P}(H) |$.
\begin{definition}[Strong-mixing] \label{def:alpha_mixing} The strictly stationary sequence $e_t$ is said to be $\alpha$-mixing, i.e. strong mixing, if the $\alpha$-mixing coefficients $\alpha_{m} = \alpha(\mathcal{F}_t,\mathcal{F}_{t-m})$ satisfy $\alpha_m \to 0$ as $m \to \infty$.   
\end{definition}

\begin{definition}[Near-epoch dependence] \label{def:NED} The sequence $y_t$ is said to be near-epoch dependent (NED) in $L_p$-norm on $\{e_t\}_{t=-\infty}^{+\infty}$, for $p>0$ if: $(\mathbb{E}[ \|y_t - \mathbb{E}(y_t|\mathcal{F}_{t-m}^{t+m})\|^p ])^{1/p} \leq d_t \nu_m$,
where $\nu_m \to 0$ as $m \to \infty$, $d_t$ is a sequence of positive constants, and $\mathcal{F}_{t-m}^{t+m} = \sigma(e_{t+m},\dots,e_{t-m})$.   
\end{definition}

\section{Preliminary Results} \label{apx:prelim}

\begin{lemma}[Dependence] \label{lem:NED} Suppose Assumptions \ref{ass:Causal}, \ref{ass:DGPdata}, \ref{ass:InnovModel}, and \ref{ass:NED} hold. Let $y_t(\theta;\psi_{k0}) = \mu(\theta) + \sum_{\ell=0}^\infty \Lambda_\ell(\theta)P(\theta;\tilde{\Sigma})e_{t-\ell}$, computed from $(e_{t-\ell})_{\ell \geq 0}$ instead of $(e_{t-\ell,k})_{t > \ell \geq 0}$ for $y_t(\theta;\psi_{k})$, and $\phi_{j} = \text{vec}(\Psi_{j})$. Then: (1) For all $\theta \in \Theta$, the sequences $\tilde{y}_t$, $y_t(\theta;\psi_0)$, $y_t(\theta;\psi_k)$, $\partial_{\theta} y_t(\theta;\psi_0)$, and $\partial_{\theta} y_t(\theta;\psi_k)$ are NED in $L_{q}$-norm of size $-b$ on $e_t$ for some $q \leq 2r$, with $r$ defined in Assumption \ref{ass:DGPdata} (iii). (2) For all $\theta$ and any $j \geq 1$, $\partial_{ \phi_j } y_t(\theta;\psi_{k0})$ is NED in $L_{q}$-norm of size $-b$; $\partial_{\tilde{\mu}} y_t(\theta;\psi_{k0})$ is constant and deterministic; and $\partial_{\text{vec}(\tilde{\Sigma})} y_t(\theta;\psi_{k0})$ is NED in $L_q$-norm with size $-b$ if $\tilde{\Sigma}$ is invertible and $\text{rank}[\Sigma(\theta)]$ is constant.
  \end{lemma}

The partial derivatives in the Lemma are expressed as follows and will be used in subsequent proofs: $\partial_{ \phi_j } y_t(\theta;\psi_{k0}) =  - \sum_{\ell=0}^\infty (\tilde{y}_{t-\ell - j -1} - \tilde{\mu})^\prime \otimes [\Lambda_\ell(\theta)P(\theta;\tilde{\Sigma})]$, $\partial_{\tilde{\mu}} y_t(\theta;\psi_{k0}) = I+( \sum_{\ell=0}^\infty \Lambda_\ell(\theta)P(\theta;\tilde{\Sigma}) ) ( I_d - \sum_{\ell=1}^\infty \Psi_\ell )$, and $\partial_{\text{vec}(\tilde{\Sigma})} y_t(\theta;\psi_{k0}) =  \sum_{\ell=0}^\infty (e_{t-\ell}^\prime \otimes \Lambda_\ell(\theta)) \partial_{\text{vec}(\tilde{\Sigma})}\text{vec}[P(\theta;\tilde{\Sigma})]$.

\section{Proofs for the Main Results} \label{apx:proofs}

\paragraph{Proof of Theorem \ref{th:cons}.} We have:
$Q_n(\theta;\hat{\psi}_{nk}) = \frac{1}{n} \sum_{t=1}^n \|y_{t}(\theta;\hat{\psi}_{nk}) - \tilde{y}_t\|^2_{W_n}$. 
Assumption \ref{ass:InnovModel} implies $Q(\cdot;\psi_0)$ is continous on $\Theta$. Given the identification assumption, it is sufficient to derive uniform equivalence and uniform law of numbers, stated as follows:
\[ \sup_{\theta \in \Theta} |Q_n(\theta;\hat{\psi}_{nk}) - Q_n(\theta;\psi_{0})| = o_p(1), \quad \sup_{\theta \in \Theta} |Q(\theta;\psi_{0}) - Q_n(\theta;\psi_{0})| = o_p(1).\]

\noindent\textbf{Step 1. Uniform Equivalence: $\sup_{\theta \in \Theta} |Q_n(\theta;\hat{\psi}_{nk}) - Q_n(\theta;\psi_{0})| = o_p(1)$.}\leavevmode\\
We first establish some properties related to upper bounds and stochastic orders to be used in the proof. By Lemma \ref{lem:VARinf}, we have $\sup_{j=1,\dots,k}\|\hat{\Psi}_j - \Psi_j\| = O_p(\sqrt{\log(n)/n})$, $\|\tilde{\mu}_n - \tilde{\mu}\| = O_p(n^{-1/2})$, and $\|\tilde{\Sigma}_{nk} - \tilde{\Sigma}\| = O_p(n^{-1/2})$. 
Let $e_{t,k} = \tilde{y}_t - \tilde{\mu} - \sum_{j=1}^k \Psi_j (\tilde{y}_{t-j-1}-\tilde{\mu})$, $\hat{e}_{t} = \tilde{y}_t - \tilde{\mu}_n - \sum_{j=1}^k \hat{\Psi}_j (\tilde{y}_{t-j-1}-\tilde{\mu}_n)$. For $t-j-1 \leq 0$, set $\tilde{y}_{t-j-1}-\tilde{\mu} =0$ and $\tilde{y}_{t-j-1}-\tilde{\mu}_n = 0$. This does not affect the properties of the VAR estimates since $k/n = o(n^{-1/2})$, by assumption.

For $q = 2r$, $\mathbb{E}(\|e_t\|^q) < \infty$ and Assumption \ref{ass:DGPdata} imply $\mathbb{E}(\|\tilde{y}_t\|^q) < \infty$. Thus, $\sup_{t=1,\dots,n} \|\tilde{y}_t\| = O_p(n^{1/q})$ \citep[Lem2.2.2]{vdvW1996}. Likewise, $\sup_{t=1,\dots,n} \|\hat{e}_t - e_{t,k}\|^2 \leq (k+1) \max ( \|\hat{\Psi}_1 - \Psi_1\|^2,\dots,\|\hat{\Psi}_k - \Psi_k\|^2,\|\tilde{\mu}_{n}-\tilde{\mu}\|^2 ) \sup_{t=1,\dots,n} \|\tilde{y}_{t}\|^2 = O_p( k \log(n) n^{2/q - 1} )$ and
$\sup_{t=1,\dots,n} \|\hat{e}_t - e_{t,k}\| = O_p( \sqrt{k \log(n)}n^{1/q - 1/2}) = o_p(1)$, 
since $q > 8$ and $k = o(n^{1/3})$. Note that $[\mathbb{E}(\|e_{t,k} - e_t\|^q)]^{1/q} \leq [\sum_{j = k+1}^\infty \|\Psi_j\|] [\mathbb{E}(\|\tilde{y}_t-\tilde{\mu}\|^q)]^{1/q} = O(n^{-1/2})$, for $t \geq k+1$, by assumption. This implies that $\sup_{t=k+1,\dots,n} \|e_{t,k} - e_t\| = O_p(n^{1/q - 1/2}) = o_p(1)$.
Also, $\mathbb{E}( \|e_{t,k}-e_t\|^q ) < \infty$ implies $\sup_{t=1,\dots,k} \|e_{t,k}-e_t\| = O_p(k^{1/q})$.\footnote{Simply use $[\mathbb{E}( \|e_{t,k}-e_t\|^q )]^{1/q} \leq [\mathbb{E}( \|e_{t,k}\|^q )]^{1/q} + [\mathbb{E}( \|e_t\|^q )]^{1/q}$ and $\sup_{t=1,\dots,k} \|e_{t,k}-e_t\| \leq O_p(k^{1/q})$ by \citet[Lem2.2.2]{vdvW1996}.}

Also, $\tilde{\Sigma}_{nk} = \tilde{\Sigma} + O_p(n^{-1/2})$ implies $0 < \underline{\lambda}/2 \preceq \tilde{\Sigma}_{nk} \preceq  2\overline{\lambda} < \infty$ with probability approaching one. Because $\theta \to \Sigma(\theta)$ has constant rank, $(\theta,\Sigma) \to P(\theta;\Sigma)$ is continuously differentiable with bounded derivative on $\Theta \times \{ \Sigma \text{ s.t. } \underline{\lambda}/2 \preceq \Sigma \preceq  2\overline{\lambda} \}$ by Lemma \ref{lem:ConstantRank}. Then, for any $\theta$ and any $\Sigma_1$ and $\Sigma_2$ in this set: $\|P(\theta;\Sigma_1)-P(\theta;\Sigma_2)\| \leq \sup_{\theta\in\Theta,\underline{\lambda}/2 \preceq \Sigma \preceq  2\overline{\lambda}}\|\partial_{\text{vec}(\Sigma)}\text{vec}[P(\theta;\Sigma)]\|_{\infty} \|\Sigma_1-\Sigma_2\|$. This implies $\sup_{\theta \in \Theta} \|P(\theta;\tilde{\Sigma}_{nk}) - P(\theta;\tilde{\Sigma})\| = O_p(n^{-1/2})$.

Now, we apply these bounds to study $y_t(\theta;\hat{\psi}_{nk}) - y_t(\theta;\psi_{0})$. Recall that $y_t(\theta;\hat{\psi}_{nk}) = \mu(\theta)+\sum_{j=0}^\infty \Lambda_j(\theta)P(\theta;\tilde{\Sigma}_{nk}) \hat{e}_{t-j}$ with $\hat{e}_t=0$ for $t \leq 0$; $y_t(\theta;\psi_{k}) = \mu(\theta)+\sum_{j=0}^\infty \Lambda_j(\theta)P(\theta;\tilde{\Sigma}) e_{t-j,k}$ with $e_{t-j,k} = 0$ for $t \leq 0$; and $y_t(\theta;\psi_{0}) =\mu(\theta)+ \sum_{j=0}^\infty \Lambda_j(\theta)P(\theta;\tilde{\Sigma}) e_{t-j}$. We have
\begin{align}
  &\sup_{\theta \in \Theta} \sup_{t=1,\dots,n} \|y_t(\theta;\hat{\psi}_{nk}) - y_t(\theta;\psi_{0})\| \notag\\ &\leq \sup_{\theta \in \Theta}\|P(\theta;\tilde{\Sigma}_{nk})\|_{op}\left( \sum_{j=0}^{\infty} \sup_{\theta \in \Theta} \|\Lambda_j(\theta)\|_{op} \right)\sup_{t=1,\dots,n}\|\hat{e}_t-e_{t,k}\| \tag{A1}\label{eq:A*}\\& + \left( \sum_{j=0}^{\infty} \sup_{\theta \in \Theta} \|\Lambda_j(\theta)\|_{op} \right) \sup_{t=1,\dots,n}\|e_{t,k}\| \sup_{\theta \in \Theta}\|P(\theta;\tilde{\Sigma}_{nk}) - P(\theta;\tilde{\Sigma})\| \tag{B1}\label{eq:B*}\\ &+ \sup_{\theta \in \Theta}\|P(\theta;\tilde{\Sigma})\|_{op}\left( \sum_{j=0}^{\infty} \sup_{\theta \in \Theta} \|\Lambda_j(\theta)\|_{op} \right)\sup_{t=1,\dots,n}\|e_t-e_{t,k}\| \tag{C1}\label{eq:C*}\\ &+  \sup_{t=1,\dots,n} \sup_{\theta \in \Theta} \|\sum_{j = k+1}^\infty \Lambda_j(\theta;\tilde{\Sigma})P(\theta;\tilde{\Sigma})e_{t-j}\|. \tag{D1}\label{eq:D*}
\end{align}
From the above and $\sup_{t=1,\dots,n}\|e_{t,k}\| \leq  \sup_{t=1,\dots,n}\|e_{t,k} -e_t\| + \sup_{t=1,\dots,n}\|e_t\|$:
\begin{alignat*}{2}
 &(\ref{eq:A*}) \leq O_p( \sqrt{k \log(n)}n^{1/q - 1/2}),  
 &&(\ref{eq:B*}) \leq O_p(n^{1/q-1/2}),\\
 &(\ref{eq:C*}) \leq \mathbbm{1}_{t \geq k + 1}O_p(n^{1/q-1/2}) + \mathbbm{1}_{t \leq k}O_p(k^{1/q}), \quad
 &&(\ref{eq:D*}) \leq O_p(n^{1/q-1/2}),
\end{alignat*}
using $\sqrt{n}\sum_{j=k+1} \sup_{\theta \in \Theta} \|\Lambda_j(\theta)\|_{op} [\mathbb{E}(\|e_t\|^q)]^{1/q} = o(1)$ for the last inequality. Together, these imply: $\sup_{\theta \in \Theta} \sup_{t=k+1,\dots,n} \|y_t(\theta;\hat{\psi}_{nk}) - y_t(\theta;\psi_{0})\| \leq O_p( \sqrt{k \log(n)}n^{1/q - 1/2})$, which is $o_p(1)$ when $k^3/n = o(1)$ and $q>8$. Also, $\sup_{\theta \in \Theta} \sup_{t=1,\dots,k} \|y_t(\theta;\hat{\psi}_{nk}) - y_t(\theta;\psi_{0})\| \leq O_p(k^{1/q})$. 

We apply the above results to evaluate $Q_n(\theta;\hat{\psi}_{nk}) - Q_n(\theta;\psi_{0})$. Let $\langle y,\tilde{y}\rangle_{W_n} = y^\prime W_n \tilde{y}$, we have: $Q_n(\theta;\hat{\psi}_{nk}) = Q_n(\theta;\psi_{0}) + (2/n) \sum_{t=1}^n \langle y_t(\theta;\hat{\psi}_{nk}) - y_t(\theta;\psi_0),  \tilde{y}_t - y_t(\theta;\psi_0) \rangle_{W_n} + (2/n) \sum_{t=1}^n \|y_t(\theta;\hat{\psi}_{nk})-y_t(\theta;\psi_0)\|^2_{W_n}$. By the Cauchy–Schwarz inequality and $\overline{\lambda}_{n,W} = \lambda_{\max}(W_n)$:
\begin{align*} &\sup_{\theta \in \Theta}|Q_n(\theta;\hat{\psi}_{nk}) - Q_n(\theta;\psi_{0})| \\ &\leq 2 \overline{\lambda}_{n,W} \{\sup_{t=k+1,\dots,n} \sup_{\theta \in \Theta} \|y_t(\theta;\hat{\psi}_{nk}) - y_t(\theta;\psi_0)\| + O_p(k^{1+1/q}/n)\} \sup_{\theta \in \Theta} (1/n)\sum_{t=1}^n \|\tilde{y}_t - y_t(\theta;\psi_0)\| \\ &+ 2 \overline{\lambda}_{n,W}\{\sup_{t=1,\dots,n} \sup_{\theta \in \Theta} \|y_t(\theta;\hat{\psi}_{nk}) - y_t(\theta;\psi_0)\|\}^2.\end{align*}
Of the three terms after the inequality: the last term is $o_p(1)$ since $W_n \overset{p}{\to} W$, with $W$ finite, which implies $\overline{\lambda}_{n,W} = O_p(1)$. The second term is $o_p(1)$ if $\sup_{\theta \in \Theta} 1/n\sum_{t=1}^n \|\tilde{y}_t - y_t(\theta;\psi_0)\| =O_p(1)$. This is the case because $\sum_{j=0}^\infty [\|\tilde{\Lambda}_j\|_{op} + \sup_{\theta \in \Theta} \|\Lambda_j(\theta)\|_{op} \sup_{\theta \in \Theta}\|P(\theta;\tilde{\Sigma})\|_{op}]\mathbb{E}(\|e_{t-j}\|) < \infty$. The first term is $o_p(1)$ because $k^{1+1/q}/n = o(k^3/n) = o(1)$, using $\sum_{t=1}^{k}  \sup_{\theta \in \Theta} \|y_t(\theta;\hat{\psi}_{nk}) - y_t(\theta;\psi_0)\| \leq k \sup_{t=1,\dots,k} \sup_{\theta \in \Theta} \|y_t(\theta;\hat{\psi}_{nk}) - y_t(\theta;\psi_0)\| \leq O_p(k^{1+1/q})$. Altogether, this implies the result:
$\sup_{\theta \in \Theta} |Q_n(\theta;\hat{\psi}_{nk}) - Q_n(\theta;\psi_{0})| = o_p(1)$.

\noindent\textbf{Step 2. Uniform Law of Large Numbers: $\sup_{\theta \in \Theta} |Q(\theta;\psi_{0}) - Q_n(\theta;\psi_{0})| = o_p(1)$}. \leavevmode\\
Using similar arguments as above:
$|Q_n(\theta;\psi_0) - \frac{1}{n} \sum_{t=1}^n \|\tilde{y}_t-y_t(\theta;\psi_0)\|_W^2| \leq \|W_n-W\|_{op} O_p(1) = o_p(1)$. Lemma \ref{lem:NED} implies that, for each $\theta \in \Theta$, $\tilde{y}_t-y_t(\theta;\psi_0)$ is  near-epoch dependent (NED) in $L_q$-norm with size $-b$ for $b>2$. Theorems 18.8, 18.9 in \citet{Davidson2021} imply that $\|\tilde{y}_t-y_t(\theta;\psi_0)\|_W^2$ is NED in $L_{q/2}$-norm with size $-b$. Thus, using Assumption \ref{ass:DGPdata} and \citet[Th18.6]{Davidson2021}, it is an $L_2$-mixingale of size $-\min(b, r/(r-2)[1/2 - 2/q] )$ and a weak law of large numbers applies: $\frac{1}{n} \sum_{t=1}^n \|\tilde{y}_t-y_t(\theta;\psi_0)\|_W^2 \overset{p}{\to} \lim_{n\to\infty} \frac{1}{n} \sum_{t=1}^n \mathbb{E}[|\tilde{y}_t-y_t(\theta;\psi_0)\|_W^2] = Q(\theta;\psi_0)$, which holds pointwise in $\theta$. 
Take any two $\theta_1,\theta_2 \in \Theta$, apply the mean-value theorem:
\[ Q_n(\theta_1;\psi_0) - Q_n(\theta_2;\psi_0) = -(2/n) \sum\nolimits_{t=1}^n (\tilde{y}_t - y_t(\tilde{\theta};\psi_0))^\prime W \partial_\theta y_t(\tilde{\theta};\psi_0)[\theta_1 - \theta_2] + o_p(1),  \]
where the $o_p(1)$ is due to $||W_n-W||=o_p(1)$ and is uniform in $\theta$ as above. We have: \[\mathbb{E}[ \sup_{\theta \in \Theta} \|\tilde{y}_t - y_t(\tilde{\theta};\psi_0))^\prime W \partial_\theta y_t(\tilde{\theta};\psi_0)\|_W ] \leq (\mathbb{E}[ \sup_{\theta \in \Theta} \|\tilde{y}_t - y_t(\theta;\psi_0)\|_W^2])^{1/2} (\mathbb{E}[ \sup_{\theta \in \Theta} \|\partial_{\theta}y_t(\theta;\psi_0)\|_W^2])^{1/2},\] which is bounded under Assumption \ref{ass:InnovModel}. Then, uniformly in $\theta_1,\theta_2$: $|Q_n(\theta_1;\psi_0) - Q_n(\theta_2;\psi_0)| \leq O_p(1)\|\theta_1-\theta_2\| + o_p(1)$. This implies stochastic equicontinuity and the uniform langle. 

\noindent\textbf{Step 3. Consistency.} 
The objective $Q(\cdot;\psi_0)$ is continuous is $\theta$ and uniquely minimized at $\theta = \theta_0$. $Q_n(\cdot;\hat{\psi}_{nk})$ converges uniformly to $Q(\cdot;\psi_0)$ in probability. By standard arguments, this implies consistency: $\hat{\theta}_n \overset{p}{\to} \theta_0$. \qed

\paragraph{Proof of Theorem \ref{th:asym_normal}.} 
By minimization, $\theta_0$ and $\hat{\theta}_n$ satisfy: 
$\mathbb{E}\left( \partial_{\theta} y_t(\theta_0;\psi_0)^\prime W [\tilde{y}_t - y_t(\theta_0;\psi_0)] \right) = 0$ and $
  (1/n) \sum\nolimits_{t=1}^n \partial_{\theta} y_t(\hat{\theta}_n;\hat{\psi}_{nk})^\prime W_n [\tilde{y}_t - y_t(\hat{\theta}_n;\hat{\psi}_{nk})] = 0.$
The score can be decomposed into:
\begin{align} 0 &= (1/n) \sum\nolimits_{t=1}^n \partial_{\theta} y_t(\hat{\theta}_n;\hat{\psi}_{nk})^\prime W_n [\tilde{y}_t - y_t(\hat{\theta}_n;\hat{\psi}_{nk})] \notag\\
  &= (1/n) \sum\nolimits_{t=1}^n \partial_\theta y_t(\hat{\theta}_n;\hat{\psi}_{nk}) ^\prime W_n [\tilde{y}_t - y_t(\theta_0;\hat{\psi}_{nk})] \tag{A} \label{eq:An}\\
  &+ (1/n) \sum\nolimits_{t=1}^n \partial_\theta y_t(\hat{\theta}_n;\hat{\psi}_{nk})^\prime W_n [y_t(\theta_0;\hat{\psi}_{nk}) - y_t(\hat{\theta}_n;\hat{\psi}_{nk})]. \tag{B} \label{eq:Bn} \end{align}
Define $\theta_n(\omega) = \omega \theta_0 + (1-\omega)\hat{\theta}_n$ for any $\omega \in [0,1]$. An integration and change of variable implies:
$(\ref{eq:Bn}) = - \frac{1}{n} \sum_{t=1}^n \int_{0}^1 \partial_\theta y_t(\hat{\theta}_n;\hat{\psi}_{nk})^\prime W_n \partial_\theta y_t(\theta_n(\omega);\hat{\psi}_{nk})d\omega [\hat{\theta}_n - \theta_0].$
Likewise:
$(\ref{eq:An}) = \frac{1}{n} \sum_{t=1}^n \partial_\theta y_t(\theta_0;\hat{\psi}_{nk})^\prime W_n [\tilde{y}_t - y_t(\theta_0;\hat{\psi}_{nk})] + \frac{1}{n} \sum_{t=1}^n \int_{0}^1 ( [\tilde{y}_t - y_t(\theta_0;\hat{\psi}_{nk})]^\prime W_n \otimes I ) \partial_\theta G_t(\theta_n(\omega);\hat{\psi}_{nk})d\omega [\hat{\theta}_n - \theta_0]$. 
The expansions of (\ref{eq:An}) and (\ref{eq:Bn}) imply:
\begin{align*}
&\hat{\theta}_n - \theta_0 = M_n^{-1} \left( (1/n) \sum\nolimits_{t=1}^n \partial_\theta y_t(\theta_0;\hat{\psi}_{nk})^\prime W_n [\tilde{y}_t - y_t(\theta_0;\hat{\psi}_{nk})] \right), \\
&M_n = \frac{1}{n} \sum_{t=1}^n \int_{0}^1 \{ \partial_\theta y_t(\hat{\theta}_n;\hat{\psi}_{nk})^\prime W_n \partial_\theta y_t(\theta_n(\omega);\hat{\psi}_{nk}) + ( [y_t(\theta_0;\hat{\psi}_{nk}) - \tilde{y}_t]^\prime W_n \otimes I ) \partial_\theta G_t(\theta_n(\omega);\hat{\psi}_{nk})d\omega\}.\end{align*}
\noindent\textbf{Step 1. Consistency of $M_n$.} 
It was shown in the proof of Theorem \ref{th:cons}: $\sup_{t=1,\dots,n}\|y_t(\theta_0;\hat{\psi}_{nk})-y_t(\theta_0;\psi_{0})\| = o_p(1)$; similar derivations yield $\sup_{t=1,\dots,n}\|\partial_{\theta} y_t(\theta_0;\hat{\psi}_{nk})-\partial_{\theta} y_t(\theta_0;\psi_{0})\| = o_p(1)$. 
The absolute summability of the third derivative (Assumption \ref{ass:InnovModel}), $\sup_{t=k+1,\dots,n} \|\hat{e}_t-e_t\| = o_p(1)$, and $\|\tilde{\Sigma}_{nk} - \tilde{\Sigma}\| = O_p(n^{-1/2})$ imply $\sup_{\theta \in \Theta}\sup_{t=k+1,\dots,n}\|\partial_{\theta} G_t(\theta;\hat{\psi}_{nk})-\partial_{\theta} G_t(\theta;\psi_{0})\| = o_p(1)$. Also, $\sup_{\omega \in [0,1]} \|\theta_n(\omega) - \theta_0\| = o_p(1)$ by consistency of $\hat{\theta}_n$. The first $k$ terms can be analyzed as in Theorem \ref{th:cons}. Thus,
$(1/n) \sum\nolimits_{t=1}^n \int_{0}^1 \left( [\tilde{y}_t - y_t(\theta_0;\hat{\psi}_{nk})]^\prime W_n \otimes I \right) \partial_\theta G_t(\theta_n(\omega);\hat{\psi}_{nk})d\omega = (1/n) \sum\nolimits_{t=1}^n \left( [\tilde{y}_t - y_t(\theta_0;\psi_0)]^\prime W \otimes I \right) \partial_\theta G_t(\theta_0;\psi_0) + o_p(1)$.
Assumption \ref{ass:InnovModel} implies $\partial_{\theta} G_t(\theta_0;\psi_{0})$ is NED in $L_q$-norm with size $-b\leq-2$. By Lemma \ref{lem:NED}, $\tilde{y}_t - y_t(\theta_0;\psi_0)$ and $\partial_{\theta} y_t(\theta_0;\psi_0)$ are also NED. Hence, $\left( [\tilde{y}_t - y_t(\theta_0;\psi_0)]^\prime W \otimes I \right) \partial_\theta G_t(\theta_0;\psi_0)$ is NED in $L_{q/2}$-norm with size $-b$. Therefore, a weak law of large numbers applies to the second term in the definition of $M_n$. Likewise, $(1/n) \sum_{t=1}^n \int_{0}^1 \partial_\theta y_t(\hat{\theta}_n;\hat{\psi}_{nk})^\prime W_n \partial_\theta y_t(\theta_n(\omega);\hat{\psi}_{nk})d\omega = (1/n) \sum_{t=1}^n \partial_\theta y_t(\theta_0;\psi_0)^\prime W \partial_\theta y_t(\theta_0;\psi_0) + o_p(1)$, an average of a NED process in $L_{q/2}$-norm with size $-b$. A weak law of large numbers applies to this term as well. Altogether: $M_n \overset{p}{\to} M$.

\noindent\textbf{Step 2. Asymptotic Normality.}
Let $\psi_k = (\tilde{\mu}^\prime,\text{vech}(\tilde{\Sigma})^\prime,\text{vec}(\Psi_1)^\prime,\dots,\text{vec}(\Psi_k)^\prime)^\prime$; recall $\hat{\psi}_{nk} = (\tilde{\mu}_{n}^\prime,\text{vech}(\tilde{\Sigma}_{nk})^\prime,\text{vec}(\hat{\Psi}_1)^\prime,\dots,\text{vec}(\hat{\Psi}_k)^\prime)^\prime$ with $\tilde{\mu}_n$ equal to the sample average.
Let $u_t = y_t(\theta_0;\psi_0) - \tilde{y}_t$, and $u_{t,k} = y_t(\theta_0;\psi_k) - \tilde{y}_t$. Note that Lemma \ref{lem:VARinf} implies that $\|\hat{\psi}_{nk} - \psi_k\|_{\infty} = O_p( \sqrt{\log(n)/n})$.
An integration identity and a change of variable with respect to $\psi$ imply:
\begin{align}
  &\frac{1}{n} \sum\nolimits_{t=1}^n \partial_\theta y_t(\theta_0;\hat{\psi}_{nk})^\prime W_n [\tilde{y}_t - y_t(\theta_0;\hat{\psi}_{nk})] \notag\\ = &-\frac{1}{n} \sum\nolimits_{t=1}^n \partial_\theta y_t(\theta_0;\psi_k)^\prime W_n u_{t,k} \tag{C} \label{eq:C}\\
  &- \frac{1}{n} \sum\nolimits_{t=1}^n \int_{0}^1 \left( u_{t,k}^\prime W_n \otimes I \right) \partial_{\psi} G_t(\theta_0;\psi_{nk}(\omega))d\omega [\hat{\psi}_{nk} - \psi_k] \tag{D} \label{eq:D}\\
  &- \frac{1}{n} \sum\nolimits_{t=1}^n \int_{0}^1  \partial_\theta y_t(\theta_0;\hat{\psi}_{nk})^\prime W_n \partial_{\psi} y_t(\theta_0;\psi_{nk}(\omega))d\omega[\hat{\psi}_{nk} - \psi_k], \tag{E} \label{eq:E}
\end{align}
where $\psi_{nk}(\omega) = \omega \hat{\psi}_{nk} + (1-\omega) \psi_k$ are intermediate values of $\hat{\psi}_{nk}$ and $\psi_k$. 

Because $W_n = W +o_p(1)$ and $W$ is invertible: $W_n = W(I_d + o_p(1))$. This implies:
$(C) = -\left[\frac{1}{n} \sum_{t=1}^n \partial_\theta y_t(\theta_0;\psi_k)^\prime W u_{t,k}\right](I_d + o_p(1))$. We now show: 
\[(C) = -(1/n) \sum\nolimits_{t=1}^n \partial_\theta y_t(\theta_0;\psi_k)^\prime W u_{t,k} + o_p(n^{-1/2}).\]
For this, note that $\partial_\theta y_t(\theta_0;\psi_k)^\prime$ and $u_{t,k}$ are NED in $L_{q}$-norm with size $-b$. Their product is NED in $L_{q/2}$-norm with size $-b$, with absolutely summable autocovariances, following from the same arguments as in the proof of Theorem \ref{th:cons}. This means $\left[\frac{1}{n} \sum_{t=1}^n \partial_\theta y_t(\theta_0;\psi_k)^\prime W u_{t,k}\right] = \mathbb{E}\left[\frac{1}{n} \sum_{t=1}^n \partial_\theta y_t(\theta_0;\psi_k)^\prime W u_{t,k}\right]+O_p(n^{-1/2})$, using Chebyshev's inequality. Next, because $u_{t,k} - u_t = y_t(\theta_0;\psi_k) - y_t(\theta_0;\psi_0)$, we have: $(\mathbb{E}[\|u_{t,k} - u_t\|^2])^{1/2} \leq \sum_{j=0}^\infty \|\Lambda_j(\theta_0)\|_{op} \|P(\theta_0;\tilde{\Sigma})\|_\infty (\mathbb{E}[\|e_{t,k} - e_t\|^2])^{1/2}$ for $t=k+1,\dots,n$, where $(\mathbb{E}[\|e_{t,k} - e_t\|^2])^{1/2} \leq \sum_{j=k+1}^\infty \| \Psi_j\|_{op} (\mathbb{E}[\|\tilde{y}_{t} - \tilde{\mu}\|^2])^{1/2} = o(n^{-1/2})$, from the condition in Lemma \ref{lem:VARinf}. Similar derivations also imply $(\mathbb{E}[\|\partial_{\theta} y_{t}(\theta_0;\psi_k) - \partial_{\theta} y_{t}(\theta_0;\psi_0)\|^2])^{1/2} = o(n^{-1/2})$, for $t=k+1,\dots,n$. Apply the Cauchy-Schwarz inequality to find $\mathbb{E}\left[\frac{1}{n} \sum_{t=1}^n \partial_\theta y_t(\theta_0;\psi_k)^\prime W u_{t,k}\right] = \mathbb{E}\left[\frac{1}{n} \sum_{t=1}^n \partial_\theta y_t(\theta_0;\psi_0)^\prime W u_{t}\right] + O(k/n) + o(n^{-1/2}) = o(n^{-1/2})$ since the moments are bounded uniformly for $t=1,\dots,k$. Putting everything together:
$(C) = -\frac{1}{n} \sum_{t=1}^n \partial_\theta y_t(\theta_0;\psi_k)^\prime W u_{t,k} + o_p(n^{-1/2})$. For notation, let $S_{1,t} = \partial_\theta y_t(\theta_0;\psi_k)^\prime W u_{t,k}$. Also, define the sample mean: $S_{1,n} = 1/n \sum_{t=1}^n S_{1,t}$.

Derivations for (\ref{eq:D}) and (\ref{eq:E}) have a similar outline, the following will consider (\ref{eq:E}).
The derivation for (\ref{eq:A*})-(\ref{eq:C*}) in the proof of Theorem \ref{th:cons} imply that $\sup_{t=k+1,\dots,n}\|\partial_\theta y_t(\theta_0;\hat{\psi}_{nk}) - \partial_\theta y_t(\theta_0;\psi_{k})\| = O_p(\sqrt{k\log(n)}n^{1/q-1/2})$ which implies:
\[ (\ref{eq:E}) = - \frac{1}{n} \sum_{t=1}^n \partial_\theta y_t(\theta_0;\psi_{k})^\prime W_n \partial_{\psi} y_t(\theta_0;\psi_{k})(\hat{\psi}_{nk} - \psi_k) + O_p(\max[\sqrt{k}\log(n)n^{1/q-1},\frac{\sqrt{\log(n)}}{n^{3/2}}k^{1+1/q}]), \] 
the last term is $o_p(n^{-1/2})$ because $q > 8$ and $k = o(n^{1/3})$. 

The next step is to look at $\partial_{\psi} y_t(\theta_0;\psi_{k})$ more closely in order to derive consistency for (\ref{eq:E}). For this, we will evaluate $\partial_{\tilde{\mu}} y_t(\theta_0;\psi_k)$, $\partial_{\phi_j} y_t(\theta_0;\psi_k)$, and $\partial_{ \text{vech}(\tilde{\Sigma})}y_t(\theta_0;\psi_k)$ separately. 

Recall that $e_{t,k} = \tilde{y}_t - \tilde{\mu} - \sum_{j=1}^k \Psi_j( \tilde{y}_{t-j-1} - \tilde{\mu} )$ with $e_{t,k} = 0$ for $t\leq 0$ and $\tilde{y}_{t-j-1} - \tilde{\mu} =0$ for $t-j-1 \leq 0$. The coupled sample is constructed as $y_t(\theta;\psi_{k}) = \mu(\theta) + \sum_{j=0}^\infty \Lambda_j(\theta)P(\theta;\tilde{\Sigma}) e_{t-j,k}$. Thus, the derivative with respect to $\tilde{\mu}$ is given by: 
$ \partial_{ \tilde{\mu}_n } e_{t,k} = I_d - \sum_{j=1}^k \Psi_j  = I_d - \sum_{j=1}^\infty \Psi_j - \sum_{j=k+1}^\infty \Psi_j  = I_d - \sum_{j=1}^\infty \Psi_j + o_p(n^{-1/2})$,
uniformly in $t$, using the condition $\sqrt{n} \sum_{j=k+1}^\infty \Psi_j = o(1)$. The absolute summability of the $\Lambda_j$ then implies that:
$\partial_{\tilde{\mu}} y_t(\theta_0;\psi_k) = I+\sum_{j=0}^\infty \Lambda_j(\theta)P(\theta;\tilde{\Sigma})(I_d - \sum_{\ell = 1}^\infty \Psi_\ell) + o(n^{-1/2})$,
uniformly in $t$, here using $\sqrt{n} \sum_{j = k+1}^\infty \sup_{\theta \in \Theta} \|\Lambda_j(\theta)\|_{op} = o(1)$. Let $\phi_j = \text{vec}(\Psi_j)$ for $j=1,\dots$ and re-write: $e_{t,k} = \tilde{y}_t - \tilde{\mu} - \sum_{j=1}^k \left( (\tilde{y}_{t-j-1}-\tilde{\mu})^\prime \otimes I \right) \phi_j$. The partial derivative is $\partial_{ \phi_j }e_{t,k} = - (\tilde{y}_{t-j-1}-\tilde{\mu})^\prime \otimes I $ for $t-j-1 \in \{1,\dots,n\}$ and $\partial_{ \phi_j }e_{t,k} = 0$ for $t-j-1 \leq 0$ and $j \in \{1,\dots,k\}$. Then, using $\sqrt{n} \sum_{\ell = k+1}^\infty \sup_{\theta \in \Theta}\|\Lambda_\ell(\theta)\|_{op} = o(1)$ and $\mathbb{E}(\|\tilde{y}_t\|^q) < \infty$, we get
$ \sup_{t=k+1,\dots,n} \|\partial_{\phi_j} y_t(\theta;\psi_k) + \sum_{\ell = 0}^\infty (\tilde{y}_{t-\ell-j-1}-\tilde{\mu})^\prime \otimes [\Lambda_{\ell}(\theta) P(\theta;\tilde{\Sigma})] \| = o_p(n^{1/q-1/2})$, here using the actual $\tilde{y}_{t-\ell-j-1}$ for all $t,j,\ell$. Using similar derivations as for Theorem \ref{th:cons}, $\sup_{t=1,\dots,k} \|\partial_{\phi_j} y_t(\theta;\psi_k) + \sum_{\ell = 0}^\infty (\tilde{y}_{t-\ell-j-1}-\tilde{\mu})^\prime \otimes [\Lambda_{\ell}(\theta) P(\theta;\tilde{\Sigma})] \| = o_p(n^{1/q-1/2}) + O_p(k^{1/q})$, since $e_{t,k}$ sets $\tilde{y}_{t-\ell-j-1}-\tilde{\mu}=0$ for $t-\ell-j-1 \leq 0$ and $e_{t}$ does not. Note that $P(\theta;\cdot)$ is twice continuously differentiable under the constant rank assumption for $\Sigma(\theta)$ and $\tilde{\Sigma}$ invertible. 
Similar to the above, $\sup_{t=k+1,\dots,n}\|\partial_{ \text{vech}(\tilde{\Sigma}) } y_t(\theta;\psi_k) - \sum_{j=0}^\infty ( e_{t-j}^\prime \otimes \Lambda_j(\theta) )\partial_{ \text{vech}(\tilde{\Sigma}) } \text{vec}[P(\theta;\tilde{\Sigma})]\| = o_p(n^{1/q-1/2})$ and $ \sup_{t=1,\dots,k}\|\partial_{ \text{vech}(\tilde{\Sigma}) } y_t(\theta;\psi_k) - \sum_{j=0}^\infty ( e_{t-j}^\prime \otimes \Lambda_j(\theta) )\partial_{ \text{vech}(\tilde{\Sigma}) } \text{vec}[P(\theta;\tilde{\Sigma})]\| = o_p(n^{1/q-1/2}) + O_p(k^{1/q}).$ 
Let $y_t(\theta;\psi_{k0})$ and $\partial_{\psi} y_t(\theta;\psi_{k0})$ be as in Lemma \ref{lem:NED}.
Recall $\hat{\psi}_{nk} - \psi_k = O_p(\sqrt{\log(n)/n})$, then:
$ \frac{1}{n} \sum_{t=1}^n \partial_\theta y_t(\theta_0;\psi_{k})^\prime W \partial_\psi y_t(\theta_0;\psi_{k})(\hat{\psi}_{nk} - \psi_k) = \frac{1}{n} \sum_{t=1}^n \partial_\psi y_t(\theta_0;\psi_{k})^\prime W \partial_\theta y_t(\theta_0;\psi_{k0})(\hat{\psi}_{nk} - \psi_k) + o_p(n^{-1/2}),$
using $o_p(n^{1/q-1/2}) = o_p(1/\sqrt{\log(n)})$, $k^{1+1/q} \sqrt{\log(n)}/n = o(n^{-1/2})$ and similar derivations as in the proof of Theorem \ref{th:cons}. Given the stated assumption, similar derivations to (\ref{eq:B*})-(\ref{eq:D*}) in Theorem \ref{th:cons} for $\partial_\theta y_t(\theta;\psi_k)$ further imply:
$\frac{1}{n} \sum_{t=1}^n \partial_\theta y_t(\theta_0;\psi_{k})^\prime W \partial_\psi y_t(\theta_0;\psi_{k})(\hat{\psi}_{nk} - \psi_k) = \frac{1}{n} \sum_{t=1}^n \partial_\theta y_t(\theta_0;\psi_{0})^\prime W \partial_\psi y_t(\theta_0;\psi_{k0})(\hat{\psi}_{nk} - \psi_k) + o_p(n^{-1/2}).$

The next step is to show: $\frac{1}{n} \sum_{t=1}^n \partial_\theta y_t(\theta_0;\psi_{0})^\prime W \partial_\psi y_t(\theta_0;\psi_{k0}) = \mathbb{E}[ \partial_\theta y_t(\theta_0;\psi_{0})^\prime W \partial_\psi y_t(\theta_0;\psi_{k0}) ] + O_p(kn^{-1/2})$. The proof for this part is long as it involves non-standard arguments.

As shown in Lemma \ref{lem:NED}, $\partial_{\theta} y_t(\theta;\psi_0)$ and $\partial_{\mu} y_t(\theta;\psi_{k0}),\partial_{\psi_j} y_t(\theta;\psi_{k0})$, and $\partial_{\text{vec}(\tilde{\Sigma})} y_t(\theta;\psi_{k0})$ are NED in $L_q$-norm with size $-b$, for each $j=1,\dots$. Consequently, $\partial_{\theta} y_t(\theta;\psi_0)^\prime W \partial_{\mu} y_t(\theta;\psi_{k0})$ and $\partial_{\theta} y_t(\theta;\psi_0)^\prime W \partial_{\text{vec}(\tilde{\Sigma})} y_t(\theta;\psi_{k0})$ are NED in $L_{p}$-norm with size $-b$ \citep[Th18.9]{Davidson2021}, with $b \geq 2$ for any $p < q/2$. They are $L_{q/2}$-mixingales of size $-\min(b,a (1/p-2/q)) = 1/2$. Pick $p=2$ and $q \geq 8$ implies $\min(b,a (1/p-2/q))\geq \min(2,2 (1/2-2/8)) = 1/2$ so that the autocovariances are absolutely summable \citep[][Th17.16]{Davidson2021}. Therefore, a weak law of large numbers applies to the sample average of $\partial_{\theta} y_t(\theta;\psi_0)^\prime W \partial_{\mu} y_t(\theta;\psi_{k0})$ and $\partial_{\theta} y_t(\theta;\psi_0)^\prime W \partial_{\text{vec}(\tilde{\Sigma})} y_t(\theta;\psi_{k0})$, using Chebyshev's inequality.

We also need to derive a weak law of large numbers for $\partial_{\theta} y_t(\theta;\psi_0)^\prime W \partial_{\phi_j} y_t(\theta;\psi_{k0})$, with $j=1,\dots,k$. This is nonstandard because $k$ diverges to infinity. Recall from Lemma \ref{lem:NED} that for all $j=1,\dots,k$, $\partial_{\phi_j} y_t(\theta_0;\psi_{k0}) = - \sum_{\ell = 0}^\infty [\tilde{y}_{t-\ell - j -1} - \tilde{\mu}]^\prime \otimes \Lambda_{j}(\theta_0)P(\theta_0;\tilde{\Sigma})$, where $[\tilde{y}_{t-\ell - j -1} - \tilde{\mu}] \in \mathcal{F}_{t-j-1}$. Apply the law of iterated expectations:
\begin{align*} 
    &\mathbb{E}[ \partial_\theta y_t(\theta_0;\psi_k)^\prime W \partial_{\phi_j} y_t(\theta_0;\psi_{k0})]
    = \mathbb{E}\left[ \mathbb{E}\{ \partial_\theta y_t(\theta_0;\psi_k)^\prime W \partial_{\phi_j} y_t(\theta_0;\psi_{k0}) | \mathcal{F}_{t-j-1}\} \right]\\
    &= -\mathbb{E}\left[ \{ \sum\nolimits_{\ell = j + 1}^\infty (e_{t-\ell}^\prime \otimes I) \partial_{\theta}\text{vec}[\Lambda_\ell(\theta_0)P(\theta_0;\tilde{\Sigma})] \}^\prime W \{ \sum\nolimits_{\ell = 0}^\infty [\tilde{y}_{t-\ell - j -1} - \tilde{\mu}]^\prime \otimes \Lambda_{j}(\theta_0)P(\theta_0;\tilde{\Sigma}) \} \right].
\end{align*}
Takes norms on both sides, and apply the Cauchy-Schwarz inequality to find:
\[ \|\mathbb{E}[ \partial_\theta y_t(\theta_0;\psi_k)^\prime W \partial_{\phi_j} y_t(\theta_0;\psi_{k0})]\| \leq \|e_t\|_2 \|\tilde{y}_t\|_2 \|P(\theta_0;\tilde{\Sigma})\|_{\infty} \sum\nolimits_{\ell = 0}^\infty \|\tilde{\Lambda}_\ell(\theta_0)P(\theta_0;\tilde{\Sigma})\|_\infty C {(j+1)^{-b-\varepsilon}}, \]
using Assumption \ref{ass:InnovModel} with $\|e_t\|_2 = (\mathbb{E}[\|e_t\|^2])^{1/2}$ and $b + \varepsilon > 2$.
Take $1 \leq j \leq k$ and $s \geq 1$, let $w_{t} = \text{vec}\{\partial_{\theta} y_{t}(\theta;\psi_0)^\prime W \partial_{\phi_j} y_{t}(\theta;\psi_{k0})\}$. From the bound above, we have:
$
  \| \mathbb{E}[ (w_{t+s} - \mathbb{E}[w_{t+s}])w_t^\prime ] \| =  \| \mathbb{E}[ \mathbb{E}(w_{t+s} - \mathbb{E}[w_{t+s}] | \mathcal{F}_t)w_t^\prime ] \|
 \leq \|e_t\|_4 \|\tilde{y}_t\|_4 \|P(\theta_0;\tilde{\Sigma})\|_{\infty} \sum\nolimits_{\ell = 0}^\infty \|\tilde{\Lambda}_\ell(\theta_0)P(\theta_0;\tilde{\Sigma})\|_\infty \\C{(j+1)^{-b-\varepsilon}} \|w_t\|_2,
$
using the same approach, where $\|w_t\|_2$ can be bounded independently of $j$. Use this bound for $s=0,\dots,8j$. Note that $(8j + 1)(j+1)^{-(b+\varepsilon)} = o(1)$ as $j \to \infty$, so that the contribution of this bound is uniformly finite in $j \geq 1$. 

Take $s > 8j$, we now bound $\|\mathbb{E}(w_{t+s} - \mathbb{E}[w_{t+s}] | \mathcal{F}_t)\|_p$ for some $p \geq 2$. Using a similar approach as \citet[p374]{Davidson2021}, divide it into two terms: $\|\mathbb{E}(w_{t+s} - \mathbb{E}[w_{t+s}] | \mathcal{F}_t)\|_p \leq \|\mathbb{E}(w_{t+s} - \mathbb{E}[w_{t+s} | \mathcal{F}_{t+s - m}^{t+s+m}] | \mathcal{F}_t)\|_p + \|\mathbb{E}( \mathbb{E}[w_{t+s}] - \mathbb{E}[w_{t+s} | \mathcal{F}_{t+s - m}^{t+s+m}] | \mathcal{F}_t)\|_p$ for any $1 \leq m \leq s$. Since $\mathbb{E}[w_{t+s} | \mathcal{F}_{t+s - m}^{t+s+m}]$ is $\alpha$-mixing with size $-a$, $\|\mathbb{E}( \mathbb{E}[w_{t+s}] - \mathbb{E}[w_{t+s} | \mathcal{F}_{t+s - m}^{t+s+m}] | \mathcal{F}_t)\|_p \leq (1+s-m)^{-a(1/p-1/r)}\|w_t\|_r$, for $r > p$. This is absolutely summable for $m = [s/2]$, over $s=1,\dots$ for any $1/p - 1/r \geq 1/2$ since $a > 2$ (Assumption \ref{ass:DGPdata}). Next, $\|\partial_{\theta} y_{t+s}(\theta;\psi_0) - \mathbb{E}[\partial_{\theta} y_{t+s}(\theta;\psi_0)|\mathcal{F}_{t+s-m}^{t+s+m}]\|_{2p} \leq \| \sum_{\ell = m+1}^\infty (I \otimes e_{t-\ell}) \partial_\theta \text{vec}[ P(\theta;\tilde{\Sigma})\Lambda_\ell(\theta) ] \|_{2p} \leq (1+m)^{-(b+\varepsilon)} C[ \|P(\theta;\tilde{\Sigma})\|_\infty + \|\partial_\theta \text{vec}[P(\theta;\tilde{\Sigma})]\|_\infty ] \|e_t\|_{2p}$. This is absolutely summable over $s \geq 1$ with $m = [s/2]$. 
Using similar arguments, $\| \tilde{y}_{t+s-j-\ell} - \mathbb{E}[\tilde{y}_{t+s-j-\ell}|\mathcal{F}_{t+s-m}^{t+s+m}] \|_{2p} \leq C(1+[m-j-\ell]^+)^{-(b+\varepsilon)}\|e_t\|_{2p}$, where $[m-j-\ell]^+ = \max(0,m-j-\ell) \geq \max(0,3/4m - \ell)$ for $s > 8j$ and $m = [s/2]$. Using the formula for $\partial_{\phi_j} y_t(\theta;\psi_{k0})$ this yields: $\| \partial_{\phi_j} y_t(\theta;\psi_{k0}) - \mathbb{E}[\partial_{\phi_j} y_t(\theta;\psi_{k0})|\mathcal{F}_{t+s-m}^{t+s+m}] \|_{2p} \leq \sum_{\ell = 0}^\infty C^2 (1+\ell)^{-(b+\varepsilon)}(1+[3/4m -\ell]^+)^{-(b+\varepsilon)}\|e_t\|_{2p}$. For $\ell = 0,\dots,[m/2] = [s/4]$, the partial sum is bounded by: $\sum_{\ell = 0}^\infty C^2 (1+\ell)^{-(b+\varepsilon)}(1+[1/4m]^+)^{-(b+\varepsilon)}\|e_t\|_{2p}$ which is absolutely summable in $s$. For $\ell > [m/2]$, the remainder is bounded by: $(1+[m/2])^{-(b+\varepsilon)/2}\sum_{\ell = 0}^\infty C^2 (1+\ell)^{-(b+\varepsilon)/2}\|e_t\|_{2p}$, also absolutely summable because $(b+\varepsilon)/2 > 1$. Then, we can bound $\| \mathbb{E}[w_{t+s} - \mathbb{E}(w_{t+s}) | \mathcal{F}_t]\|_{p} \leq C_p (1+s)^{-(b+\varepsilon)/2}$, where $C_p$ depends on $p$ via $\|e_t\|_{2p}$. This is absolutely summable. 

Then, use H\"older's inequality with $1 = 1/p + 1/r$ such that $\|e_t\|_{2p}$ and $\|w_t\|_r$ are finite, and $s > 8j$:
$\| \mathbb{E}[ (w_{t+s} - \mathbb{E}[w_{t+s}])w_t^\prime ] \| \leq C_p (1+s)^{-(b+\varepsilon)/2}\|w_t\|_{r}$,
which is absolutely summable. The sum over $s \leq 8j$ is of order $j (1+j)^{-(b+\varepsilon)}$, which is bounded. Thus the autocovariances are absolutely summable and the resulting sum is bounded uniformly in $j$. Chebyshev's inequality can be applied uniformly in $j$ and yields the rate:
$\sup\nolimits_{j=1,\dots,k} \Big\| (1/n)\sum\nolimits_{t=1}^n \partial_\theta y_t(\theta;\psi_0)^\prime W \partial_{\psi_j} y_t(\theta;\psi_{k0}) - \mathbb{E}[\partial_\theta y_t(\theta;\psi_0)^\prime W \partial_{\psi_j} y_t(\theta;\psi_{k0})] \Big\| \leq O_p(kn^{-1/2})$.
As $\|\hat{\psi}_{nk} - \psi_k\| \leq O_p(\sqrt{\log(n)/n})$ and $O_p(kn^{-1/2})O_p(\sqrt{\log(n)/n})= o_p(n^{-1/2})$, this implies:
$ \frac{1}{n} \sum_{t=1}^n \partial_\theta y_t(\theta_0;\psi_{k})^\prime W \partial_\psi y_t(\theta_0;\psi_{k})(\hat{\psi}_{nk} - \psi_k) = \mathbb{E}\left[ \partial_\theta y_t(\theta_0;\psi_{0})^\prime W \partial_\psi y_t(\theta_0;\psi_{k0})\right](\hat{\psi}_{nk} - \psi_k) + o_p(n^{-1/2}).$ This proves (\ref{eq:E}).
Similar derivations can be applied to (\ref{eq:D}). Combining them:
\begin{align*}
  &\frac{1}{n} \sum_{t=1}^n \left[  \left( u_{t,k}^\prime W_n \otimes I \right) \partial_{\psi} G_t(\theta_0;\bar{\psi}_{nk}) + \partial_\theta y_t(\theta_0;\hat{\psi}_{nk})^\prime W_n \partial_{\psi} y_t(\theta_0;\tilde{\psi}_{nk}) \right][\hat{\psi}_{nk} - \psi_k]\\
  &= \mathbb{E}\left[  \left( u_{t}^\prime W \otimes I \right) \partial_{\psi} G_t(\theta_0;\psi_{k0}) + \partial_\theta y_t(\theta_0;\psi_{k})^\prime W \partial_{\psi} y_t(\theta_0;\psi_{k0}) \right](\hat{\psi}_{nk}-\psi_k) + o_p(n^{-1/2}),
\end{align*}
where the expected matrix has full rank for $k \geq \underline{k}$, sufficiently large, by assumption.

The next step is to derive a CLT for the leading term on the right hand side of the preceding expression, where $\hat{\psi}_{nk} - \psi_k$ has increasing dimension. Note:
\begin{align*} &\mathbb{E}[ \partial_\theta y_t(\theta_0;\psi_k)^\prime W \partial_\psi y_t(\theta_0;\psi_{k0})](\hat{\psi}_{nk}-\psi_k)\\ =& \mathbb{E}[ \partial_\theta y_t(\theta_0;\psi_k)^\prime W \partial_{\tilde{\mu}} y_t(\theta_0;\psi_{k0})](\tilde{y}_{n}-\tilde{\mu})+\mathbb{E}[ \partial_\theta y_t(\theta_0;\psi_k)^\prime W \partial_{\vech{\tilde{\Sigma}}} y_t(\theta_0;\psi_{k0})]\text{vech}(\tilde{\Sigma}_{n}-\tilde{\Sigma})\\
+&\mathbb{E}[ \partial_\theta y_t(\theta_0;\psi_k)^\prime W \partial_{\phi_1} y_t(\theta_0;\psi_{k0})](\hat{\phi}_{1}-\phi_1)
+\dots+\mathbb{E}[ \partial_\theta y_t(\theta_0;\psi_k)^\prime W \partial_{\phi_k} y_t(\theta_0;\psi_{k0})](\hat{\phi}_{k}-\phi_k).  \end{align*}
The partial derivatives $\mathbb{E}[ \partial_\theta y_t(\theta_0;\psi_k)^\prime W \partial_{\phi_j} y_t(\theta_0;\psi_{k0})]$ are absolutely summable, as seen from the upper bound derived above. Similar arguments imply that the same holds for $\mathbb{E}[ (u_t^\prime W \otimes I) \partial_{\phi_j}G_t(\theta_0;\psi_{k0}) ]$ since $\partial_{\phi_j}G_t(\theta_0;\psi_{k0}) \in \mathcal{F}_{t-j}$ as well, $\mathbb{E}(u_t - \mathbb{E}[u_t]|\mathcal{F}_{t-j})$ are absolutely summable, and $\mathbb{E}[\partial_{\phi_j}G_t(\theta_0;\psi_{k0}) ]=0$. As a result: $\|D_{\theta,\psi}(k)\| \leq c_2 < \infty$ for some constant $c_2 > 0$. Together with the singular value condition, this implies that:
$0 < c_1 \leq \sigma_{\min}[D_{\theta,\psi}(k)] \leq \sigma_{\max}[D_{\theta,\psi}(k)] \leq c_2 < \infty$, 
for all $k \geq \underline{k} \geq 1$; that is, for $k$ large enough $D_{\theta,\psi}(k)$ is a matrix with $d_{\theta}$ rows, each with norm bounded away from zero and infinity.

The following relies on \citet[Th2]{lewis1985}, which holds when the innovations are non-iid, under the stated assumptions; see e.g. \citet[Ch7]{hannan2012}, \citet[Th2.6]{kuersteiner2005}, \citet[LemA.6]{gonccalves2007} and \citet[Ch15]{lutkepohl2005}. Let $\tilde{Y}_{t-1,k} = (\tilde{y}_{t-1}-\tilde{\mu},\dots,\tilde{y}_{t-k}-\tilde{\mu})^\prime$, with mean zero, and $\Gamma_k = \mathbb{E}[\tilde{Y}_{t,k} \tilde{Y}_{t,k}^\prime]$ is the autocovariance matrix. For any sequence of vectors $l(k)$, such that $0 < c_1 \leq \|l(k)\| \leq c_2 < \infty$:
$ \sqrt{n}l(k)^\prime[ (\hat{\phi}_1,\dots,\hat{\phi}_k) - (\phi_1,\dots,\phi_k) ]^\prime = l(k)^\prime \text{vec} \left[ (1/\sqrt{n}) \sum\nolimits_{t=1}^n e_{t} \tilde{Y}_{t-1,k}^\prime \Gamma_k^{-1} \right] + o_p(1), $
where the sum is taken from $t=1$ rather than $t=1+k$, the difference being $\sqrt{n}$-negligible. Note that $\Gamma_k^{-1}$ is finite and uniformly bounded in $k \geq 1$ \citep{lewis1985}. 

The next step is to derive the dependence properties of the right-hand-side of the last display. Because $e_t$ is the Wold innovation, it is a martingale difference sequence. A central limit Theorem for $\hat{\phi}_{nk}$ is usually derived from this property. However, here $\hat{\psi}_{nk}$ also involves $\tilde{y}_t$ and $e_te_t^\prime$, which are not martingale differences. Using the notation of \citet{lewis1985}, let $\Gamma(j) = \mathbb{E}[ (\tilde{y}_t -\tilde{\mu})(\tilde{y}_{t-j} -\tilde{\mu})^\prime ]$. The first set of rows of $\Gamma_k$ is given by $\Gamma(0),\Gamma(1),\dots,\Gamma(k-1)$, the second row consists of $\Gamma(1)^\prime,\Gamma(0),\Gamma(1),\Gamma(2),\dots,\Gamma(k-2)$, etc. Using the white noise property of the innotations, we have:
\begin{align*} \|\Gamma(j)\| = \|\mathbb{E} \left( [\sum\nolimits_{\ell=j}^\infty \tilde{\Lambda}_j e_{t-\ell}][\sum\nolimits_{\ell=0}^\infty \tilde{\Lambda}_j e_{t-\ell - j}]^\prime \right)\| &\leq \sum\nolimits_{\ell = j}^\infty \|\tilde{\Lambda}_j\|_{op} \|e_t\|^2 \|\tilde{y}_t- \tilde{\mu}\|_2 \\ &\leq C \|e_t\|^2 \|\tilde{y}_t- \tilde{\mu}\|_2 (j-1)^{-(b+\varepsilon)}.\end{align*}
This implies that $\|\Gamma(j)\|_\infty \leq C_1 j^{-(b+\varepsilon)}$, for some constant $C_1$ and the elements of $\Gamma_k$ decay polynomially away from the diagonal, i.e. $(\Gamma_{k})_{\ell,s} \leq C_2 (1 + |\ell-s|)^{-(b+\varepsilon)}$ for some constant $C_2$ and for all $1 \leq \ell,s \leq k$, for all $k \geq 1$. Proposition 3 and the ``window lemma'' in \citet{jaffard1990} then imply:
$(\Gamma_{k}^{-1})_{\ell,s} \leq C_3 (1 + |\ell-s|)^{-(b+\varepsilon)}$, 
for some other constant $C_3$, with the same polynomial rate of decay. Denote by $(\Gamma_k^{-1})_{j}$, $1 \leq j \leq k$, the $j$-th block of columns of the matrix $\Gamma_k^{-1}$; that is for $j=1$, $(\Gamma_k^{-1})_{1}$ contains columns $1$ to $\text{dim}(\tilde{y}_t)$. Let $l_j(k)$ denote the coefficients of $l(k)$ which are multiplied by $\hat{\phi}_j - \phi_j$ above. Then, for each $j=1,\dots,k$:
\[ \sqrt{n}l_j(k)^\prime( \hat{\phi}_j - \phi_j) = l_j(k)^\prime \text{vec}\left[ (1/\sqrt{n}) \sum\nolimits_{t=1}^n e_t \tilde{Y}_{t-1,k}^\prime(\Gamma_k^{-1})_j \right] + o_p(1), \]
where $\|l_j(k)\| \leq (1+j)^{-(b+\varepsilon)}$, up to some constant, as shown above. 
The next step is to investigate the dependence properties of each $e_t \tilde{Y}_{t-1,k}^\prime(\Gamma_k^{-1})_j$, $1 \leq j \leq k$. Recall that $\tilde{Y}_{t-1,k} = (\tilde{y}_{t-1}-\tilde{\mu},\dots,\tilde{y}_{t-k}-\tilde{\mu})$, $\mathcal{F}_{t-m}^{t+m}$ is the filtration generated from $(e_{t-m},\dots,e_{t+m})$. Because $e_t \in \mathcal{F}_{t-m}^{t+m}$, we can write using H\"older's inequality:
\( \|e_t\tilde{y}_{t-j}- \mathbb{E}(e_t \tilde{y}_{t-j} | \mathcal{F}_{t-m}^{t+m})\|_{p/2} \leq \|e_t\|_{p}\|\tilde{y}_{t-j} - \mathbb{E}(\tilde{y}_{t-j} | \mathcal{F}_{t-m}^{t+m})\|_p \leq  d  \|e_t\|_{p} \nu( (m-j)^+ ),\) 
where $d$ and $\nu$ are the NED coefficients satisfying:
$\|\tilde{y}_{t-j} - \mathbb{E}(\tilde{y}_{t} | \mathcal{F}_{t-m}^{t+m})\|_p \leq  d \nu( m )$, 
for $m \geq 0$ and $(m-j)^+ = \max(m-j,0)$. The NED coefficients are derived in the proof of Lemma \ref{lem:NED}. Their decay factor satisfies $\nu(m) \leq (1+m)^{-(b+\varepsilon)}$. 
Together, we get:
\begin{align*}
  &\Big\|l(k)^\prime \text{vec}\left[ e_t \tilde{Y}_{t-1,k}^\prime(\Gamma_k^{-1}) - \mathbb{E}[ e_t \tilde{Y}_{t-1,k}^\prime(\Gamma_k^{-1}) | \mathcal{F}_{t-m}^{t+m}] \right] \Big\|_{p/2}\\ \leq& \sum\nolimits_{j=1}^k \Big\|l_j(k)^\prime \text{vec}\left[ e_t \tilde{Y}_{t-1,k}^\prime(\Gamma_k^{-1})_{j} - \mathbb{E}[ e_t \tilde{Y}_{t-1,k}^\prime(\Gamma_k^{-1})_{j} | \mathcal{F}_{t-m}^{t+m}] \right] \Big\|_{p/2}\\
  \leq & C_5 \sum\nolimits_{j=1}^k \sum\nolimits_{\ell = 1}^k (1+j)^{-(b+\varepsilon)}(1+|\ell-j|)^{-(b+\varepsilon)}(1+(m-\ell)^+)^{-(b+\varepsilon)},
\end{align*}
for some constant $C_5$. We now split the double sum into three terms and study them separately. The first term is: $
  \sum\nolimits_{j=1}^k \sum\nolimits_{\ell = 1}^{[m/2]} (1+j)^{-(b+\varepsilon)}(1+|\ell-j|)^{-(b+\varepsilon)}(1+(m-\ell)^+)^{-(b+\varepsilon)} \\ \leq  (1+[m/2])^{-(b+\varepsilon)} \sum\nolimits_{j=1}^\infty \sum\nolimits_{\ell = -\infty}^\infty (1+j)^{-(b+\varepsilon)}(1+|\ell|)^{-(b+\varepsilon)},
$
where the right-hand side is summable since $b+\varepsilon>2$. The second term is: $
\sum\nolimits_{j=1}^{[m/4]} \sum\nolimits_{\ell = [m/2] + 1}^{k} (1+j)^{-(b+\varepsilon)}(1+|\ell-j|)^{-(b+\varepsilon)}(1+(m-\ell)^+)^{-(b+\varepsilon)} \leq  (1+[m/4])^{-(b+\varepsilon)/2} \sum\nolimits_{j=1}^\infty \sum\nolimits_{\ell = -\infty}^\infty (1+j)^{-(b+\varepsilon)}(1+|\ell|)^{-(b+\varepsilon)/2},
$
which is summable since $(b+\varepsilon)/2 > 1$. The third term is: $
\sum\nolimits_{j=1+[m/4]}^{k} \sum\nolimits_{\ell = [m/2] + 1}^{k} (1+j)^{-(b+\varepsilon)}(1+|\ell-j|)^{-(b+\varepsilon)}(1+(m-\ell)^+)^{-(b+\varepsilon)} \leq \sum\nolimits_{j=1 + [m/4]}^\infty \sum\nolimits_{\ell = -\infty}^\infty (1+j)^{-(b+\varepsilon)}(1+|\ell|)^{-(b+\varepsilon)} \leq \int_{[m/4]}^\infty (1+x)^{-(b+\varepsilon)}dx \sum\nolimits_{\ell = -\infty}^\infty (1+|\ell|)^{-(b+\varepsilon)} \leq \frac{1}{b+\varepsilon-1}(1+[m/4])^{1-(b+\varepsilon)} \sum\nolimits_{\ell = -\infty}^\infty (1+|\ell|)^{-(b+\varepsilon)}.
$
From these, we get that the sum of the three terms is bounded above by:
\begin{align*}
  &\Big\|l(k)^\prime \text{vec}\left[ e_t \tilde{Y}_{t-1,k}^\prime(\Gamma_k^{-1}) - \mathbb{E}[ e_t \tilde{Y}_{t-1,k}^\prime(\Gamma_k^{-1}) | \mathcal{F}_{t-m}^{t+m}] \right] \Big\|_{p/2} \leq C_6 (1+m)^{-(b+\varepsilon)/2},
\end{align*}
since $b \geq 2$ implies $1-b \leq -b/2$, it is NED in $L_{p/2}$-norm with size $-b/2\leq-1$. Similarly,
\[ - (1/n) \sum\nolimits_{t=1}^n D_{\theta,\psi}(k) ((\tilde{y}_t-\tilde{\mu})^\prime,\text{vec}[ e_t \tilde{Y}_{t-1,k}^\prime\Gamma_k^{-1} ]^\prime,\text{vech}[e_te_t^\prime - \tilde{\Sigma}]^\prime)^\prime = (\ref{eq:D}) + (\ref{eq:E}) + o_p(n^{-1/2}) \]
is NED in $L_{q/2}$-norm with size $-\min(b-1,a)$ where $a$ is the mixing size of $e_t$ in Assumption \ref{ass:DGPdata}. Altogether, we get:
$ \sqrt{n}(\hat{\theta}_n - \theta_0) = - (1/\sqrt{n}) \sum\nolimits_{t=1}^n M^{-1}[ \partial_{\theta}y_t(\theta_0;\psi_k)^\prime W u_{t,k} + D_{\theta,\psi}(k)Z_{t,k}] + o_p(1), $
where the leading term is NED in $L_{q/2}$-norm, $q/2>2$, with size $-\min(b/2,a)$, $\min(b/2,a) > 1/2$ on $e_t$, which is strongly-mixing with size $-a$, $a > r/(r-2)$ for $r>4$. The NED derivations imply that $V_{n,k} $ is bounded. With the normalization $V_{n,k}^{-1/2}$, the conditions for Corollary 4.2 in \citet{wooldridge1988} hold and:
$\sqrt{n}V_{n,k}^{-1/2}(\hat{\theta}_n-\theta_0) \overset{d}{\to} \mathcal{N}(0,I)$. \qed

\newpage

\begin{titlingpage} 
  \emptythanks
  \title{ {Supplement to\\ \lQ {\bf Fitting Dynamically Misspecified Models:
  An Optimal Transportation Approach}''}}
  \author{ Jean-Jacques Forneron\thanks{Department of Economics, Boston University, 270 Bay State Road, Boston, MA 02215 USA.\newline Email: \href{mailto:jjmf@bu.edu}{jjmf@bu.edu}, Website: \href{http://jjforneron.com}{http://jjforneron.com}}. \and Zhongjun Qu\thanks{Department of Economics, Boston University, 270 Bay State Road, Boston, MA 02215 USA.\newline Email: \href{qu@bu.edu}{qu@bu.edu}.}} 
  \setcounter{footnote}{0}
  \setcounter{page}{0}
  \setcounter{figure}{0}
  \setcounter{table}{0}

  \clearpage 
  \maketitle 
  \thispagestyle{empty} 
  \begin{center}
  This Supplemental Material consists of Appendices \ref{apx:Extra_Algo}, \ref{apx:proof_prelim}, \ref{apx:Spec}, \ref{apx:motivating}, \ref{apx:emp} to the main text.
  \end{center}
\end{titlingpage}

\setcounter{page}{1}

\section{Additional details for Algorithm \ref{algo:OTF}} \label{apx:Extra_Algo}

The following describes an implementation of Algorithm \ref{algo:OTF} which combines the Bootstrap filter \citep[Ch10.3.1]{chopin2020} with optimal transport. 

\begin{algorithm}[H]
  \caption{An Implementation of the Optimal Transport Filter}\label{algo:OTF_PF_SK} { \small
  \begin{algorithmic}[1]
      \Procedure{\textsc{otf}}{}\newline
      \textbf{Inputs:} 1) Sample: $\tilde{y}_1,\dots,\tilde{y}_n$, predictive distribution $\tilde{p}(\tilde{y}_t|\tilde{y}_{t-1},\dots)$.\newline\hphantom{\textbf{Inputs:}} 2) Model: $p(y_t,z_t|z_{t-1};\theta)$. Initial beliefs $z_0 \sim p_{0|0}(z_0)$. Number of draws $B$.\newline
      \textbf{Outputs:} 1) Mapped data $y_1,\dots,y_n$.\newline\hphantom{\textbf{Outputs:}} 2) Filtered states $z_{t|t} \sim p( z_t|y_t,\dots,y_1 )$.\newline
      \textbf{Initialize:} For $b = 1,\dots,B$, do 1) draw $z_0^b \sim p_{0|0}(z_0;\theta)$, 2) weight $w_0^b = 1/B$.
      \For{$t \in \{1,\dots,n\}$} 
      \If{ESS($w_{t-1}^{1:B}$) $<$ ESS$_{\min}$}
        \State{resample $(z_{t-1}^b)_{b=1,\dots,B}$ with replacement and weight $w_{t-1}^b$, set $w_{t-1}^b = 1/B$.}
      \EndIf
      \State{\textbf{Predict:} Draw $(y_{t}^b,z_{t}^b) \sim p(y_t,z_t|z_{t-1}^b)$ and $\tilde{y}_t^b \sim \tilde{p}(\tilde{y}_t|\tilde{y}_{t-1})$.}  
      \State{\textbf{Transport plan:} Find a joint distribution $\pi_{t|t-1} = (p_{i,j})_{1 \leq i,j \leq B}$ which solves}  
      \[ \min_{p_{i,j}} \left\{ \sum_{i,j} p_{i,j}\|y_t^i - \tilde{y}_t^j\|^2 - \varepsilon \sum_{i,j} p_{i,j} \log( p_{i,j} ) \right\} \] 
      where $0 \leq p_{i,j} \leq 1$, $\sum_{i} p_{i,j} = 1/B$, $\sum_{j} p_{i,j} = w_{t-1}^i$.
      \State{\textbf{Barycentric projection:}} find $j^\star$ such that $\|\tilde{y}_t - \tilde{y}_t^{j^\star}\|$ is minimal. Compute $p_i = p_{i,j^\star} / [\sum_{i=1}^B p_{i,j^\star}]$. Compute $y_t = \sum_{i=1}^B p_i y_{t}^i$.
      \State{\textbf{Update:} Set $w_{t}^b = w_{t-1}^b p(y_t | z_{t}^b)$. Normalize $w_{t}^b = w_{t}^b / [\sum_i w_{t}^i]$.} 
      \EndFor
      \EndProcedure
  \end{algorithmic}}
  \label{alg_3}
\end{algorithm}
Algorithm \ref{algo:OTF_PF_SK} combine particle filtering steps with an additional draw $\tilde{y}_t^j$ and an optimal transport projection (steps 7,8) to construct the model-consistent sample $(y_t)_{t=1,\dots,n}$. 
Step 7 can be approximated using the Sinkhorn algorithm \citep[Ch4]{peyre2019}.


\section{\large{Proofs for Lemmas and Preliminary Results}} \label{apx:proof_prelim}

\paragraph{Proof of Lemma  \ref{lem:Qbound}.} For any $m \in \mathcal{M}$:
$\|\mathbb{E}\left( m({\bf{\tilde{y}}}_t)\right) - \mathbb{E} \left( m({\bf{{y}}}_t(\theta;\psi_0)) \right) \| \leq \mathbb{E}_{\pi}( \|m({\bf{\tilde{y}}}_t) - m({\bf{{y}}}_t(\theta;\psi_0))\| )$, where $\pi$ any joint distribution with marginals given by the respective VMA($\infty$) representations. Then: $
\mathbb{E}_{\pi}( \|m({\bf{\tilde{y}}}_t) - m({\bf{{y}}}_t(\theta;\psi_0))\| ) \leq \sqrt{\sum_{j\geq 0}\|A_j\|^2 \mathbb{E}_{\pi}(\|\tilde{y}_{t}-y_{t}(\theta;\psi_0)\|^2)} + \sum_{j,j^\prime \geq 0} \|B_{j,j^\prime}\|_{\infty} \mathbb{E}_{\pi}(\|\tilde{y}_{t}-y_{t}(\theta;\psi_0)\|^2) + \sum_{j,j^\prime \geq 0} \|B_{j,j^\prime}\|_{\infty} \sqrt{ \mathbb{E}_{\pi}(\|\tilde{y}_{t}-y_{t}(\theta;\psi_0)\|^2) \mathbb{E}(\|\tilde{y}_{t}\|^2) }],$
using the Cauchy-Schwarz inequality and covariance stationarity to remove the dependence on $j,j^\prime$. Take $C_W = \lambda_{\min}(W)^{-1}$ and $\pi$ be the coupling constructed using Algorithm \ref{algo:KOTF}. If $\tilde{\mu}=\mu(\theta)$, then the linear terms cancel out. This concludes the proof. \qed

\paragraph{Proof of Lemma \ref{lem:DI}.} It suffices to note that the underlying innovations used to compute both forecasts are the same: $\hat{e}_1,\dots,\hat{e}_n$ and $\hat{e}_{n+h}=0$ for all $h \geq 1$. Since the VMA($\infty$) coefficients are numerically identical, the forecasts are identical. \qed

\paragraph*{Proof of Lemma \ref{lem:SStoVMA}.} Condition i) implies Assumption \ref{ass:DGPdata} (i). For Assumption \ref{ass:DGPdata} (ii), $\mu(\cdot)$ is continuous on $\Theta$ compact, hence bounded. Then, $\Lambda_j(\cdot) = A(\cdot)C(\cdot)^j K(\cdot)$ is the product of $A(\cdot)C(\cdot)$, continuous, with $K(\cdot)$. The mapping $\theta \to \Sigma(\theta)$ is continuously differentiable, $\Sigma$ has constant rank, so the pseudo-inverse $\Sigma(\cdot)^\dagger$ is continuously differentiable with bounded derivative \citep[Prop8.2]{magnus2019}. This implies $V(\cdot)$ is continuously differentiable by the Implicit Function Theorem, and $K(\cdot)$ is continuously differentiable as a product of continuously differentiable matrices. Hence, $A(\cdot)$, $C(\cdot)$, $K(\cdot)$ are continuously differentiable on $\Theta$. We then have $\|\Lambda_j(\cdot)\|_{op} \leq \|A(\cdot)\|_{op}\|C(\cdot)\|^j_{op} \|K(\cdot)\|_{op}$. Condition (iii) implies $\sup_{\theta \in \Theta} \|C(\theta)\|_{op} \leq \overline{c}$ for some $\overline{c} \in [0,1)$. Then, we get: $\sum_{j=0}^\infty \sup_{\theta \in \Theta} \|\Lambda_j(\theta)\|_{op} \leq \sup_{\theta \in \Theta}\|A(\theta)\|_{op} \sup_{\theta \in \Theta}\|K(\theta)\|_{op}\sum_{j =0}^{\infty} \overline{c}^j < \infty$. For the last Assumption \ref{ass:DGPdata} (iii), direct differentiation with respect to $\theta_\ell$ yields $\partial_{\theta_\ell} \Lambda_j(\theta) = \partial_{\theta_\ell} A(\theta) C(\theta)^j K(\theta) + A(\theta) C(\theta)^j \partial_{\theta_\ell} K(\theta) + A(\theta) \partial_{\theta_\ell} C(\theta) C(\theta)^{j-1} K(\theta) + \dots + A(\theta) C(\theta)^{j-1} \partial_{\theta_\ell} C(\theta) K(\theta).$ Then, as above: $ \|\partial_{\theta_\ell} \Lambda_j(\theta)\|_{op} \leq \overline{c}^{j-1} \sup_{\theta \in \Theta}\left( \|\partial_\theta \text{vec}[K(\theta)]\|_{\infty} + (j-1) \|K(\theta)\|_{\infty} \|\partial_\theta \text{vec}[C(\theta)]\|_{\infty} \right)$ is summable since $\overline{c} < 1$. Equivalence between norms implies $\sup_{\theta \in \Theta} \|\partial_\theta \text{vec}[\Lambda_j(\theta)]\|_{\infty} \leq \text{C}\sum_{\ell = 1}^{d_\theta} \sup_{\theta \in \Theta}\|\partial_{\theta_\ell} \Lambda_j(\theta)\|_{op}$, where $\text{C}$ only depends on the size of the matrix, which is summable. Similar derivations yield the results for derivatives of order $s=2,3$.\qed

\noindent\textbf{Proof of Lemma \ref{lem:ConstantRank}.} Take any $\theta_0 \in \Theta$. Because $A(\theta_0) \geq 0$, there exists $U_0$ and $U_1$ with sizes $m \times d$ and $r \times d$, respectively, where $m + r = d$, such that:
\begin{align} &A(\theta_0)U_0 = 0_{d,m}, \quad U_0^\prime U_0 = I_m\\
  &U^\prime_1A(\theta_0)U_1 > 0, \quad U^\prime_1A(\theta_0)^\dagger A(\theta_0)U_1 > 0, \quad U_1^\prime U_1 = I_r, \quad U_0^\prime U_1 = 0_{r,m}. \end{align}
They can be found using the eigendecomposition: 
\[ A(\theta_0) = U \left( \begin{array}{cc} \Lambda(\theta_0) & 0_{r,m} \\ 0_{m,r} & 0_{m,m} \end{array} \right) U^\prime, \]
where $U U^\prime = U^\prime U = I_d $, $\Lambda_0 > 0$ is diagonal, and $U = (U_0, U_1)$.
These are not unique but their spans, and the associated projection matrices, are unique. Note that $U_1^\prime A(\theta_0) U_1 = \Lambda(\theta_0)$ with $0 < \underline{\lambda} \preceq \Lambda(\theta_0) \preceq \overline{\lambda} < \infty$. 
Define: $U_0(\theta) = [I_d - A(\theta)[A(\theta)^\prime A(\theta)]^{\dagger} A(\theta)]U_0(\theta_0)$, $U_1(\theta) = [A(\theta)[A(\theta)^\prime A(\theta)]^{\dagger} A(\theta)]U_1(\theta_0)$.
They are continuously differentiable in $\theta$ because $\text{rank}[A(\theta)^\prime A(\theta)]=r$ is constant \citep[using][Prop8.2]{magnus2019}. By construction: $U_1(\theta)^\prime U_0(\theta) = 0$ for $\forall \theta \in \Theta$. Because $A(\theta)$ is Hermitian, we have $A(\theta)[A(\theta)^\prime A(\theta)]^{\dagger} A(\theta) = A(\theta) A(\theta)^\dagger$. Then, by the identities for the pseudo-inverse and $A(\theta_0)U_0 = 0_{d,m}$, we have $U_0(\theta_0)^\prime U_0(\theta_0) = I_m$. By construction, $A(\theta)[I_d - A(\theta) A(\theta)^\dagger] = 0_{d,d}$, which implies $A(\theta)U_0(\theta) = 0_{m,n}$ for all $\theta$. By the definition of a singular value: $\lambda_{\min}[U(\theta)^\prime U(\theta)] = \sigma_{\min}[U(\theta)]^2$. Weyl's inequality for singular values \citep[Th3.3.16]{horn1991} implies:
$\sigma_{\min}[U_0(\theta)] \geq \sigma_{\min}[U_0(\theta_0)] - \|U_0(\theta_0) - U_0(\theta)\|_{op} = 1 - \|U_0(\theta_0) - U_0(\theta)\|_{op}$, 
since $U_0(\theta_0)^\prime U_0(\theta_0) = I_m$. Next, $\theta \to A(\theta)A(\theta)^\dagger$ is $s$-times continuously differentiable on $\Theta$, so it is globally Lipschitz continuous with constant $ 0 \leq L_A < \infty$. This implies:
\[ \|U_0(\theta_0) - U_0(\theta)\|_{op} = \| [A(\theta_0)A^\dagger(\theta_0) - A(\theta_0)A^\dagger(\theta_0)]U_0 \|_{op} \leq L_A \|\theta-\theta_0\|. \]
Pick $0 < \delta \leq [2L_A]^{-1}$, then for all $\|\theta-\theta_0\| \leq \delta$, we have:
$\sigma_{\min}[U_0(\theta)] \geq 1/2 > 0 \Rightarrow \lambda_{\min}[U_0(\theta)^\prime U_0(\theta)] \geq 1/4 > 0$. 
By composition and invertibility, $\theta \to \tilde{U}_0(\theta) = U_0(\theta)[U_0(\theta)^\prime U_0(\theta)]^{-1/2}$ is $s$-times continously differentiable, $\tilde{U}_0(\theta)^\prime \tilde{U}_0(\theta) = I_m$, and $U_1(\theta)^\prime \tilde{U}_0(\theta) = 0_{r,m}$. For the same $\delta$, $\tilde{U}_1(\theta)$ constructed the same way is $s$-times continously differentiable with  $\tilde{U}_1(\theta)^\prime \tilde{U}_1(\theta) = I_r$ and $\tilde{U}_1(\theta)^\prime \tilde{U}_0(\theta) = 0_{r,m}$. 
It remains to show $\lambda_{\min}[\tilde{U}_1(\theta)^\prime A(\theta) \tilde{U}_1]$ is bounded below. Apply Weyl's inequality:
\[ \lambda_{\min}[\tilde{U}_1(\theta)^\prime A(\theta) \tilde{U}_1] \geq \lambda_{\min}[\tilde{U}_1(\theta_0)^\prime A(\theta_0) \tilde{U}_1(\theta_0)] - \|\tilde{U}_1(\theta)^\prime A(\theta) \tilde{U}_1 - \tilde{U}_1(\theta_0)^\prime A(\theta_0) \tilde{U}_1(\theta_0)\|_{op},\]
where the last term is $s$-time continuously differentiable on $\mathcal{B}_\delta(\theta_0)$, hence Lipschitz continuous with finite constant $L_{U,A}$. Eventually, we get:
$\lambda_{\min}[\tilde{U}_1(\theta)^\prime A(\theta) \tilde{U}_1] \geq \lambda_{\min}[\tilde{U}_1(\theta_0)^\prime A(\theta_0) \tilde{U}_1(\theta_0)] - L_{U,A}\|\theta-\theta_0\|$.
By choosing $0 < \delta \leq \min[ (2L_A)^{-1}, \underline{\lambda}[2L_{U,A}]^{-1}]$, we find:
$ \lambda_{\min}[\tilde{U}_1(\theta)^\prime A(\theta) \tilde{U}_1] \geq \underline{\lambda}/2 > 0$. Now we have $M(\theta) = (U_1(\theta),U_0(\theta))$ continuously differentiable on $\mathcal{B}_\delta(\theta_0)$, invertible and $M(\theta)M(\theta)^\prime=I_d$. By composition, $M(\theta)^\prime A(\theta) M(\theta) = \text{blockdiag}(B(\theta),0_{m,m})$ is continuously differentiable and $0 < \underline{\lambda}/2 \preceq B(\theta) \preceq \overline{\lambda} < \infty$. Then, for $\mathcal{B}_\delta(\theta_0)$, $A(\theta)^{1/2} = M(\theta) \text{blockdiag}(B(\theta)^{1/2},0_{m,m}) M(\theta)^\prime$ is $s$-times continuously differentiable by composition and $B(\theta)$ strictly positive definite.

Since $\Theta$ is compact and finite-dimensional, we can take a finite $\delta$-cover $\{\theta_1,\dots,\theta_N\}$ of $\Theta$. The mapping $\theta \to A(\theta)^{1/2}$ is continuously differentiable on each $\mathcal{B}_\delta(\theta_j) \cap \Theta$ and $\Theta = \cup_{j=1}^N \mathcal{B}_\delta(\theta_j) \cap \Theta$ so, by the gluing lemma \citep[Lem3.23]{lee2010}, $\theta \to A(\theta)^{1/2}$ is $s$-times continuously differentiable on $\Theta$.
\qed

\paragraph{Proof of Lemma \ref{lem:VARinf}.} The existence of the infinite order VAR representation follows from the properties of the VMA polynomial stated in Assumption \ref{ass:InnovModel}(i); $\text{det}( \sum_{j=0}^\infty \Psi_j z^j ) \neq 0$ for any $|z| \leq 1$ is a result of the same property of the VMA polynomial.

Lemma \ref{lem:VARinf}(i) is proved in \citet[Th7.4.5]{hannan2012}; estimating the VAR coefficients on de-meaned data does not affect their results \citep[Lem4.1]{kuersteiner2005}. \citet{hannan2012} require that $\lim_{s \to \infty}\mathbb{E}(e_te_t^\prime -\tilde{\Sigma} | \mathcal{F}_{t-s}) = 0$, this is implied by the mixing condition and \citet[Th15.2]{Davidson2021}. Lemma \ref{lem:VARinf}(ii) is shown in \citet[Prop15.1-15.3]{lutkepohl2005}. For Lemma \ref{lem:VARinf}(iii), note that Assumption \ref{ass:DGPdata} (i) and the mixing condition implies $\tilde{y}_t$ is Near-Epoch Dependent with size $-b$ in $L_{q}$-norm for $q \geq 8$ (Lemma \ref{lem:NED}). Its autocovariances are absolutely summable, using Theorem 17.16 with Corollary 18.7 in \citep{Davidson2021}. This and Chebyshev's inequality imply Lemma \ref{lem:VARinf}(iii) for $\tilde{y}_n$. Finally, $e_te_t^\prime$ is $\alpha$-mixing with size $-a$ and finite fourth moment. Thus its autocovariances are absolutely summable. This and Chebyshev's inequality yields Lemma \ref{lem:VARinf}(iii) for $\tilde{\Sigma}_n$. \qed

\paragraph{Proof of Lemma \ref{lem:NED}.} The first part of the Lemma follows from \citet[Example 18.3]{Davidson2021} with a multivariate one-sided VMA($\infty$) process. For the second part, use the double series representation:
$ \partial_{\phi_j} y_t(\theta;\psi_{k0})= - \sum\nolimits_{\ell=0}^\infty \sum\nolimits_{s=0}^\infty [e_{t-\ell-j-1-s}^\prime \tilde{\Lambda}_s^\prime] \otimes [\Lambda_\ell(\theta)P(\theta;\tilde{\Sigma})]. $
Recall that an absolutely summable double series can be reparameterized as an absolutely summable single series. A single series is a one-sided VMA($\infty$) representation, which in our setting can be used to derive NED properties. It is sufficient to upper-bound the coefficients of the latter and bound their rate of decay to get the result. Using $\|e_t\|_q = [\mathbb{E}(\|e_t\|^q)]^{1/q}$ and $\|A \otimes B\|_{op} = \|A\|_{op} \|B\|_{op}$, we have, for any $\theta$, by the triangle inequality followed by the Cauchy-Schwarz inequality:
\[\|\partial_{\phi_j} y_t(\theta;\psi_{k0})\|_q \leq \left[ \sum\nolimits_{s=0}^\infty \|\tilde{\Lambda}_{s}\|_{op} \right]\sum\nolimits_{\ell=0}^\infty \sup_{\theta \in \Theta}\|\Lambda_\ell(\theta)\|_{op} \sup_{\theta \in \Theta}\|P(\theta;\tilde{\Sigma})\|_{op}\|e_{t-\ell-j-1-s}\|_q\leq C< \infty.\] The majoration absolute summability and the NED property in $L_q$-norm. Also, the summation starting with $m+1$ satisfies:
$\left[ \sum_{s=0}^\infty \|\tilde{\Lambda}_{s}\|_{op} \right]\sum_{\ell=m+1}^\infty \sup_{\theta \in \Theta}\|\Lambda_\ell(\theta)\|_{op} \sup_{\theta \in \Theta}\|P(\theta;\tilde{\Sigma})\|_{op} \leq  C m^{-(b+\varepsilon)},$
implying that the size is $-b$. The constant rank assumption for $\Sigma(\theta)$ and invertibility condition for $\tilde{\Sigma}$ imply continuous differentiability of $P(\cdot;\cdot)$ using Lemma \ref{lem:ConstantRank} for the matrix square root. The VMA($\infty$) representation then implies the NED property.
\qed

\section{Proofs for the Specification Test} \label{apx:Spec}

\begin{lemma}[Strong Approximation] \label{lem:SA}
  Let $Z_{t,k}$ and $S_{n,k}$ be as Theorem \ref{th:asym_normal}. Suppose the Assumptions in Theorem \ref{th:spec} holds. Then, there exists $\mathcal{Z}_{n,k} \sim \mathcal{N}(0,S_{n,k}/n)$ such that:
  $\sqrt{n}\|\overline{Z}_{n,k} - \mathcal{Z}_{n,k}\| =o_p([\log(n)]^{-2})$.  
\end{lemma}

\paragraph{Proof of Lemma \ref{lem:SA}.} There are four main steps: Step 1. Use the NED properties to approximate $Z_{t,k}$ by a strong-mixing sequence, in blocks of observations; Step 2. Use coupling results for strong-mixing sequences to further approximate the blocks with iid blocks; Step 3. Apply Yurinskii's method to approximate the iid blocks with Gaussian random variables; Step 4. Adjust the covariance of the Gaussian from Step 3. to get the coupling with the same covariance structure. 

\noindent\textbf{Step 1. Approximate $Z_{t,k}$ by a strong-mixing process $Z^m_{t,k}$.}
Recall:
$Z_{t,k} = ( (\tilde{y}_t - \tilde{\mu})^\prime,\text{vec}[ e_t \tilde{Y}_{t-1,k}^\prime \Gamma_k^{-1} ]^\prime, \text{vech}[e_te_t^\prime]^\prime )^\prime$, 
with $\mathbb{E}[Z_{t,k}] = 0$, a column vector of dimension $d(k) = O(k)$. Note that when analyzing the properties of $Z_{t,k}$ below, we often consider its elements $(\tilde{y}_t - \tilde{\mu})$, $\text{vec}[ e_t \tilde{Y}_{t-1,k}$, and $\text{vech}[e_te_t^\prime]^\prime )^\prime$ separately.

For any $m$ satisfying $m \geq 2k$ and $m/n = o(1)$, define $Z_{t,k}^m = \mathbb{E}[ Z_{t,k} | \mathcal{F}_{t-m}^{t} ]$, a strong-mixing sequence with coefficients $\alpha_m(s) = \alpha([s-m]^+)$, with $\alpha(\cdot)$ the mixing coefficients of $e_t$. Note that $\mathbb{E}[e_te_t^\prime | \mathcal{F}_{t-m}^t] = e_te_t^\prime$. Since $\Gamma_k^{-1}$ is bounded, we only need to bound $\| e_t \tilde{y}_{t-j} - e_t \mathbb{E}[\tilde{y}_{t-j}|\mathcal{F}_{t-m}] \|_{q/2}$ for $j \in \{1,\dots,k\}$. For this, we have $\|\tilde{y}_{t-j} - \mathbb{E}[\tilde{y}_{t-j}|\mathcal{F}_{t-m}]\|_q \leq \sum_{\ell \geq m-k} \|\tilde{\Lambda}_{\ell}\|_{op} \|e_t\|_q \leq C [m-k]^{-(b+\varepsilon)} \|e_t\|_q$, uniformly in $j \leq k$; consequently, $\| e_t \tilde{y}_{t-j} - e_t \mathbb{E}[\tilde{y}_{t-j}|\mathcal{F}_{t-m}] \|_{q/2} \leq C [m-k]^{-(b+\varepsilon)} \|e_t\|_q^2$ using the Cauchy-Schwarz inequality.

Note that $\mathbb{E}[\overline{Z}_{n,k} - \mathbb{E}(\overline{Z}_{n,k}^m) ] = 0$, and $\mathbb{E}[ \text{vec}(e_t \tilde{Y}_{t-1,k}^\prime) |\mathcal{F}_{t-1}] = 0$ so that the elements of $Z_{t,k}$ and $Z_{t,k}^m$ corresponding to $\text{vec}(e_t \tilde{Y}_{t-1,k}^\prime \Gamma_k^{-1})$ are serially uncorrelated. By the NED and mixing properties, the autocovariances of $\tilde{y}_t$ and $\text{vech}[e_{t}e_t^\prime]$ are absolutely summable. For each row $\ell$ of $\overline{Z}_{t,k}$ we have: $\text{var}[ \overline{Z}_{n,k,\ell} - \overline{Z}^m_{n,k,\ell} ] = 0$ for indices corresponding to $\text{vech}[e_{t}e_t^\prime]$, and $\text{var}[ \overline{Z}_{n,k,\ell} - \overline{Z}^m_{n,k,\ell} ] = \frac{1}{n}\text{var}[ Z_{t,k,\ell} - Z^m_{t,k,\ell} ] \leq n^{-1} C [m-k]^{-2(b+\varepsilon)}\|e_t\|_4^2$ for indices corresponding to $\text{vec}(e_t \tilde{Y}_{t-1,k}^\prime\Gamma_k^{-1})$. 
For indices $\ell$ corresponding to $\tilde{y}_t$, the autocovariance of order $s \geq 1$ satisfies: $|\text{cov}( Z_{t,k,\ell} - Z^m_{t,k,\ell}, Z_{t-s,k,\ell} - Z^m_{t-s,k,\ell} )| \leq \{ |\text{cov}( Z_{t,k,\ell} - Z^m_{t,k,\ell}, Z_{t-s,k,\ell} - Z^m_{t-s,k,\ell} )| \}^{1/2}[ \|Z_{t,k,\ell} - Z^m_{t,k,\ell}\|_2 \|Z_{t,k,\ell} - Z^m_{t,k,\ell}\|_2 ]^{1/2} = \{ |\text{cov}( Z_{t,k,\ell} - Z^m_{t,k,\ell}, Z_{t-s,k,\ell} - Z^m_{t-s,k,\ell} )| \}^{1/2}\|Z_{t,k,\ell} - Z^m_{t,k,\ell}\|_2$.\footnote{Here we use $|\text{cov}(X,Y)| = |\text{cov}(X,Y)|^{1/2} |\text{cov}(X,Y)|^{1/2}$ and Cauchy–Schwarz $|\text{cov}(X,Y)| \leq \|X\|_2 \|Y\|_2$ if $X,Y$ have mean zero.} Next, $|\text{cov}( Z_{t,k,\ell} - Z^m_{t,k,\ell}, Z_{t-s,k,\ell} - Z^m_{t-s,k,\ell} )| = |\mathbb{E}[ \mathbb{E}\{ Z_{t,k,\ell} - Z^m_{t,k,\ell}  |\mathcal{F}_{t-s} \}(Z_{t-s,k,\ell} - Z^m_{t-s,k,\ell}) ]| \leq \| \mathbb{E}\{ Z_{t,k,\ell} - Z^m_{t,k,\ell}  |\mathcal{F}_{t-s} \} \|_2 \|Z_{t,k,\ell} - Z^m_{t,k,\ell}\|_2 \leq 2C (1+s)^{-(b+\varepsilon)} \|e_t\|_2 \|Z_{t,k,\ell} - Z^m_{t,k,\ell}\|_2$, using derivations from Lemma \ref{lem:NED}. Taking the summation, we get:
\begin{align*} \sum\nolimits_{s=1}^\infty |\text{cov}( Z_{t,k,\ell} - Z^m_{t,k,\ell}, Z_{t-s,k,\ell} - Z^m_{t-s,k,\ell} )| &\leq [ \sqrt{2C} \sum\nolimits_{s \geq 1} (1+s)^{-(b+\varepsilon)/2}\|e_t\|^{1/2}_2] \|Z_{t,k,\ell} - Z^m_{t,k,\ell}\|^{3/2}_2,
\end{align*}
which is a $O([m-k]^{-(3/2)(b+\varepsilon)})$ since $(b+\varepsilon)/2 \geq 1+\varepsilon/2$.
From this, we get that $\text{var}[ \overline{Z}_{n,k,\ell} - \overline{Z}^m_{n,k,\ell} ] = O( [m-k]^{-(3/2)(b+\varepsilon)} n^{-1} )$.

In sum, we have $\max_{\ell=1,\dots,d(k)}\mathbb{E}[ \| \overline{Z}_{n,k,\ell} - \overline{Z}^m_{n,k,\ell}\|_{2}] \leq O( n^{-1/2} m^{-(3/4)(b+\varepsilon)})$. Then \citet{pisier1983}'s and Markov's inequalities yield:
$\sqrt{n}\|\overline{Z}_{n,k} - \overline{Z}^m_{n,k}\|_{\infty} = O_p( k^{1/2}m^{-(3/4)(b+\varepsilon)} )$,
which is a $o_p(1)$ for $m \geq 2k$ since $(3/4)b > 1/2$.

\noindent\textbf{Step 2. Coupling for the strong-mixing process $Z^m_{t,k}$.}

For this step, we will use the method of Bernstein sums. Using the notation of \citet[Ch15]{Davidson2021}, the sample $(Z_{t,k}^m)_{t=1,\dots,n}$ will be divided into large blocks of size $b_n$, separated by small blocks of size $l_n$. We will set $2m \leq l_n = o(b_n)$ and $b_n = o(n)$; $r_n = [n/b_n]$ is the number of large blocks. The last block has $n - r_n b_n - (r_n-1)l_n = o(1)$ observations. 

The first large block consists of $(Z_{1,k}^m,\dots,Z_{b_n,k}^m)$, the second $(Z_{b_n+l_n+1,k}^m,\dots,Z_{2b_n+l_n+1,k}^m)$. The dependence between the first and second block is given by $\alpha( [l_n-m] ) \leq \alpha( m ) \to 0$, since the blocks are separated by $l_n \geq 2m$ time periods. For $i = 1,\dots,r_n$, let $X_{i,k}^m = \sum_{t= (i-1)b_n +(i-1) l_n + 1}^{ i b_n + (i-1) l_n} Z_{t,k}^m$ denote the partial sum over the i-th large block. 

For each row of $X_{i,k}^m$, say the $\ell$-th row, $X_{i,k,\ell}^m$, we can apply Theorem 1 in \citet{peligrad2002}, to find $\tilde{X}_{i,k,\ell}^m \overset{d}{=} X_{i,k,\ell}^m$, iid over $i=1,\dots,r_n$, such that $\mathbb{E}|X_{i,k,\ell}^m-\tilde{X}_{i,k,\ell}^m| \leq \sqrt{\alpha(l_n-m)}\|X_{i,k,\ell}^m\|_2$. Being a sum of $b_n$ terms with absolutely summable autocovariances (cf. step 1), $\|X_{i,k,\ell}^m\|_2 = O( \sqrt{b_n} )$. Then, we have $\mathbb{E}| \sum_{i=1}^{r_n} [X_{i,k,\ell}^m-\tilde{X}_{i,k,\ell}^m]  | \leq O( r_n \sqrt{b_n \alpha(l_n-m)} )$ by the triangle inequality on the $r_n$ blocks. 

The difference $\|n \overline{Z}_{n,k,\ell}^m - \sum_{i=1}^{r_n} X_{i,k,\ell}^m\|_2$ is of order $O(\sqrt{n - r_nb_n})$, as there are $n - r_n b_n$ observations in the smaller and last blocks of observations with absolutely summable autocovariances. These inequality holds uniformly in $\ell = 1,\dots,d(k)$, so this implies:
\begin{align*}
    &\mathbb{E} \left[ \max_{\ell =1,\dots,d(k)}|n\overline{Z}_{n,k,\ell}^m - \sum_{i=1}^{r_n} \tilde{X}_{i,k,\ell}^m | \right]\\ & \leq \mathbb{E} \left[ \max_{\ell =1,\dots,d(k)}| \sum_{i=1}^{r_n} X_{i,k,\ell}^m- \sum_{i=1}^{r_n} \tilde{X}_{i,k,\ell}^m | \right] + \mathbb{E} \left[ \max_{\ell =1,\dots,d(k)}| \sum_{i=1}^{r_n} X_{i,k,\ell}^m- n \overline{Z}_{n,k,\ell}^m |^2 \right]^{1/2}\\ 
    &\leq O(k r_n\sqrt{b_n\alpha(l_n-m)}) + O( k^{1/2}\sqrt{ n-r_nb_n } ),
\end{align*}
Let $\tilde{X}^m_{n,k} = 1/n \sum_{i=1}^{r_n} \tilde{X}_{i,k}^m$, then:
\[ \sqrt{n}\|\overline{Z}_{n,k}^m - \tilde{X}_{n,k}^m \|_{\infty} \leq O_p( \max[n^{-1/2}k r_n \sqrt{ b_n \alpha(l_n-m) },(k/n)^{1/2}\sqrt{n-r_nb_n}] ),\]
where $(k/n)^{1/2}\sqrt{n-r_nb_n} = O(\sqrt{k r_n l_n /n })$ by construction.

\noindent\textbf{Step 3. Gaussian Approximation.}
Here, the idea is to apply Yurinskii's coupling to the iid sequence $\tilde{X}_{i,k}^m$, see \citet[Ch10]{pollard2002} for an introduction. For this, we need a bound on $\mathbb{E}|\tilde{X}_{i,k,\ell}^m|^3$, this will rely on moment bounds for strong-mixing random variables in \citet[Ch2]{rio1999}. From step 2. we have $\tilde{X}^m_{i,k,\ell} \overset{d}{=} X^m_{i,k,\ell} =  \sum_{t=(i-1)b_n + (i-1)l_n}^{ib_n + (i-1)l_n} Z^m_{t,k,\ell}$, where $Z^m_{t,k,\ell}$ are strong-mixing with coefficients $\alpha( [\cdot - m]^+ ) = \alpha_m(\cdot)$ to simplify notation below. Using the notation from \citet{rio1999}, let $\alpha_m^{-1}(u) = \inf \{ j \in \mathbb{N}, \alpha_m(j) \leq u \} = \sum_{j=0}^\infty \mathbbm{1}_{ u < \alpha_m(j) }$ for any $u \in [0,1]$. \citet[Th2.2]{rio1999} implies:\footnote{The bound is only available for even moments $2p$ with $p \geq 1$.}
\begin{align*}
  \mathbb{E}[ |X^m_{i,k,\ell}|^{4} ] \leq 12 a_4 b_n^2 \left[ \int_{0}^1 \min[\alpha_m^{-1}(u),b_n] Q_m^2(u)du \right]^2 + 12 b_4 b_n  \int_{0}^1 \min[\alpha_m^{-1}(u),b_n]^{3} Q_m^4(u)du,
\end{align*}
where $Q_m$ is the quantile function of $|Z^m_{t,k,\ell}|$ and $a_4,b_4$ are universal constants. Take $p \in \{2,4\}$, we now want to bound: 
\begin{align*} 
    \int_0^1 \min[\alpha_m^{-1}(u),b_n]^{p-1} Q_m^p(u)du & \leq [\int_0^1 [\alpha_m^{-1}(u)]^{\vartheta (p-1) } du]^{1/\vartheta}[\int_0^1 Q_m(u)^{ p \vartheta/(\vartheta-1)} du]^{(\vartheta-1)/\vartheta}\\
    &= [\int_0^1 [\alpha_m^{-1}(u)]^{\vartheta (p-1) } du]^{1/\vartheta} \|Z_{t,k,\ell}^m\|_{p \vartheta/(\vartheta-1)}^{p},
\end{align*}
using $\int_{0}^1 Q_m(u)^p du = \mathbb{E}[|Z_{t,k,\ell}^m|^p]$. Apply inequality (C.5) in \citet[p156]{rio1999} to find:
\begin{align*} \int_0^1 [\alpha_m^{-1}(u)]^{\vartheta (p-1) } du &\leq 2 \sum_{j=0}^\infty (j+1)^{\vartheta (p-1) - 1} \alpha_m(j)\\ &= 2 \sum_{j=0}^{m-1} (j+1)^{\vartheta (p-1) - 1} \alpha(0) + 2 \sum_{j = 0}^\infty (m+j+1)^{\vartheta (p-1) - 1} \alpha(j)\\
&\leq 2 \frac{\alpha(0)}{\vartheta(p-1)-1} m^{\vartheta(p-1)} + 2 C (1+m)^{\vartheta(p-1)-1} \sum_{j = 0}^\infty (1 + j)^{\vartheta(p-1)-1 - a - \varepsilon }, \end{align*}
if $\alpha(j) \leq C (1+j)^{-(a+\varepsilon)}.$ Take $\vartheta = 2$, since $a \geq 6$ the series on the right-hand-side is summable. Then we get $\int_0^1 [\alpha_m^{-1}(u)]^{\vartheta (p-1) } du \leq  C_{a,p} m^{2(p-1)}$, for a constant which depends on $a+\varepsilon$, $C$ and $p$. Apply this to the fourth moment bound:
\begin{align*}
  \mathbb{E}[ |X^m_{i,k,\ell}|^{4} ] \leq 12 a_4 C_{a,2}\|Z_{t,k,\ell}^m\|_{4}^{4} (mb_n)^2 +  12 b_4 C_{a,4}^{1/2}\|Z_{t,k,\ell}^m\|_8^4  b_n m^{3} \leq O(\max[(mb_n)^2,b_nm^3]),
\end{align*}
and note that $\|Z_{t,k,\ell}^m\|_8$ is finite if $\|e_t\|_{16}$ is finite.
Let $L_{n,k}^m=\sum_{i=1}^{r_n} \mathbb{E}[\|n^{-1/2}X^m_{i,k}\|^4] = O(n^{-2} kr_n\max[(mb_n)^2,b_nm^3])$ and $V_{k}^m = \text{var}[n^{-1/2}\tilde{X}_{i,k}^m]$. By Theorem 6 in \citet{zaitsev2013}, there exists $\mathcal{Z}_{i,k}^m \overset{iid}{\sim} \mathcal{N}(0,V_{k}^m)$ such that:
$\sqrt{n}\|\tilde{X}_{n,k}^m - \mathcal{Z}_{n,k}^m\|_4 \leq C_0 16 [L_{n,k}^m]^{1/4}$,  
where $\mathcal{Z}_{n,k}^m = 1/n\sum_{i=1}^{r_n} \mathcal{Z}_{i,k}^m$ so that $\sqrt{n}\mathcal{Z}_{n,k}^m \sim \mathcal{N}(0, r_n/n V_{k}^m )$.

\noindent\textbf{Step 4. Overall Approximation Error.}
Now, we need to combine the steps to examine how well $\sqrt{n}\overline{Z}_{n,k}^m$ is approximated by the Gaussian vector $\sqrt{n}\mathcal{Z}_{n,k}^m$:
\begin{align*}
 \sqrt{n}\|\overline{Z}_{n,k} - \mathcal{Z}_{n,k}^m\| & \leq  \sqrt{n}\|\overline{Z}_{n,k} - \overline{Z}_{n,k}^m\| + \sqrt{n}\|\overline{Z}_{n,k}^m - \tilde{X}_{n,k}^m\| + \sqrt{n}\|\tilde{X}_{n,k}^m - \mathcal{Z}_{n,k}^m\|\\ &\leq O_p\Big( \max\Big[ k^{1/2}m^{-3/4(b+\varepsilon)},(n/b_n)^{1/2} k [1+l_n-m]^{-(a+\varepsilon)/2},\\ &\quad\quad (k l_n /b_n)^{1/2}, k^{1/2} n^{-1/4} \max[ m^{1/2} b_n^{1/4}, m^{3/4} ] \Big] \Big),
\end{align*}
using $\alpha(j) \leq (1+j)^{-(a+\varepsilon)}$ and $r_n \leq n/b_n$. Set $k = n^{1-\delta_1}$, $m = 2 n^{1-\delta_2}$, $b_n = n^{1-\delta_3}$ with $1 > \delta_1 \geq \delta_2 > \delta_3 > 0$ and $l_n = 2m$. The main idea here is to check, given the restrictions on $k,b$, and $a$, whether there are feasible choices of $(\delta_2,\delta_3)$ such that the above approximation error is negligible. For this, we have
\begin{align*}
  \sqrt{n}\|\overline{Z}_{n,k} - S_{n,k}^m\| & \leq  \sqrt{n}\|\overline{Z}_{n,k} - \overline{Z}_{n,k}^m\| + \sqrt{n}\|\overline{Z}_{n,k}^m - \tilde{X}_{n,k}^m\| + \sqrt{n}\|\tilde{X}_{n,k}^m - S_{n,k}^m\|\\ &\leq O_p\Big( \max\Big[ n^{ 1/2-3/4(b+\varepsilon) - \delta_1/2 + 3/4(b+\varepsilon)\delta_2},n^{1 - (a+\varepsilon)/2 + \delta_3/2 - \delta_1 + \delta_2(a+\varepsilon)/2},\\ &\quad\quad  n^{(1-(\delta_1+\delta_2) + \delta_3)/2}, n^{1 - (\delta_1 + \delta_2)/2}\max[ n^{ -\delta_3/4}, n^{-\delta_2/4} ] \Big] \Big),
 \end{align*}
 which is a $o_p([\log(n)]^{-2})$ if: 
 \begin{align*} 
    \delta_1 + \frac{3 (b+\varepsilon) }{2}  &> 1+\frac{3(b+\varepsilon)}{2} \delta_2 , && \delta_1 + \frac{a+\varepsilon}{2} > 1 + \frac{a+\varepsilon}{2}\delta_2 + \frac{1}{2}\delta_3 \\
    \delta_1+\delta_2 &> 1 + \delta_3,  && \delta_1 + \delta_2 + \frac{1}{2}\delta_3 > 2.
 \end{align*}
The last row of inequalities implies $1 > \delta_1 \geq \delta_2 \geq \delta_3 > 2/3$ and $\delta_1 > 1 + \delta_3-\delta_2$. The top left inequality is not binding for $b + \varepsilon > 2$, and the top right inequality further yields: $\delta_1 > [ 5- (a+\varepsilon) ]/3$ which is not binding for $a+\varepsilon >6$. This implies that there is a feasible solution for which $\delta_1 > 3/4 > 2/3$, i.e. $k = o(n^{1/4})$, such that $\sqrt{n}\|\overline{Z}_{n,k} - S_{n,k}^m\|_{\infty} = o_p( n^{-\delta} )$ for some $\delta \in (0,1)$ which depends on $(a,b,\varepsilon)$. It can be found by minimizing the rates above with respect to $(\delta_1,\delta_2,\delta_3)$ over the feasible set.

The covariance matrix of $\sqrt{n}\overline{Z}_{n,k}$ is given by $S_{n,k} = \text{var}[ n^{-1/2} \sum_{t=1}^n Z_{t,k} ]$, whereas the variance of $\sqrt{n}\mathcal{Z}_{n,k}^m$ is, by construction, equal to $\text{var}[ n^{-1/2} \sum_{i=1}^{r_n} \tilde{X}^m_{i,k} ] = n^{-1}r_n \text{var}[X_{i,k}^m]:= S_{n,k}^m$ by independence and equality in distribution. Let $X_{i,k} = \sum_{t = (i-1)b_n + (i-1)l_n}^{i_b + (i-1)l_n}Z_{t,k}$, we have:
\[ \|S_{n,k} - S_{n,k}^m\| \leq \|S_{n,k} - n^{-1}r_n \text{var}[X_{i,k}]\| + n^{-1}r_n\|\text{var}[X_{i,k}] - \text{var}[X_{i,k}^m]\|. \]
Starting with the first term on the right-hand-side, standard calculations imply for $\Gamma_{k,j} = \mathbb{E}(Z_{t,k}Z_{t-j,k}^\prime)$:
\begin{align}
  S_{n,k} - n^{-1}r_n \text{var}[X_{i,k}] &= \Gamma_{k,0} + \sum_{j=1}^{n-1} \frac{n-j}{n}\left[ \Gamma_{k,j} + \Gamma_{k,j}^\prime \right] - \frac{r_nb_n}{n}\left( \Gamma_{k,0} + \sum_{j=1}^{b_n-1} \frac{b_n-j}{b_n}\left[ \Gamma_{k,j} + \Gamma_{k,j}^\prime \right] \right) \notag\\
  &= \sum_{j=1}^{b_n-1}(b_n^{-1} - n^{-1})j\left[ \Gamma_{k,j} + \Gamma_{k,j}^\prime \right] \tag{S1} \label{eq:V1}\\ &+ \sum_{j=b_n+1}^{n-1} \frac{n-j}{n}\left[ \Gamma_{k,j} + \Gamma_{k,j}^\prime \right] \tag{S2} \label{eq:V2}\\
  &+ [b_nr_n/n-1] \left( \Gamma_{k,0} + \sum_{j=1}^{b_n-1} \frac{b_n-j}{b_n}\left[ \Gamma_{k,j} + \Gamma_{k,j}^\prime \right] \right). \tag{S3} \label{eq:V3}
\end{align}
Begin with (\ref{eq:V2}). Recall from step 1 that elements of $Z_{t,k}$ corresponding to $\text{vec}(e_t \tilde{Y}_{t-1,k}^\prime \Gamma_k^{-1})$ are serially uncorrelated. The remaining terms correspond to $\tilde{y}_t - \tilde{\mu}$ and $\text{vech}[ e_te_t^\prime - \tilde{\Sigma} ]$. Using previous calculations, for any $j \geq 1$: $\|\Gamma_{k,j}\| = \| \mathbb{E}( Z_{t,k}| \mathcal{F}_{t-j} )Z_{t-j,k}^\prime \| \leq \|\mathbb{E}( Z_{t,k}| \mathcal{F}_{t-j} )\|_{2}\|Z_{t,k}\|_2.$ There are two bounds to compute here: $\|\mathbb{E}( \tilde{y}_{t} - \tilde{\mu}| \mathcal{F}_{t-j} )\| \leq C (1+j)^{-(b+\varepsilon)}\|e_t\|_2$ and $\|\mathbb{E}(\text{vech}[ e_te_t^\prime - \tilde{\Sigma} ]|\mathcal{F}_{t-j})\|_2 \leq 6 d^2 (1+j)^{-(a+\varepsilon)3/8}\|e_t\|_{16}$ using \citet[Th15.2]{Davidson2021} with $p=2$ and $r=8$; $d = \text{dim}(e_t)$. Using these two bounds, we find:
\begin{align*} \|(\ref{eq:V2})\| &\leq 4[C + 6d^2]\|e_t\|_{16}\|Z_{t,k}\|_2 \sum_{j=b_n+1}^{\infty} (1+j)^{-\min((3/8)[a+\varepsilon],(b+\varepsilon))}\\ &\leq \frac{4[C + 6d^2]\|e_t\|_{16}\|Z_{t,k}\|_2 b_n^{1-\min((3/8)[a+\varepsilon],(b+\varepsilon))}  }{\min((3/8)[a+\varepsilon],(b+\varepsilon))},\end{align*}
which is a $o(b_n^{-5/4}k^{1/2})$ since $a,b \geq 6$ and $\varepsilon > 0$ by assumption. Since $b_n > k$, this implies that $\|(\ref{eq:V2})\| = o(k^{-3/4})$. For (\ref{eq:V1}), using $b_n < n$ and the same bounds:
\[ \|(\ref{eq:V1})\| \leq 4b_n^{-1} 4[C + 6d^2]\|e_t\|_{16}\|Z_{t,k}\|_2 \sum_{j=1}^\infty (1+j)^{1-\min(3/8[a+\varepsilon],(b+\varepsilon))} = O(b_n^{-1}k^{1/2}). \]
It is possible to pick $b_n^{-1} = o(k^{-1}[\log(n)]^{-6})$, from the constraints above.
For (\ref{eq:V3}), recall that $n = b_nr_n + (r_n-1)l_n + o(1)$ so that $b_nr_n/n - 1 = (r_n-1)/n l_n + o(n^{-1}) = O(l_n/b_n) + o(n^{-1})$. Looking at the feasible values for $(\delta_1,\delta_2,\delta_3)$ above, we have $1-\delta_1 < \delta_2-\delta_3$ which implies $l_n/b_n = o(k^{-1}[\log(n)]^{-6})$ is feasible, which implies:
\[ \|(\ref{eq:V3})\| \leq o(k^{-1}[\log(n)]^{-6}) \sum_{j=0}^\infty\|\Gamma_{k,j}\| = o(k^{-1/2}[\log(n)]^{-6}). \]
Altogether, we find:
\[ \|S_{n,k} - n^{-1}r_n \text{var}(X_{i,k})\| \leq o_p(k^{-1/2}[\log(n)]^{-6}). \]

Now, we consider $n^{-1}r_n \|\text{var}[X_{i,k}]-\text{var}[X^m_{i,k}]\| \leq 2 n^{-1}r_n\| \mathbb{E}[(X_{i,k}-X_{i,k}^m)(X_{i,k}-X_{i,k}^m)^\prime] \| \leq n^{-1}r_nb_n^2 \|Z_{t,k}-Z_{t,k}^m\|_2 \|Z_{t,k} + Z_{t,k}^m\|_2 \leq O( b_n k^{1/2}m^{-3/4(b+\varepsilon)} ).$ Using the strategy from above, set $b_n = n^{1-\delta_3}$, $k = n^{1-\delta_1},$ and  $m = n^{1-\delta_2}$. Using the inequalities $\delta_1 > 3/4$, $\delta_3 > 2/3$, $b+\varepsilon > 4$, we get an upper bound $\delta_2 < 61/72$, which is within the feasible set so that we can further set $b_n,m$, such that $n^{-1}r_n \|\text{var}[X_{i,k}]-\text{var}[X^m_{i,k}]\| = o_p([\log(n)]^{-6})$. This implies that overall we have:
$\|S_{n,k} - S_{n,k}^m\| = o_p([\log(n)]^{-6})$.

We can write $\sqrt{n}\mathcal{Z}_{n,k}^m = (S_{n,k}^m)^{1/2} \mathcal{Z}_{k}$ where $\mathcal{Z}_{k} \sim \mathcal{N}(0,I)$ and let $\sqrt{n}\mathcal{Z}_{n,k} = (S_{n,k})^{1/2} \mathcal{Z}_{k} \sim \mathcal{N}(0,S_{n,k})$ have the desired covariance structure. Now, apply the inequality $\|S^{1/2}_1-S^{1/2}_2\| \leq \|S_1-S_2\|^{1/2}$ and H\"older's inequality to find $\| [(S_{n,k}^m)^{1/2}-(S_{n,k})^{1/2}]\mathcal{Z}_{n,k} \| \leq \|S_{n,k}^m-S_{n,k}\|^{1/2} \|\mathcal{Z}_{n,k}\|_{\infty}$, which is less than $[\log(n)]^{-3} \|\mathcal{Z}_{n,k}\|_{\infty}$ with probability approaching one since $\|S_{n,k}^m-S_{n,k}\| = o_p([\log(n)]^{-6})$. Combine this with Chernoff's bound:
\[ \mathbb{P} \Big( \| [(S_{n,k}^m)^{1/2}-(S_{n,k})^{1/2}]\mathcal{Z}_{n,k} \| > [\log(n)]^{-2} \Big) \leq 2d(k)\exp\left( - [\log(n)]^2\right) + o(1) = o(1),  \]
since $d(k) = O(k)$ and $\log(k) = o([\log(n)]^2)$. This implies that $\| [(S_{n,k}^m)^{1/2}-(S_{n,k})^{1/2}]\mathcal{Z}_{n,k} \| = o_p([\log(n)]^{-2})$. Combining all the results together, we find:
$ \sqrt{n}\| \overline{Z}_{n,k} - \tilde{Z}_{n,k} \| = o_p([\log(n)]^{-2}),  $
where $\sqrt{n}\tilde{Z}_{n,k} \sim \mathcal{N}(0,S_{n,k})$ as desired, which implies the strong approximation result.\qed 

\paragraph{Proof of Theorem \ref{th:spec}.} For correctly specified models, we have $\tilde{y}_t = y_t(\theta_0;\psi_0)$ and thus
\begin{align} nQ_n(\hat{\theta}_n;\hat{\psi}_{nk}) = \sum_{t=1}^n (y_t(\hat{\theta}_n;\hat{\psi}_{nk})-y_t(\theta_0;\psi_0))^\prime W_n (y_t(\hat{\theta}_n;\hat{\psi}_{nk})-y_t(\theta_0;\psi_0)). \label{eq:nQn} \end{align}

\newcommand{\ybar}{\pmb{\bar{y}_t}} 

To simplify notation, define $\partial_\theta \ybar(\hat{\theta}_n,\theta_0;\psi) = \int_{0}^1 \partial_\theta y_t( \omega \hat{\theta}_n + (1-\omega)\theta_0;\psi )d\omega$, for any $\psi$, and  $\partial_\psi \ybar(\theta;\hat{\psi}_{nk},\psi_k) = \int_{0}^1 \partial_\theta y_t( \theta;\omega \hat{\psi}_{nk} + (1-\omega) \psi_{k} )d\omega$, for any $\theta$. The proof consists of five steps as follows.

\noindent\textbf{Step 1. Expansion for each $y_t(\hat{\theta}_n;\hat{\psi}_{nk})-y_t(\theta_0;\psi_0)$.}
For each $t = 1,\dots,n$, we have:
\[ y_t(\hat{\theta}_n;\hat{\psi}_{nk}) - y_t(\theta_0;\psi_k) = \partial_\psi \ybar(\theta_0;\hat{\psi}_{nk},\psi_k)(\hat{\psi}_{nk}-\psi_k) + \partial_{\theta} \ybar(\hat{\theta}_{n},\theta_0;\hat{\psi}_{nk})(\hat{\theta}_{n} - \theta_0). \]
Before expanding on (\ref{eq:nQn}), we express $\hat{\theta}_{n} - \theta_0$ in terms of $\hat{\psi}_{nk}-\psi_k$. 
We begin with the following representation, derived in the proof of Theorem \ref{th:asym_normal}:
\[ \hat{\theta}_n - \theta_0 = M_n^{-1}[ (\ref{eq:C}) + (\ref{eq:D}) + (\ref{eq:E})], \]
where (\ref{eq:D}) and (\ref{eq:E}) involve $\hat{\psi}_{nk} - \psi_k$, while (\ref{eq:C}) only involves $u_{t,k} = y_t(\theta_0;\psi_k) - \tilde{y}_t$. Let $D_n = -  \frac{1}{n} \sum_{t=1}^n (u_{t,k}^\prime W_n \otimes I) \partial_\psi \pmb{\bar{G}_t}(\theta_0;\hat{\psi}_{nk},\psi_k)$ and $E_n = - \frac{1}{n} \sum_{t=1}^n \partial_\theta y_t(\theta_0;\hat{\psi}_{nk})^\prime W_n \partial_\psi \ybar(\theta_0;\hat{\psi}_{nk},\psi_k)$. Then, $\hat{\theta}_n - \theta_0 = M_n^{-1}(\ref{eq:C}) + M_n^{-1}[D_n + E_n](\hat{\psi}_{nk} - \psi_{k})$. From these expressions, we derive:
\begin{align*} y_t(\hat{\theta}_n;\hat{\psi}_{nk}) - y_t(\theta_0;\psi_0) &= \{\partial_\psi y_t(\theta_0;\tilde{\psi}_{k,t}) + \partial_{\theta} y_t(\tilde{\theta}_{n,t};\hat{\psi}_{nk})M_n^{-1}[D_n+E_n]\}(\hat{\psi}_{nk}-\psi_k) \\ &+ \partial_{\theta} y_t(\tilde{\theta}_{n,t};\hat{\psi}_{nk})M_n^{-1}(\ref{eq:C}) + y_t(\theta_0;\psi_k) - y_t(\theta_0;\psi_0), \end{align*}
where $y_t(\theta_0;\psi_k) - y_t(\theta_0;\psi_0) = u_{t,k}$ because the model is correctly specified.

We now establish an upper bound for (\ref{eq:C}) to be used subsequently. For $t=k+1,\dots,n$:
$ u_{t,k} = y_t(\theta_0;\psi_k) - y_t(\theta_0;\psi_0) = \sum_{\ell = 0}^\infty \Lambda_\ell(\theta_0)P(\theta_0;\tilde{\Sigma})[ e_{t,k} - e_t ],
$
where $e_{t,k} = \tilde{y}_t - \tilde{\mu} - \sum_{j=1}^k \Psi_j [\tilde{y}_{t-j} - \tilde{\mu}] = e_t + \sum_{j=k+1}^\infty \Psi_j [\tilde{y}_{t-j} - \tilde{\mu}]$. Assumption \ref{ass:stronger} implies $\sqrt{n}[\log(n)]^2(\mathbb{E}[\|e_t-e_{t,k}\|^2])^{1/2} = o(1)$, and in combination with Assumption \ref{ass:InnovModel} (ii), it yields $\sqrt{n}[\log(n)]^2(\mathbb{E}[\|u_t-u_{t,k}\|^2])^{1/2} = o(1)$. Thus, for  (\ref{eq:C}), we have:
\begin{align*} \|(\ref{eq:C})\| &\leq \|W_n-W\|_{op}(1/n) \sum_{t=1}^n \|\partial_{\theta}y_t(\theta_0;\psi_k)\|_{\infty}  \|u_{t,k}\| + \overline{\lambda}_W (1/n) \sum_{t=1}^n \|\partial_{\theta}y_t(\theta_0;\psi_k)\|_{\infty}  \|u_{t,k}\|\\
& \leq [1+o_p(1)][o_p( n^{-1/2} [\log(n)]^{-2} ) + O_p(k n^{-1})]  =o_p( n^{-1/2} [\log(n)]^{-2} ), \end{align*}
using the Cauchy-Schwarz inequality and the above results. The first $t=1,\dots,k$ terms in each summation have finite moments, contributing $O_p(k/n)$ to the average, as in the proofs of Theorems \ref{th:cons} and \ref{th:asym_normal}, where $k = o(n^{1/4}).$ They yield the $O_p(kn^{-1})$ on the right hand side.\\

\noindent\textbf{Step 2. Expanding $n Q_n(\hat{\theta}_n;\hat{\psi}_{nk})$ to show: $n Q_n(\hat{\theta}_n;\hat{\psi}_{nk}) = n(\hat{\psi}_{nk}-\psi_k)^\prime M_{k} (\hat{\psi}_{nk}-\psi_k) + o_p(1)$, with $M_{k}$ defined below in equation (\ref{equ:Mk}). }

Square and sum the terms from Step 1:
\begin{align}
 & n Q_n(\hat{\theta}_n;\hat{\psi}_{nk}) = n (\hat{\psi}_{nk}-\psi_k)^\prime M_{n,k} (\hat{\psi}_{nk}-\psi_k) \tag{Q1} \label{eq:Q1}\\
  &+ 2 (\hat{\psi}_{nk}-\psi_k)^\prime \sum_{t=1}^n  B_{n,t}^\prime W_n \{ \partial_{\theta} \ybar(\hat{\theta}_{n},\theta_0;\hat{\psi}_{nk})M_n^{-1}(\ref{eq:C}) + u_{t,k} \}\tag{Q2} \label{eq:Q2}\\
  &+ \sum_{t=1}^n \{ \partial_{\theta} \ybar(\hat{\theta}_{n},\theta_0;\hat{\psi}_{nk})M_n^{-1}(\ref{eq:C}) + u_{t,k} \}^\prime W_n \{ \partial_{\theta} \ybar(\hat{\theta}_{n},\theta_0;\hat{\psi}_{nk})M_n^{-1}(\ref{eq:C}) + u_{t,k} \} \tag{Q3} \label{eq:Q3}
\end{align}
where $B_{n,t} = \partial_\psi \ybar(\theta_0;\hat{\psi}_{nk},\psi_k) + \partial_{\theta} \ybar(\hat{\theta}_{n},\theta_0;\hat{\psi}_{nk})M_n^{-1}[D_n+E_n]$, and 
$M_{n,k} = \frac{1}{n}\sum_{t=1}^n B_{n,t}^\prime  W_n B_{n,t}$. 

Next, we will show that (\ref{eq:Q2}) and (\ref{eq:Q3}) are negligible. Recall that $\|\hat{\psi}_{nk}-\psi_k\|_{\infty} \leq O_p(\sqrt{\log(n)/n})$ (cf. Lemma \ref{lem:VARinf}), and $\|(\ref{eq:C})\| \leq o_p(n^{-1/2}[\log(n)]^{-2})$. For (\ref{eq:Q2}):
\begin{align*}
  (\ref{eq:Q2}) &= 2\sqrt{n}(\hat{\psi}_{nk}-\psi_k)^\prime \Big\{ \frac{1}{n} \sum_{t=1}^n B_{n,t}^\prime W_n \partial_{\theta} y_t(\tilde{\theta}_{n,t};\hat{\psi}_{nk}) \Big\} M_n^{-1} \sqrt{n} (\ref{eq:C})\\
  &+ \sqrt{n}(\hat{\psi}_{nk}-\psi_k)^\prime \Big\{ \frac{1}{n} \sum_{t=1}^n B_{n,t}^\prime W_n \sqrt{n}u_{t,k} \Big\}.
\end{align*}
The proof of Theorem \ref{th:asym_normal} implies $||\frac{1}{n} \sum_{t=1}^n B_{n,t}^\prime W_n \partial_{\theta} \ybar(\hat{\theta}_{n},\theta_0;\hat{\psi}_{nk})|| = ||\mathbb{E}[ \{\partial_\psi y_t(\theta_0;\psi_k) + \partial_\theta y_t(\theta_0;\psi_k)M^{-1}\{ \mathbb{E}[(u_{t,k}^\prime W \otimes I) \partial_\psi G_t(\theta_0;\psi_k) + \partial_\theta y_t(\theta_0;\psi_k) \}^\prime W \partial_\theta y_t(\theta_0;\psi_k)] \}]|| + o_p(1)$. The expectation on the right hand side is absolutely summable, hence bounded. Thus, for the first part of (\ref{eq:Q2}), we get: $\|\sqrt{n}(\hat{\psi}_{nk}-\psi_k)^\prime \Big\{ \frac{1}{n} \sum_{t=1}^n B_{n,t}^\prime W_n \partial_{\theta} \ybar(\hat{\theta}_{n},\theta_0;\hat{\psi}_{nk}) \Big\} M_n^{-1} \sqrt{n} (\ref{eq:C})\| \leq O_p(\sqrt{\log(n)})o_p([\log(n)]^{-2}) = o_p([\log(n)]^{-3/2})$. For the second part of (\ref{eq:Q2}), following the proof of Theorems \ref{th:cons} and \ref{th:asym_normal}, we have $||\frac{1}{n} \sum_{t=1}^n B_{n,t}^\prime W_n \sqrt{n} u_{t,k}|| = ||\frac{1}{n} \sum_{t=1}^n [\partial_\psi y_t(\theta_0;\psi_{k}) + \partial_{\theta} y_t(\theta_{0};\psi_{k})M^{-1}\{ \mathbb{E}[(u_{t,k}^\prime W \otimes I) \partial_\psi G_t(\theta_0;\psi_k) + \partial_\theta y_t(\theta_0;\psi_k)^\prime W \partial_\psi y_t(\theta_0;\psi_k)] \}]^\prime W \sqrt{n}u_{t,k}|| + o_p(1)$.
From the expression for $\partial_{\phi_j} y_t(\theta_0;\psi_k)$ in Lemma \ref{lem:NED}, the summability conditions in Assumption \ref{ass:InnovModel}, and the moment condition in Assumption \ref{ass:DGPdata}, we have: $\mathbb{E}[\|\partial_{\phi_j} y_t(\theta_0;\psi_k)\|^q]$ is bounded uniformly in $j \geq 1$ for $q = 2r > 8$, and $\mathbb{E}[\|\partial_{\psi} y_t(\theta_0;\psi_k)\|^2_{\infty}] \leq O(k^{2/q})$. We have $\|\partial_{\psi} y_t(\theta_0;\psi_k)^\prime W \sqrt{n}u_{t,k}\| \leq \overline{\lambda}_W \|\partial_{\psi} y_t(\theta_0;\psi_k)\|_{\infty} \| \sqrt{n}u_{t,k}\|$ so that with the Cauchy-Schwarz inequality: $\mathbb{E}[\|\partial_{\psi} y_t(\theta_0;\psi_k)^\prime W \sqrt{n}u_{t,k}\|] \leq O(k^{1/q})o_p( n^{-1/(4r)}[\log(n)]^{-2})$, where $k = o(n^{1/4})$ by Assumption \ref{ass:stronger}. Since $\partial_\theta y_t(\theta_0;\psi_k)$ is of fixed dimension, the same holds. Eventually, we get: $\|\sqrt{n}(\hat{\psi}_{nk}-\psi_k)^\prime \Big\{ \frac{1}{n} \sum_{t=1}^n B_{n,t}^\prime W_n \sqrt{n}u_{t,k} \Big\}\| \leq O_p(\sqrt{\log(n)})o_p([\log(n)]^{-2}) = o_p([\log(n)]^{-3/2})$. Overall, we find $ (\ref{eq:Q2}) = o_p([\log(n)]^{-3/2})$. Following the same steps as for (\ref{eq:Q2}), we can show that $\|(\ref{eq:Q3})\| \leq o_p([\log(n)]^{-3/2})$. Overall, we get:
$nQ_n(\hat{\theta}_n;\hat{\psi}_{nk}) = n(\hat{\psi}_{nk}-\psi_k)^\prime M_{n,k} (\hat{\psi}_{nk}-\psi_k) + o_p([\log(n)]^{-3/2})$. 
The last part of this step is to show $M_{n,k} = M_k + o_p( [\log(n)]^{-1} )$. This is similar to the derivations for $M_n$ in the proof of Theorem \ref{th:asym_normal}, but now the matrix has $O(k^2)$ elements instead of a fixed dimension. Write $M_{n,k}$ as follows:
\begin{align}
  M_{n,k} &= \frac{1}{n} \sum_{t=1}^n \partial_{\psi} \ybar(\theta_0;\hat{\psi}_{nk},\psi_k)^\prime W_n \partial_{\psi} \ybar(\theta_0;\hat{\psi}_{nk},\psi_k) \tag{M1} \label{eq:M1}\\
  &+ [D_n + E_n]^\prime M_n^{-1}\left\{ \frac{1}{n} \sum_{t=1}^n \partial_{\theta} \ybar(\hat{\theta}_n,\theta_0;\hat{\psi}_{nk})^\prime W_n \partial_{\theta} \ybar(\hat{\theta}_n,\theta_0;\hat{\psi}_{nk}) \right\} M_n^{-1}[D_n + E_n] \tag{M2} \label{eq:M2}\\
  &+ [D_n + E_n]^\prime M_n^{-1}\left\{ \frac{1}{n} \sum_{t=1}^n \partial_{\theta} \ybar(\hat{\theta}_n,\theta_0;\hat{\psi}_{nk})^\prime W_n \partial_{\psi} \ybar(\theta_0;\hat{\psi}_{nk},\psi_k) \right\} \tag{M3} \label{eq:M3}\\
  &+  \left\{ \frac{1}{n} \sum_{t=1}^n \partial_{\psi} \ybar(\theta_0;\hat{\psi}_{nk},\psi_k)^\prime W_n \partial_{\theta} \ybar(\hat{\theta}_n,\theta_0;\hat{\psi}_{nk}) \right\} M_n^{-1} [D_n + E_n]. \tag{M4} \label{eq:M4}
\end{align}

Note that $M_{n,k}$ depends on estimates $\hat{\theta}_n,\hat{\psi}_{nk}$, and $\bar{\psi}_k = \omega \hat{\psi}_{nk} + (1-\omega)\psi_k$, while $M_k$ depends on the true values $\theta_0$ and $\psi_k$. We now show that the former parameter values can be replaced by the later with a bounded error term. Specifically, for any intermediate value $\bar{\psi}_k$ of $\hat{\psi}_{nk}$ and $\psi_k$, we have, for $t \geq k+1$: $\partial_{\tilde{\mu}}\text{vec}[\partial_{\tilde{\mu}} y_t(\theta;\bar{\psi}_k)] = 0$, $\partial_{\phi_j}\text{vec}[\partial_{\tilde{\mu}} y_t(\theta;\bar{\psi}_k)] = - I_d \otimes [\sum_{\ell = 0}^\infty \Lambda_{\ell}(\theta_0)P(\theta_0;\bar{\Sigma})]$, which is bounded; $\partial_{\text{vech}[\tilde{\Sigma}]}\text{vec}[\partial_{\tilde{\mu}} y_t(\theta;\bar{\psi}_k)] = [ I_d - \sum_{\ell=1}^k \bar{\Psi}_\ell ]^\prime \otimes \left[ \sum_{\ell=0}^{t-1} \Lambda_{\ell}(\theta_0) \right] \partial_{\text{vech}}\text{vec}[P(\theta_0;\bar{\Sigma})]$, which is also bounded with probability approaching $1$ since $\|\bar{\psi}_k-\psi_k\|_{\infty} = O_p(\sqrt{\log(n)/n})$. For any $ 1 \leq j,\ell \leq k$, 
$\partial_{\phi_j} \text{vec}[\partial_{\phi_\ell} y_t(\theta;\bar{\psi}_k)] = 0$ and 
$\partial_{\text{vech}[\tilde{\Sigma}]_{i}} \text{vec}[\partial_{\phi_j} y_t(\theta;\bar{\psi}_k)] = - \sum_{\ell=0}^{t-1} [\tilde{y}_{t-\ell - j - 1}-\tilde{\mu}]^\prime \otimes [\Lambda_{\ell}(\theta_0) \partial_{\text{vech}[\tilde{\Sigma}]_{i}}P(\theta_0;\bar{\Sigma})]$. Using $(AC )\otimes (BD) = (A \otimes B)(C \otimes D)$ for conformable matrices, we get $\partial_{\text{vech}[\tilde{\Sigma}]_{i}} \text{vec}[\partial_{\phi_j} y_t(\theta;\bar{\psi}_k)] =  \{ - \sum_{\ell=0}^{t-1} [\tilde{y}_{t-\ell - j - 1}-\tilde{\mu}]^\prime [\Lambda_{\ell}(\theta_0) \}  \{ I_d \otimes \partial_{\text{vech}[\tilde{\Sigma}]_{i}}P(\theta_0;\bar{\Sigma})] \}$, and thus $\|\partial_{\text{vech}[\tilde{\Sigma}]_{i}} \text{vec}[\partial_{\phi_j} y_t(\theta;\bar{\psi}_k)]\| \leq \| \sum_{\ell=0}^{t-1} [\tilde{y}_{t-\ell - j - 1}-\tilde{\mu}]^\prime [\Lambda_{\ell}(\theta_0) \| \sup_{ \tilde{\Sigma} } \partial_{\text{vech}[\tilde{\Sigma}]_{i}} \|P(\theta;\tilde{\Sigma})\|_{\infty}$, which has bounded $q$-th moment uniformly in $t$. The same steps reveal that $\|\partial_{\text{vech}[\tilde{\Sigma}]_{i}} \text{vec}[\partial_{\text{vech}[\tilde{\Sigma}]} y_t(\theta;\bar{\psi}_k)]\|$ also has bounded $q$-th moment, uniformly in $t$. Let $\Psi_{nk} = \{\bar{\psi}_k, \|\bar{\psi}_k - \psi_k\|_{\infty} \leq n^{-1/2} \log(n)\}.$ The above results and \citet{pisier1983}'s inequality imply:\footnote{\citet{pisier1983}:  $(\mathbb{E}[\sup_{j=1,\dots,k} |X_j|^q ])^{1/q} \leq  k^{1/q} (\sup_{j=1,\dots,k}\mathbb{E}[|X_j|^q ])^{1/q}$ for any random variables $X_1,\dots,X_k$ with finite $q$-th moment.} $ (\mathbb{E}[ \sup_{ \bar{\psi}_k \in \Psi_{nk} }\|\partial^2_{\psi} y_t(\theta_0;\bar{\psi}_k)\|_{\infty}^q ])^{1/q} \leq O(k^{1/q}),$
as the number of non-zero derivatives is linear in $k$. We now apply this upper bound to analyze the following decomposition for (\ref{eq:M1}), which holds with probability approaching 1:
\begin{align*} &\|(\ref{eq:M1}) - \frac{1}{n} \sum_{t=1}^n \partial_{\psi} y_t(\theta_0;\psi_k)^\prime W_n \partial_{\psi} y_t(\theta_0;\psi_k) \| \\ \leq &2\lambda_{\max}[W_n] \|\hat{\psi}_{nk}-\psi_k\|_{\infty} \frac{1}{n} \sum_{t=1}^n \Big\{  \big[\sup_{ \bar{\psi}_k \in \Psi_{nk} }\|\partial^2_{\psi} y_t(\theta_0;\bar{\psi}_k)\|_{\infty} \big] \big[  \sup_{ \bar{\psi}_k \in \Psi_{nk}  } \|\partial_\psi y_t(\theta_0;\bar{\psi}_{k})\|_{\infty} \big] \Big\}. \end{align*}
Among the right hand side terms, $\|\hat{\psi}_{nk}-\psi_k\|_{\infty}$ is $O_p(n^{-1/2}k^{-1/2})$, $(\mathbb{E}[\|\partial^2_\psi y_t(\theta_0;\bar{\psi}_{k})\|_{\infty}^{q}])^{1/q}$ is $O(k^{1/q})$ as we have shown, and $(\mathbb{E}[\|\partial_\psi y_t(\theta_0;\bar{\psi}_{k})\|_{\infty}^{q}])^{1/q} = O(k^{1/q})$ using similar derivations as for $\partial^2_\psi y_t(\theta_0;\bar{\psi}_{k})$. Applying these results and the Cauchy-Schwarz inequality, the right-hand side of the inequality above is $O_p( n^{-1/2}k^{1/2} k^{1/2}\log(n) ) = o_p( n^{-1/4}\log(n) )$, since $k = o(n^{1/4})$ by Assumption. Similar results hold for (\ref{eq:M2}), (\ref{eq:M3}), and (\ref{eq:M4}) using the fact that $\|\hat{\theta}_n - \theta_0\| \leq n^{-1/2} \log(n)$ with probability approaching 1 (cf. Theorem \ref{th:asym_normal}). These results imply that replacing $\hat{\theta}_n,\hat{\psi}_{nk}$, and $\bar{\psi}_k = \omega \hat{\psi}_{nk} + (1-\omega)\psi_k$ by their true values $\theta_0,$, $\psi_k$, and $\psi_k$ impacts $M_{n,k}$ by no more than $o_p( n^{-1/4}\log(n) )$.

An additional difference between $M_{n,k}$ and $M_k$ is that $M_{n,k}$ depends on $W_n$, while $M_k$ depends on $W$. Using similar arguments as above, the effect of this substitution is at most $o_p( n^{-1/4}\log(n) )$ because $\|W_n-W\| = O_p(n^{-1/2})$. Therefore, we have: $M_{n,k}=\bar{M}_{n,k}+o_p( n^{-1/4}\log(n) )$,
with
\begin{align}
  \bar{M}_{n,k} &= (1/n) \sum\nolimits_{t=1}^n \partial_{\psi} y_t(\theta_0;\psi_k)^\prime W \partial_{\psi} y_t(\theta_0;\psi_k) \tag{\ref{eq:M1}'} \label{eq:M1p}\\
  &+ [D_n + E_n]^\prime M_n^{-1}\{ (1/n) \sum\nolimits_{t=1}^n \partial_{\theta} y_t(\theta_0;\psi_k)^\prime W \partial_{\theta} y_t(\theta_0;\psi_k) \} M_n^{-1}[D_n + E_n] \tag{\ref{eq:M2}'} \label{eq:M2p}\\
  &+ [D_n + E_n]^\prime M_n^{-1}\{ (1/n) \sum\nolimits_{t=1}^n \partial_{\theta} y_t(\theta_0;\psi_k)^\prime W \partial_{\psi} y_t(\theta_0;\tilde{\psi}_{k,t}) \} \tag{\ref{eq:M3}'} \label{eq:M3p}\\
  &+  \{ (1/n) \sum\nolimits_{t=1}^n \partial_{\psi} y_t(\theta_0;\psi_{k})^\prime W \partial_{\theta} y_t(\theta_0;\psi_k) \} M_n^{-1} [D_n + E_n]. \tag{\ref{eq:M4}'} \label{eq:M4p}\\
  &+ o_p( n^{-1/4}\log(n) ). \notag
\end{align}
This implies: $nQ_n(\hat{\theta}_n;\hat{\psi}_{nk}) = n (\hat{\psi}_{nk}-\psi_k)^\prime \bar{M}_{n,k} (\hat{\psi}_{nk}-\psi_k) + o_p([\log(n)]^{-3/2})$.

Now, we derive a law of large numbers for (\ref{eq:M1p})-(\ref{eq:M4p}). We will only consider (\ref{eq:M1p}) since the others are similar. 
For this, we first derive a bound for $\partial_{\phi_j} y_t(\theta_0;\psi_k) - \partial_{\phi_j} y_t(\theta_0;\psi_{k0})$, so that we eventually can use $\mathbb{E}[\partial_{\psi} y_t(\theta_0;\psi_{k0})^\prime W \partial_{\psi} y_t(\theta_0;\psi_{k0})]$ to approximate (\ref{eq:M1p}). Note that $\partial_{\phi_j} y_t(\theta_0;\psi_k)$ sets $\tilde{y}_t-\tilde{\mu} = 0$ for $t \leq 0$, while the latter uses the actual, unobserved, $\tilde{y}_t-\tilde{\mu}$ for $t \leq 0$. As a result, the latter is stationary. For $t \geq 2k +1$ and any $j \in \{1,\dots,k\}$, we have
\[(\mathbb{E}[\|\partial_{\phi_j} y_t(\theta_0;\psi_k) - \partial_{\phi_j} y_t(\theta_0;\psi_{k0})\|^2])^{1/2} \leq \sum\nolimits_{\ell = k+1}^\infty \|\Lambda_j(\theta_0)\|_{\infty} \|P(\theta_0;\tilde{\Sigma})\|_\infty (\mathbb{E}[\|\tilde{y}_{t-j - \ell}\|^2])^{1/2} = o(n^{-1/2}),\] uniformly in $j$. Also, $(\mathbb{E}[\|\partial_{\psi} y_t(\theta_0;\psi_k) - \partial_{\psi} y_t(\theta_0;\psi_{k0})\|^2_{\infty}])^{1/2} = o(n^{-1/2}k^{1/2})$. By the Cauchy-Schwarz inequality: \[\frac{1}{n} \sum_{t=1}^n \partial_{\psi} y_t(\theta_0;\psi_k)^\prime W \partial_{\psi} y_t(\theta_0;\psi_k) = \frac{1}{n} \sum_{t=1}^n \partial_{\psi} y_t(\theta_0;\psi_{k0})^\prime W \partial_{\psi} y_t(\theta_0;\psi_{k0}) + o_p(n^{-1/2}k) = o_p(n^{-1/4}).\]
For any $m \geq 2k$, let $\mathcal{F}_{t-m}^t$ be the $\sigma$-field generated by $(e_t,\dots,e_{t-m})$. We have:
\begin{align*}
  \partial_{\phi_j} y_t(\theta_0;\psi_{k0}) - \mathbb{E}[\partial_{\phi_j} y_t(\theta_0;\psi_{k0}) | \mathcal{F}_{t-m}^t] &= \sum_{\ell = 0}^\infty \Lambda_\ell(\theta_0)P(\theta_0;\tilde{\Sigma}) \sum_{s=0}^\infty \tilde{\Lambda}_s [e_{t-s-j-\ell}-\mathbb{E}(e_{t-s-j-\ell}|\mathcal{F}_{t-m}^t)],
\end{align*}
for any $j \in \{1,\dots,k\}$, where $[e_{t-s-j-\ell}-\mathbb{E}(e_{t-s-j-\ell}|\mathcal{F}_{t-m}^t)] = 0$ for all $s+j+\ell \leq m$. Since $m \geq 2k$ and $j \leq k$, this representation holds for all $j + \ell \leq m - k$ with $m - k \geq k \to \infty$. Using this representation, an upper bound can be derived as:
\begin{align*}
  &(\mathbb{E}[\|\partial_{\phi_j} y_t(\theta_0;\psi_{k0}) - \mathbb{E}[\partial_{\phi_j} y_t(\theta_0;\psi_{k0}) | \mathcal{F}_{t-m}^t]\|^q])^{1/q}\\ &\leq \|P(\theta_0;\tilde{\Sigma})\|_{\infty}\sum_{\ell = 0}^\infty \{ \|\Lambda_\ell(\theta_0)\|_{\infty} \sum_{s = (m-k)-\ell}^\infty \|\tilde{\Lambda}_s\|_{\infty} \} \|e_{t}\|_q \\
  &\leq \|P(\theta_0;\tilde{\Sigma})\|_{\infty} \Big\{ \underbrace{\sum_{\ell = 0}^{[(m-k)/2]} \{ \|\Lambda_\ell(\theta_0)\|_{\infty} \sum_{s = k-\ell}^\infty \|\tilde{\Lambda}_s\|_{\infty} \} }_{ \leq \sum_{\ell = 0}^{\infty}  \|\Lambda_\ell(\theta_0)\|_{\infty}C [(m-k)/2]^{-(b+\varepsilon)} }  + \underbrace{\sum_{\ell = [(m-k)/2] + 1}^{\infty}  \|\Lambda_\ell(\theta_0)\|_{\infty} \{\sum_{s = (m-k)-\ell}^\infty \|\tilde{\Lambda}_s\|_{\infty} \} }_{ \leq   \sum_{s = 0}^\infty \|\tilde{\Lambda}_s\|_{\infty} C [(m-k)/2]^{-(b+\varepsilon)}} \Big\} \|e_{t}\|_q,
\end{align*}
which is a $O( m^{-(b+\varepsilon)}) \leq O(k^{-(b+\varepsilon)})$. Similar derivations apply to $\partial_{\tilde{\mu}}y_t(\theta_0;\psi_{k0})$ and $\partial_{\text{vech}[\tilde{\Sigma}]}y_t(\theta_0;\psi_{k0})$. Altogether, we get: $(\mathbb{E}[\|\partial_{\psi} y_t(\theta_0;\psi_{k0}) - \mathbb{E}[\partial_{\psi} y_t(\theta_0;\psi_{k0}) | \mathcal{F}_{t-m}^t]\|_{\infty}^q])^{1/q} \leq O(k^{-(b+\varepsilon)+1/q})$. Using Cauchy-Schwarz inequality, we find: 
\[(\ref{eq:M1p}) = (1/n) \sum\nolimits_{t=1}^n \mathbb{E}[\partial_{\psi} y_t(\theta_0;\psi_{k0}) | \mathcal{F}_{t-m}^t]^\prime W \mathbb{E}[\partial_{\psi} y_t(\theta_0;\psi_{k0}) | \mathcal{F}_{t-m}^t] + O_p(k^{1-(b+\varepsilon)}), \]
and $O_p(k^{1-(b+\varepsilon)}) = o_p([\log(n)]^{-2})$ since $b \geq 2$ and $[\log(n)]^{2}/k = o(1)$. Because $e_t$ are strong-mixing with coefficients $\alpha(\cdot)$, and $B_{t,m} = \mathbb{E}[\partial_{\psi} y_t(\theta_0;\psi_{k0}) | \mathcal{F}_{t-m}^t]^\prime W \mathbb{E}[\partial_{\psi} y_t(\theta_0;\psi_{k0}) | \mathcal{F}_{t-m}^t]$ is a function of $(e_t,\dots,e_{t-m})$, it is mixing with coefficients $\alpha([\cdot-m]^{+})$, where $[j-m]^{+} = \max(j-m,0)$. For any scalar element $\ell$, we can bound its autocovariances as:
$ |\text{cov}(B_{t,m,\ell},B_{t-s,m,\ell})| \leq 6 \alpha([s-m]^+)^{1-1/p-1/r} \|B_{t,m,\ell}\|_p \|B_{t,m,\ell}\|_r, $
using \citet[Cor15.3]{Davidson2021}. Since $B_{t,m,\ell}$ has $q/2$ finite moments, we can set $p=r=4$ in the above inequality. Note that $\|B_{t,m,\ell}\|_4$ is bounded uniformly in $\ell,m,k$. Apply Chebyshev's inequality:
\begin{align*} \text{var}\left( \frac{1}{n} \sum_{t=1}^n B_{t,m,\ell} \right) &\leq \frac{\text{var}(B_{t,m,\ell})}{n}  + \frac{2}{n}\sum_{s=1}^{n-1}|\text{cov}(B_{t,m,\ell},B_{t-s,m,\ell})|\\
  &\leq \frac{\text{var}(B_{t,m,\ell})}{n}  + \frac{ m }{n} 12\alpha(0)^{1/2} + \frac{12}{n} \sum_{s=m+1}^{n-1} \alpha(s-m)^{1/2} \|B_{t,m,\ell}\|_4^2
  \leq O( k/n ),
 \end{align*}
where the last inequality holds because $\text{var}(B_{t,m,\ell})$ is finite and, because $e_t$ is strong-mixing with size $-a$ where $a > 2$, the $\alpha(s)^{1/2}$ are summable over $s \geq 1$. Since $B_{t,m}$ has a $O(k^2)$ elements, we get:
$\|\frac{1}{n} \sum_{t=1}^n B_{t,m} - \mathbb{E}(B_{t,m})\|_{\infty} \leq O_p(k^{3/2} n^{-1/2}) = o_p([\log(n)]^{-2}). $ This implies 
\[ (\ref{eq:M1p})=\mathbb{E}\{ \mathbb{E}[\partial_{\psi} y_t(\theta_0;\psi_{k0}) | \mathcal{F}_{t-m}^t]^\prime W \mathbb{E}[\partial_{\psi} y_t(\theta_0;\psi_{k0}) | \mathcal{F}_{t-m}^t]\}+o_p([\log(n)]^{-2}). 
\]
The right hand side satisfies:
\begin{align*} &\|\mathbb{E}\{ \mathbb{E}[\partial_{\psi} y_t(\theta_0;\psi_{k0}) | \mathcal{F}_{t-m}^t]^\prime W \mathbb{E}[\partial_{\psi} y_t(\theta_0;\psi_{k0}) | \mathcal{F}_{t-m}^t] \} - \mathbb{E}[\partial_{\psi} y_t(\theta_0;\psi_{k0})^\prime W \partial_{\psi} y_t(\theta_0;\psi_{k0}) ]\|\\
&\leq \|\mathbb{E}\{ \| ( \partial_{\psi} y_t(\theta_0;\psi_{k0}) - \mathbb{E}[\partial_{\psi} y_t(\theta_0;\psi_{k0}) | \mathcal{F}_{t-m}^t])^\prime W ( \partial_{\psi} y_t(\theta_0;\psi_{k0}) + \mathbb{E}[\partial_{\psi} y_t(\theta_0;\psi_{k0})])\|\}\\
&\leq \overline{\lambda}_W O(k^{1-(b+\varepsilon)})O(k^{1/2}) = o(k^{-1/2}) = o([\log(n)]^{-2}). \end{align*}
Putting everything together, we find the desired result:
\[ (\ref{eq:M1p}) = \mathbb{E}[\partial_{\psi} y_t(\theta_0;\psi_{k0})^\prime W \partial_{\psi} y_t(\theta_0;\psi_{k0})] + o_p([\log(n)]^{-2}). \]
Similar derivations apply to (\ref{eq:M2p}), (\ref{eq:M3p}), (\ref{eq:M4p}) so that we have:
\[ nQ_n(\hat{\theta}_n;\hat{\psi}_{nk}) = n (\hat{\psi}_{nk}-\psi_k)^\prime M_k (\hat{\psi}_{nk}-\psi_k) + o_p([\log(n)]^{-1}), \]
where $M_k$ equals 
\begin{align}
\mathbb{E} \bigg[ 
    \big( \partial_\psi y_t(\theta_0; \psi_{k0}) 
    + \partial_\theta y_t(\theta_0; \psi_k) M^{-1}J_k \big)^\prime W 
    \big( \partial_\psi y_t(\theta_0; \psi_{k0}) 
    + \partial_\theta y_t(\theta_0; \psi_k) M^{-1}J_k \big) 
\bigg], \tag{$M_k$} \label{equ:Mk}
\end{align}
with $J_k = D_k + E_k$, $D_k = - \mathbb{E}[ (u_t^\prime W \otimes I)\partial_\psi G_t(\theta_0;\psi_{k0}) ] = 0,$ since $u_t=0$, and $E_k = - \mathbb{E}[ \partial_\theta y_t(\theta_0;\psi_{k0})^\prime W \partial_\psi y_t(\theta_0;\psi_{k0}) ]$.

\noindent\textbf{Step 3. Further expanding $n Q_n(\hat{\theta}_n;\hat{\psi}_{nk})$ to obtain: $n Q_n(\hat{\theta}_n;\hat{\psi}_{nk}) = n\bar{Z}_{n,k}^\prime M_k \bar{Z}_{n,k} + o_p(1)$, with $\bar{Z}_{n,k} = 1/n \sum_{t=1}^n Z_{k,t}$ for $Z_{k,t}$ defined in Theorem \ref{th:asym_normal}.}

In this step, the goal is to replace $(\hat{\psi}_{nk} - \psi_k)$ with $\bar{Z}_{n,k}$ in the final approximation of $n Q_n(\hat{\theta}_n;\hat{\psi}_{nk})$ in step 2. The main difference with the proof of Theorem \ref{th:asym_normal}, which relied on existing results, is that we need more refined result regarding the order of the difference $(\hat{\psi}_{nk} - \psi_k)-\bar{Z}_{n,k}$. For $\tilde{\mu}_n$ and $\text{vech}(\tilde{\Sigma}_{nk})$ this is immediate, so the main focus is on the autoregressive coefficients:
\[ (\hat{\phi}_1,\dots,\hat{\phi}_k) - (\phi_1,\dots,\phi_k)  = (1/n) \sum\nolimits_{t=1}^n \text{vec} \left[ e_{t,k} \tilde{Y}_{t-1,nk}^\prime\hat{\Gamma}^{-1}_{nk} \right], \]
where $\tilde{Y}_{t-1,nk} = ((\tilde{y}_{t-1}-\tilde{\mu}_n)^\prime,\dots,(\tilde{y}_{t-k}-\tilde{\mu}_n)^\prime)^\prime$ and $\hat{\Gamma}_{nk} = (1/n)\sum_{t=1}^n \tilde{Y}_{t-1,nk}\tilde{Y}_{t-1,nk}^\prime$. 

Let $\tilde{Y}_{t-1,k}$ and $\hat{\Gamma}_{k}$ be the same as $\tilde{Y}_{t-1,nk}$ and $\hat{\Gamma}_{nk}$, but with $\tilde{\mu}$ replacing $\tilde{\mu}_n$. We want to show that the $\tilde{Y}_{t-1,nk}$, $\hat{\Gamma}_{nk}$, and $e_{t,k}$ in the above displayed expression can be replaced by $\tilde{Y}_{t-1,k}$, $\hat{\Gamma}_{k}$, and $e_{t}$ with negligible error. To this end, note that $\|\tilde{Y}_{t-1,nk} - \tilde{Y}_{t-1,k}\|_{\infty} = \|\tilde{\mu}_n - \tilde{\mu}\|_\infty = O_p(n^{-1/2})$ since $\tilde{y}_t$ is NED in $L_2$-norm with appropriate size. Likewise, for $1_{k} = (1,\dots,1)$ of size $k$: $\hat{\Gamma}_{nk} = \hat{\Gamma}_k + [1/n \sum_{t=1}^n \tilde{Y}_{t-1,k}][ (\tilde{\mu}_n - \tilde{\mu}) \otimes 1_{k}^\prime ]^\prime$ so that $\|\hat{\Gamma}_{nk} - \hat{\Gamma}_k\|_{\infty} = O_p(n^{-1})$. Also, using similar projection and mixing arguments as in step 2, we can show that for $m-k \geq k$ we have $\|\hat{\Gamma}_k-\Gamma_k\|_{\infty} \leq O_p( k(m/n)^{1/2} ) + O_p(k m^{-(b+\varepsilon)})$ since it has $O(k^2)$ elements, all with finite $q/4 \geq 4$ moments. Then, using an inequality between operator and sup-norm:\footnote{For a matrix $A$ of size $d \times d$, $\|A\|_{op} \leq d\|A\|_{\infty}$; here $d = O(k)$} $\|\hat{\Gamma}_k-\Gamma_k\|_{op} \leq O_p( k^{2}(m/n)^{1/2} ) + O_p(k^{2}m^{-(b+\varepsilon)})= o_p([\log(n)]^{-2})$ by setting $m = 2k$ and using the Assumptions. Apply Weyl's inequality \citep[Th3.3.16]{horn1991}: $\lambda_{\min}(\hat{\Gamma}_{nk}) \geq \lambda_{\min}(\Gamma_k) - \|\hat{\Gamma}_{nk} - \Gamma_k\|_{op} = \lambda_{\min}(\Gamma_k) - o_p([\log(n)]^{-2})$. This means that $\|\hat{\Gamma}_{nk}^{-1}\|_{op}$ is bounded with probability approaching one. 

Using this results, we can examine the effects of the substitutions. When substituting $\tilde{Y}_{t-1,nk}$ for $\tilde{Y}_{t-1,k}$, we have: $\frac{1}{n} \sum_{t=1}^n  e_{t,k} \tilde{Y}_{t-1,nk}^\prime\hat{\Gamma}^{-1}_{nk} = \frac{1}{n} \sum_{t=1}^n  e_{t,k} \tilde{Y}_{t-1,k}^\prime\hat{\Gamma}^{-1}_{nk} + O_p(n^{-1})$ since $\frac{1}{n} \sum_{t=1}^n e_{t,k} = \frac{1}{n} \sum_{t=1}^n e_{t} + o_p(n^{-1/2}) = O_p(n^{-1/2})$, using previous derivations and the strong-mixing properties of $e_t$. Next, for $t \geq k+1$: $\|e_{t,k}-e_t\|_p \leq \sum_{t=k+1}^\infty \|\Psi_j\|_{op}\|\tilde{y}_t-\tilde{\mu}\|_p = o_p([nk]^{-1/2}\log(n)^{-2})$ since $\sqrt{nk}\sum_{t=k+1}^\infty \|\Psi_j\|_{op} = o_p([\log(n)]^{-2})$. Also, $(\mathbb{E}[\|\tilde{Y}_{t-1,k}\|_{\infty}^2])^{1/2} \leq O(k^{1/2}) = o(n^{1/16} \sqrt{\log(n)})$. Thus, when substituting $e_{t,k}$ for $e_t$, we have : $\frac{1}{n} \sum_{t=1}^n  e_{t,k} \tilde{Y}_{t-1,k}^\prime\hat{\Gamma}^{-1}_{nk} = \frac{1}{n} \sum_{t=1}^n  e_{t} \tilde{Y}_{t-1,k}^\prime\hat{\Gamma}^{-1}_{nk} + o_p(n^{-7/8}) + o_p(k^{1/2}[nk]^{-1/2}[\log(n)]^{-2}) = o_p(n^{-1/2}[\log(n)]^{-2}).$ The $o_p(n^{-7/8})$ term is due to the summation from $t=1$ to $k$. Finally, we substitute $\hat{\Gamma}_{nk}^{-1}$ for $\Gamma_k^{-1}$. Using projection arguments as above: $\|\frac{1}{n} \sum_{t=1}^n e_t \tilde{Y}_{t-1,k}^\prime\|_{\infty} \leq O_p(k^{1/2}(1/n)^{1/2}) + O_p(k^{1/2}m^{-(b+\varepsilon)})$. It is possible here to replace the $(m/n)$ term with $1/n$ because of a martingale property $\mathbb{E}(e_t \tilde{Y}_{t-1,k}|\mathcal{F}_{t-1})=0$, so that the autocovariances, even after projection, are zero. Recall that $\hat{\Gamma}_{nk}^{-1} - \Gamma_k^{-1} = \hat{\Gamma}_{nk}^{-1}[ \Gamma_k  - \hat{\Gamma}_{nk}]\Gamma_k^{-1} = O_p(k^2(m/n)^{1/2}) + O_p(k^2m^{-(b+\varepsilon)})$ since $\|\hat{\Gamma}_{nk}^{-1}\|_{op} \leq \|\Gamma_k^{-1}\|_{op} + o_p(1)=O_p(1)$. Thus: $\frac{1}{n} \sum_{t=1}^n  e_{t} \tilde{Y}_{t-1,k}^\prime\hat{\Gamma}^{-1}_{nk} - \frac{1}{n} \sum_{t=1}^n  e_{t} \tilde{Y}_{t-1,k}^\prime\hat{\Gamma}^{-1}_{k} = O_p(k^{5/2}m^{1/2}/n ) + O_p(k^{5/2}m^{-(b+\varepsilon)})$. Pick $m= \max(2k,n^{1/4}[\log(n)]^{-1}) = o(n^{1/4})$ to find $k^{5/2}m^{1/2}/n = o(n^{-9/16}[\log(n)]^{-3})$ and $k^{5/2}m^{-(b+\varepsilon)} = o( n^{-19/16}[\log(n)]^{-6} )$ for $b \geq 6$. Altogether, we get:
$\sqrt{n}\|(\hat{\psi}_n - \psi_k) - \overline{Z}_{n,k}\|_{\infty} = o_p([\log(n)]^{-2})$. 

Next, show that $(M_k)_{k \geq 1}$ is a sequence of bounded operators, i.e. $\|M_k\|_{op} \leq c$ for all $k \geq 1$. We will focus on (\ref{eq:M1p}), as the derivations for the others are similar. The elements of $\mathbb{E}[ \partial_{\psi} y_t(\theta_0;\psi_{k0})^\prime W \partial_{\psi} y_t(\theta_0;\psi_{k0}) ]$ take the form $\mathbb{E}[ \partial_{\phi_\ell} y_t(\theta_0;\psi_{k0})^\prime W \partial_{\phi_j} y_t(\theta_0;\psi_{k0}) ]$ or involve derivatives with respect to $\tilde{\mu}$ or $\text{vech}[\tilde{\Sigma}]]$. For $\ell < j$, apply the law of iterated expectations against the filtration $\mathcal{F}_{t-j}$. We get: $\mathbb{E}[ \partial_{\phi_\ell} y_t(\theta_0;\psi_{k0})^\prime W \partial_{\phi_j} y_t(\theta_0;\psi_{k0}) ] = \mathbb{E}[ \mathbb{E}[  \partial_{\phi_\ell} y_t(\theta_0;\psi_{k0}) | \mathcal{F}_{t-j}]^\prime W \partial_{\phi_j} y_t(\theta_0;\psi_{k0}) ]$. Repeating the derivations from the proof of Theorem \ref{th:asym_normal} to bound $\|\mathbb{E}[  \partial_{\phi_\ell} y_t(\theta_0;\psi_{k0}) | \mathcal{F}_{t-j}]\|_2$ and applying the Cauchy-Schwarz inequality, we get: $\|\mathbb{E}[ \partial_{\phi_\ell} y_t(\theta_0;\psi_{k0})^\prime W \partial_{\phi_j} y_t(\theta_0;\psi_{k0}) ]\| \leq C_M (1+|\ell-j|)^{-(b+\varepsilon)}$, for some constant $C_M$ which does not depend on $k$. Since $b \geq 2$, these upper bounds are absolutely summable over $\ell$, the sum is bounded over $j$ - and vice-versa. By Schur's Lemma \citep[Lem1]{jaffard1990}, this implies that $\|\mathbb{E}[ \partial_{\phi_1,\dots,\phi_k} y_t(\theta_0;\psi_{k0})^\prime W \partial_{\phi_1,\dots,\phi_k} y_t(\theta_0;\psi_{k0}) ]\|_{op} \leq C_M \sup_{j \geq 1} \sum_{\ell = 1}^\infty (1+|\ell-j|)^{-(b+\varepsilon)} \leq C_M \sum_{\ell = -\infty}^{+\infty} (1+|\ell|)^{-(b+\varepsilon)}$. Similar bounds apply to the other derivatives, and as a result, $\|M_k\|_{op}$ is uniformly bounded. Now we have: 
\begin{align*} 
  |n(\hat{\psi}_n - \psi_k)^\prime M_k (\hat{\psi}_n - \psi_k) - n \overline{Z}_{n,k}^\prime M_k \overline{Z}_{n,k}| &= n | ([\hat{\psi}_n - \psi_k]-\overline{Z}_{n,k})^\prime M_k ([\hat{\psi}_n - \psi_k]+\overline{Z}_{n,k})|,\\
  &\leq n \|[\hat{\psi}_n - \psi_k]-\overline{Z}_{n,k}\|_{\infty} \|M_k\|_{op} [\|\hat{\psi}_n - \psi_k\|  +\|\overline{Z}_{n,k}\|]\\
  &\leq o_p([\log(n)]^{-2})O_p(\sqrt{k \log(n)}) = o_p(k^{1/2}[\log(n)]^{-3/2}),
\end{align*} 
where the last inequality holds because $\sqrt{n}\|\overline{Z}_{n,k}\|_{\infty} = O_p(k^{1/2})$, each element having finite and bounded second moment, and $\sqrt{n}\|\hat{\psi}_n - \psi_k\| \leq O(k^{1/2})\|\hat{\psi}_n - \psi_k\|_{\infty}$ using an inequality between the $\ell_2$ and $\ell_\infty$ norms. This allows to conclude this step with:
\[ nQ_n(\hat{\theta}_n;\hat{\psi}_{nk}) = n \overline{Z}_{n,k}^\prime M_k \overline{Z}_{n,k} + o_p(k^{1/2}[\log(n)]^{-3/2}).  \]

\noindent\textbf{Step 4. Obtaining the strong Approximation: $n \overline{Z}_{n,k}^\prime M_k \overline{Z}_{n,k} = n \mathcal{Z}_{n,k}^\prime M_k \mathcal{Z}_{n,k} + o_p(1)$, for some $\sqrt{n}\mathcal{Z}_{n,k} \sim \mathcal{N}(0,S_{n,k})$.}

In Lemma \ref{lem:SA}, we constructed a normally distributed random vector $\mathcal{Z}_{n,k}$ such that:
$\sqrt{n}\|\overline{Z}_{n,k} - \mathcal{Z}_{n,k}\| = o_p([\log(n)]^{-2})$.  
Given that the elements of $\sqrt{n}\overline{Z}_{n,k}$ and $\sqrt{n}\mathcal{Z}_{n,k}$ have finite second moments, we also have:
$\sqrt{n}\|\overline{Z}_{n,k} + \mathcal{Z}_{n,k}\|_{\infty} \leq O_p( k^{1/2} )$,
using Pisier's inequality. Now, using H\"older's inequality, we find:
\begin{align*} |n \overline{Z}_{n,k}^\prime M_k \overline{Z}_{n,k} - n \mathcal{Z}_{n,k}^\prime M_k \mathcal{Z}_{n,k}| &=  |\sqrt{n} [\overline{Z}_{n,k}-\mathcal{Z}_{n,k}]^\prime M_k \sqrt{n}[\overline{Z}_{n,k}+\mathcal{Z}_{n,k}]|\\
&\leq \sqrt{n} \|\overline{Z}_{n,k}-\mathcal{Z}_{n,k}\| \times \| M_k \sqrt{n}[\overline{Z}_{n,k}+\mathcal{Z}_{n,k}] \|_{\infty}\\
&= o_p( k^{1/2}[\log(n)]^{-2} ), \end{align*}
since $\|M_k\|_{\infty}$ is uniformly bounded in $k$. Now we can conclude that:
$ nQ_n(\hat{\theta}_n;\hat{\psi}_{nk}) = \sqrt{n}\mathcal{Z}_{n,k}^\prime M_k \sqrt{n}\mathcal{Z}_{n,k} + o_p(k^{1/2}[\log(n)]^{-3/2}).  $

\noindent\textbf{Step 5. Validating the rejection rate.}

It is possible to re-write the leading terms in the preceding equation as follows \citep[p152]{buckley1988}:
$ n \mathcal{Z}_{n,k}^\prime M_k \mathcal{Z}_{n,k} = \sum_{j=1}^{d(k)} \lambda_j( S_{n,k}^{1/2}M_kS_{n,k}^{1/2} ) W_j, $
where $(W_1,\dots,W_{d(k)})$ are iid $\chi^2_1$ distributed and $\lambda_j( S_{n,k}^{1/2}M_k S_{n,k}^{1/2} )$ are the eigenvalues of $S_{n,k}^{1/2}M_k S_{n,k}^{1/2}$, with $d(k) = \text{dim}(\psi_k)$. Then, we have
$\mathbb{E}[n \mathcal{Z}_{n,k}^\prime M_k \mathcal{Z}_{n,k}] = \sum_{j=1}^{d(k)} \lambda_j( S_{n,k}^{1/2}M_kS_{n,k}^{1/2} ) = \text{trace}( S_{n,k}M_k )$ and $\text{var}[n \mathcal{Z}_{n,k}^\prime M_k \mathcal{Z}_{n,k}] = 3 \sum_{j=1}^{d(k)} \lambda^2_j( S_{n,k}^{1/2}M_kS_{n,k}^{1/2} ) = 3 \text{trace}( [S_{n,k}M_k ]^2).
$
Both $S_{n,k}$ and $M_k$ have eigenvalues bounded above by Schur's Lemma. This implies that the two traces are at most $O(k)$. If, in addition, $\text{trace}( S_{n,k}M_k ) \geq O(k)$ and $\text{trace}( [S_{n,k}M_k]^2 ) \geq O(k)$ then the Paley–Zygmund inequality implies:
\begin{align*} \mathbb{P} \left( n \mathcal{Z}_{n,k}^\prime M_k \mathcal{Z}_{n,k} > k^{1/2} \right) &\geq \left(1-\frac{k^{1/2}}{\text{trace}(S_{n,k}M_k)}\right)^2 \frac{[\text{trace}(S_{n,k}M_k)]^2}{3\text{trace}([S_{n,k}M_k]^2) + [\text{trace}(S_{n,k}M_k)]^2} \\ &\geq \left( 1 - O(k^{-1/2}) \right)^2\frac{1}{1+O(k^{-1})} =1 - o(1), \end{align*}
as $n \to\infty$ so that the $o_p(k^{1/2}[\log(n)]^{-3/2})$ term is negligible. We can write:
\begin{align*}
  \mathbb{P} \left( nQ_n(\hat{\theta}_n;\hat{\psi}_{nk}) > c_{n,k}(1-\alpha)\right) &= 
  \mathbb{P} \left( n \mathcal{Z}_{n,k}^\prime M_k \mathcal{Z}_{n,k} > c_{n,k}(1-\alpha) + o_p(c_{n,k}(1-\alpha)) \right)= \alpha + o(1),
\end{align*}
which concludes the proof.
\qed

\begin{lemma} \label{lem:cons_test} Suppose the conditions for Theorem \ref{th:spec} are satisfied but the model is misspecified, i.e. for $\theta_0$ defined in Theorem \ref{th:cons} $Q(\theta_0;\psi_0) > 0$, then:
$ \lim_{n \to \infty} \mathbb{P} \left( n Q_n(\hat{\theta}_n;\hat{\psi}_{nk}) > c_{n,k}(\alpha) \right) = 1, $
where $c_{n,k}$ is defined in Theorem \ref{th:spec}. 
\end{lemma}

\paragraph{Proof of Lemma \ref{lem:cons_test}:}  Under a fixed alternative, $Q_n(\hat{\theta}_n;\hat{\psi}_{nk}) \overset{p}{\to} Q(\theta_0;\psi_0) > 0$ by uniform convergence, derived in the proof of Theorem \ref{th:cons}. Derivations in the proof of Theorem \ref{th:spec} imply $n \mathcal{Z}_{n,k}^\prime M_k \mathcal{Z}_{n,k} \leq O_p(k) = o_p(n)$ so that $\mathbb{P}\left( nQ_n(\hat{\theta}_n;\hat{\psi}_{nk}) > c_{n,k}(1-\alpha) \right) = \mathbb{P}\left( Q_n(\hat{\theta}_n;\hat{\psi}_{nk}) > o_p(1) \right) \to 1$. This is the desired result. \qed

\section{Additional Details for Section \ref{sec:motivating}} \label{apx:motivating}

The auxiliary model used for the OTF is given by:
\[ \tilde{y}_t = \delta_0 + \delta_1 t + \delta_2 \cos(2\pi t/n) + \dots + \delta_5 \cos(2\pi 4t/n) + \eta_t, \quad \eta_t = \rho_1  \eta_{t-1} + \dots \rho_4  \eta_{t-4} + e_t. \]
The model is estimated in two steps, first the data is de-trended using OLS. Then an AR(4) is fitted to the residuals $\hat{\eta}_t$ using OLS. The residuals $\hat{e}_t$ are then used in Algorithm \ref{alg_2}. The model is then estimated by minimizing the loss $Q_n$ as described in the main text.

\section{Additional Details for Section \ref{sec:emp}} \label{apx:emp}

\subsection{Small New-Keynesian Model} \label{apx:LS}

\paragraph{Data.} Estimation and testing in Section \ref{sec:LS} use linearly detrended US log GDP, annualized inflation, and interest rates for the period
1960Q1-2007Q4.

\subsection{Smets-Wouters Model} \label{apx:SW}

\paragraph{Data.} 
We use US data from the same sample period of 1960:I-2007:IV as above. 

\subsection{Affine Term Structure Model} \label{apx:ATSM}

\paragraph{Data.} 
Estimation relies on monthly data for 6 zero-coupon bond yields, with
maturities of 1 month and 1-5 years. The bond yields (1-5 years) are
obtained from the Fama CRSP zero-coupon files, and the 1-month yields come
from the Fama CRSP Treasury Bill files. The sample period is from December
1952 to December 2000, as in \citet{ang2003}.

  \begin{table}[H]
  \caption{LS Model Parameter Interpretation and Prior Weights.}
  \centering
  \small
  \setlength\tabcolsep{0.75pt}
    \renewcommand{\arraystretch}{0.55} 
  { \small
  \begin{tabular}{ll|c|ccc||c|ccc}
  \hline \hline
  &  & \multicolumn{4}{c||}{
  Determinacy Regime} & \multicolumn{4}{c}{Indeterminacy Regime} \\ \hline
  \multirow{2}{*}{${\theta }$} & \multirow{2}{*}{Parameter Interpretation} & \multirow{2}{*}{Bounds} & \multicolumn{3}{c||}{Prior Distribution} & \multirow{2}{*}{Bounds} & \multicolumn{3}{c}{Prior Distribution} \\ 
  &  & & \textsc{density} & \textsc{mean} & \textsc{sd} &  & \textsc{density} & \textsc{mean} & \textsc{sd} \\ \hline
  ${ \tau }^{-1}$ & { risk aversion} & { [1, 10]} & { %
  Gamma} & { 2.00} & { 0.50} & { [1, 10]} & { Gamma} & 
  { 2.00} & { 0.50} \\ 
  ${ r}^{\ast }$ & { steady-state real interest rate} & { %
  [1, 4]} & { Gamma} & { 2.00} & { 1.00} & { [1, 4]} & 
  { Gamma} & { 2.00} & { 1.00} \\ 
  ${ \kappa }$ & { Phillips curve slope} & { [0.1, 1]} & 
  { Gamma} & { 0.50} & { 0.20} & { [0.1, 1]} & { %
  Gamma} & { 0.50} & { 0.20} \\ 
  ${ \psi }_{1}$ & { inflation target} & { [1.1, 5]} & 
  { Gamma} & { 1.10} & { 0.50} & { [0.1,0.9]} & 
  { Gamma} & { 1.10} & { 0.50} \\ 
  ${ \psi }_{2}$ & { output target} & { [0, 0.99]} & { %
  Gamma} & { 0.25} & { 0.13} & { [0.01,5]} & { Gamma}
  & { 0.25} & { 0.13} \\ 
  ${ \rho }_{r}$ & { interest rate smoothing} & { [0.01, 0.9]%
  } & { Beta} & { 0.50} & { 0.20} & { [0.01,0.9]} & 
  { Beta} & { 0.50} & { 0.20} \\ 
  ${ \rho }_{g}$ & { exog spending AR} & { [0.01, 0.99]}
  & { Beta} & { 0.70} & { 0.10} & { [0.01,0.99]} & 
  { Beta} & { 0.70} & { 0.10} \\ 
  ${ \rho }_{z}$ & { technology shock AR} & { [0.01, 0.99]}
  & { Beta} & { 0.70} & { 0.10} & { [0.01,0.99]} & 
  { Beta} & { 0.70} & { 0.10} \\ 
  ${ \sigma }_{r}$ & { monetary policy shock SD} & { [0.01,
  3]} & { IGamma} & { 0.31} & { 0.16} & { [0.01, 3]} & 
  { IGamma} & { 0.31} & { 0.16} \\ 
  ${ \sigma }_{g}$ & { exog spending SD} & { [0.01, 3]}
  & { IGamma} & { 0.38} & { 0.20} & { [0.01, 3]} & 
  { IGamma} & { 0.38} & { 0.20} \\ 
  ${ \sigma }_{z}$ & { technology shock SD} & { [0.01, 3]} & 
  { IGamma} & { 1.00} & { 0.52} & { [0.01, 3]} & 
  { IGamma} & { 1.00} & { 0.52} \\ 
  ${ \rho }_{gz}$ & { exog spending-technology cor} & 
  { [-0.9, -0.9]} & { Normal} & { 0.00} & { 0.40} & 
  { [-0.9, -0.9]} & { Normal} & { 0.00} & { 0.40} \\ 
  ${ M}_{r\epsilon }$ & { sunspot-monetary coef} & { --} & 
  { --} & { --} & { --} & { [-3,3]} & { Normal}
  & { 0.00} & { 1.00} \\ 
  ${ M}_{g\epsilon }$ & { sunspot-exog spending coef} & 
  { --} & { --} & { --} & { --} & { [-3,3]} & 
  { Normal} & { 0.00} & { 1.00} \\ 
  ${ M}_{z\epsilon }$ & { \ sunspot-technology coef} & { --}
  & { --} & { --} & { --} & { [-3,3]} & { Normal}
  & { 0.00} & { 1.00} \\ 
  ${ \sigma }_{\epsilon }$ & { \ sunspot shock SD} & { --} & 
  { --} & { --} & { --} & { [0.01,3]} & { IGamma}
  & { 0.25} & { 0.13} \\ 
  ${ \pi }^{\ast }$ & { steady-state inflation} & { [2, 10]}
  & { Gamma} & { 4.00} & { 2.00} & { [2,10]} & { %
  Gamma} & { 4.00} & { 2.00} \\ \hline \hline
  \end{tabular}\notes{\textbf{Legend:} The prior follows Prior 1 specification of
  \citet{lubik2004}.} }
  \label{tab:LSpar}
  \end{table}

\begin{table}[ht]
  \caption{LS Model: Parameter Estimates ($k = 4$ lags)} 
 \centering
 \setlength\tabcolsep{4.0pt}
  \renewcommand{\arraystretch}{0.85} 
 {\small \
 \begin{tabular}{ll|ccc|ccc|ccc}
 \hline\hline
 \multicolumn{2}{c|}{\multirow{2}{*}{Parameter Estimates}} & 
 \multicolumn{3}{c|}{Determinacy} & \multicolumn{3}{c|}{Indeterminacy} & 
 \multicolumn{3}{c}{Determinacy} \\ 
 &  & \multicolumn{3}{c|}{(Full Sample)} & \multicolumn{3}{c|}{(Pre-Volcker)}
 & \multicolumn{3}{c}{(Post-Volcker)} \\ \hline
 ${\theta }$ & {Parameter Interpretation} & \textsc{est} & \textsc{sd}$_{c}$
 & \textsc{sd}$_{r}$ & \textsc{est} & \textsc{sd}$_{c}$ & \textsc{sd}$_{r}$ & 
 \textsc{est} & \textsc{sd}$_{c}$ & \textsc{sd}$_{r}$ \\ \hline
 ${\tau }^{-1}$ & {risk aversion} & 2.43 & 4.12 & 2.06 & 1.29 & 3.96 & 0.77 & 
 1.56 & 1.29 & 3.60 \\ 
 ${r}^{\ast }$ & {steady state real interest rate} & 1.87 & 0.28 & 0.38 & 0.99
 & 0.56 & 0.48 & 2.49 & 0.73 & 1.45 \\ 
 ${\kappa }$ & {Phillips curve slope} & 0.31 & 0.58 & 0.11 & 0.54 & 1.85 & 
 0.44 & 0.30 & 0.48 & 4.16 \\ 
 ${\psi }_{1}$ & {inflation target} & 1.29 & 0.18 & 0.32 & 0.69 & 0.11 & 0.19
 & 1.80 & 0.67 & 14.99 \\ 
 ${\psi }_{2}$ & {output target} & 0.16 & 0.72 & 1.10 & 0.14 & 0.67 & 0.28 & 
 0.18 & 1.17 & 31.72 \\ 
 ${\rho }_{r}$ & {interest rate smoothing} & 0.68 & 0.06 & 0.19 & 0.46 & 0.18
 & 0.18 & 0.79 & 0.08 & 1.87 \\ 
 ${\rho }_{g}$ & {exog spending AR} & 0.89 & 0.06 & 0.04 & 0.74 & 0.40 & 0.16
 & 0.92 & 0.05 & 0.14 \\ 
 ${\rho }_{z}$ & {technology shock AR} & 0.82 & 0.08 & 0.04 & 0.78 & 0.10 & 
 0.12 & 0.83 & 0.07 & 0.24 \\ 
 ${\sigma }_{r}$ & {monetary policy shock SD} & 0.24 & 0.04 & 0.06 & 0.22 & 
 0.05 & 0.08 & 0.21 & 0.07 & 0.73 \\ 
 ${\sigma }_{g}$ & {exog spending SD} & 0.18 & 0.05 & 0.05 & 0.28 & 1.17 & 
 0.21 & 0.18 &  0.10 & 0.43 \\ 
 ${\sigma }_{z}$ & {technology shock SD} & 1.56 & 0.42 & 0.25 & 1.12 & 0.50 & 
 0.34 & 1.14 & 0.41 & 2.38 \\ 
 ${\rho }_{gz}$ & {exog spending-technology cor} & 0.90 & 0.27 & 0.22 & 0.18
 & 3.19 & 1.09 & 0.35 & 0.68 & 1.75 \\ 
 ${M}_{r\epsilon }$ & {sunspot-monetary coef} & -- & -- & -- & 0.43 & 1.82 & 
 1.04 & -- & -- & -- \\ 
 ${M}_{g\epsilon }$ & {sunspot-exog spending coef} & -- & -- & -- & -1.80 & 
 5.87 & 1.66 & -- & -- & -- \\ 
 ${M}_{z\epsilon }$ & {\ sunspot-technology coef} & -- & -- & -- & 0.63 & 0.91
 & 0.24 & -- & -- & -- \\ 
 ${\sigma }_{\epsilon }$ & {\ sunspot shock SD} & -- & -- & -- & 0.08 & 4.41
 & 1.33 & -- & -- & -- \\ 
 ${\pi }^{\ast }$ & {steady state inflation} & 4.07 & 0.74 & 0.73 & 5.10 & 
 1.78 & 1.55 & 3.83 & 0.82 & 0.79 \\ \hline\hline
 \end{tabular}%
 \notes{ \textbf{Legend:} \textsc{est}: parameter estimates $\hat{\theta}_n$.
 \textsc{sd}$_{c}$: standard errors assuming correct model specification.
 \textsc{sd}$_{r}$: misspecification-robust standard errors. $n = 192,78,114$ for the full, pre and post-Volcker
 samples. } } \label{tab:LSest2}
 \end{table}

\begin{table}[H]
  \caption{LS Model: Parameter Estimates, Specification Test ($k = 2$ lags)} 
 \centering
 \small
 \setlength\tabcolsep{4.0pt}
  \renewcommand{\arraystretch}{0.55} 
 {\small \ 
 \begin{tabular}{ll|ccc|ccc|ccc}
 \hline\hline
 \multicolumn{2}{c|}{\multirow{2}{*}{Parameter Estimates}} & 
 \multicolumn{3}{c|}{Determinacy} & \multicolumn{3}{c|}{Indeterminacy} & 
 \multicolumn{3}{c}{Determinacy} \\ 
 &  & \multicolumn{3}{c|}{(Full Sample)} & \multicolumn{3}{c|}{(Pre-Volcker)}
 & \multicolumn{3}{c}{(Post-Volcker)} \\ \hline
 ${\theta }$ & {Parameter Interpretation} & \textsc{est} & \textsc{sd}$_{c}$
 & \textsc{sd}$_{r}$ & \textsc{est} & \textsc{sd}$_{c}$ & \textsc{sd}$_{r}$ & 
 \textsc{est} & \textsc{sd}$_{c}$ & \textsc{sd}$_{r}$ \\ \hline
 ${\tau }^{-1}$ & {risk aversion} & 2.45 & 3.86 & 1.93 & 1.25 & 2.06 & 0.63 & 
 1.40 & 1.08 & 2.13 \\ 
 ${r}^{\ast }$ & {steady state real interest rate} & 1.86 & 0.23 & 0.30 & 0.98
 & 0.58 & 0.48 & 2.44 & 0.59 & 4.91 \\ 
 ${\kappa }$ & {Phillips curve slope} & 0.49 & 0.73 & 0.41 & 0.58 & 2.07 & 
 0.30 & 0.46 & 0.49 & 46.68 \\ 
 ${\psi }_{1}$ & {inflation target} & 1.21 & 0.20 & 0.42 & 0.67 & 0.09 & 0.13
 & 1.73 & 0.64 & 70.25 \\ 
 ${\psi }_{2}$ & {output target} & 0.15 & 0.71 & 0.76 & 0.15 & 0.67 & 0.16 & 
 0.18 & 1.45 & 192.48 \\ 
 ${\rho }_{r}$ & {interest rate smoothing} & 0.66 & 0.07 & 0.22 & 0.43 & 0.18
 & 0.15 & 0.73 & 0.10 & 1.77 \\ 
 ${\rho }_{g}$ & {exog spending AR} & 0.88 & 0.05 & 0.03 & 0.72 & 0.37 & 0.11
 & 0.91 & 0.06 & 0.79 \\ 
 ${\rho }_{z}$ & {technology shock AR} & 0.82 & 0.05 & 0.03 & 0.79 & 0.11 & 
 0.11 & 0.83 & 0.05 & 0.33 \\ 
 ${\sigma }_{r}$ & {monetary policy shock SD} & 0.28 & 0.05 & 0.05 & 0.23 & 
 0.07 & 0.06 & 0.26 & 0.07 & 16.21 \\ 
 ${\sigma }_{g}$ & {exog spending SD} & 0.16 & 0.04 & 0.05 & 0.29 & 0.87 & 
 0.16 & 0.19 & 0.09 & 0.56 \\ 
 ${\sigma }_{z}$ & {technology shock SD} & 1.33 & 0.27 & 0.18 & 1.07 & 0.52 & 
 0.29 & 0.91 & 0.21 & 9.41 \\ 
 ${\rho }_{gz}$ & {exog spending-technology cor} & 0.90 & 0.26 & 0.16 & 0.10
 & 1.83 & 0.75 & 0.33 & 0.62 & 15.67 \\ 
 ${M}_{r\epsilon }$ & {sunspot-monetary coef} & -- & -- & -- & 0.42 & 1.40 & 
 0.76 & -- & -- & -- \\ 
 ${M}_{g\epsilon }$ & {sunspot-exog spending coef} & -- & -- & -- & -1.77 & 
 4.20 & 0.96 & -- & -- & -- \\ 
 ${M}_{z\epsilon }$ & {\ sunspot-technology coef} & -- & -- & -- & 0.67 & 0.81
 & 0.19 & -- & -- & -- \\ 
 ${\sigma }_{\epsilon }$ & {\ sunspot shock SD} & -- & -- & -- & 0.09 & 3.23
 & 0.85 & -- & -- & -- \\ 
 ${\pi }^{\ast }$ & {steady state inflation} & 4.04 & 0.62 & 0.61 & 5.11 & 
 1.69 & 1.50 & 3.79 & 0.74 & 1.13 \\ \hline\hline
 \multicolumn{2}{c|}{Specification Test} & \textsc{stat} & $10\%$ & $5\%$ & 
 \textsc{stat} & $10\%$ & $5\%$ & \textsc{stat} & $10\%$ & $5\%$ \\ \hline
 \multicolumn{2}{c|}{All} & 121.6 & 66.5 & 89.7 & 58.2 & 63.6 & 97.0 & 69.4 & 
 89.0 & 127.0 \\ \hline
 \multicolumn{2}{c|}{Output} & 65.5 & 45.6 & 63.4 & 43.7 & 30.5 & 47.8 & 37.3
 & 67.3 & 98.6 \\ 
 \multicolumn{2}{c|}{Inflation} & 33.1 & 12.2 & 15.6 & 7.0 & 21.4 & 32.7 & 
 12.8 & 12.7 & 16.6 \\ 
 \multicolumn{2}{c|}{Interest Rate} & 23.1 & 12.2 & 16.7 & 7.6 & 14.1 & 22.0
 & 19.3 & 15.8 & 22.7 \\ \hline\hline
 \end{tabular}%
 \notes{ \textbf{Legend:} \textsc{est}: parameter estimates $\hat{\theta}_n$.
 \textsc{sd}$_{c}$: standard errors assuming correct model specification.
 \textsc{sd}$_{r}$: misspecification-robust standard errors. \textsc{stat}:
 $nQ_n(\hat{\theta}_n;\hat{\psi}_{nk})$. $10\%$, $5\%$: critical values for
 specification test at corresponding significance levels. All: specification
 test on all variables. Output, Inflation, Interest Rate: specification test
 on individual variables. $n = 192,78,114$ for the full, pre and post-Volcker
 samples. } } \label{tab:LSest_k2}
 \end{table}

\begin{table}[H]
  \caption{SW Model: parameters, bounds, and prior distribution} \label{tab:SWpar}
  \centering
  \setlength\tabcolsep{0.75pt}
    \renewcommand{\arraystretch}{0.85} 
  { \small
  \begin{tabular}{ll|c|ccc}
  \hline \hline
  \multirow{2}{*}{${\theta }$} & \multirow{2}{*}{Parameter Interpretation} & \multirow{2}{*}{Bounds} & \multicolumn{3}{c}{Prior Distribution} \\ 
  &  &  & { \textsc{density}} & { \textsc{mean}} & { \textsc{sd}} \\ \hline
  ${ \rho }_{ga}$ & { Corr.: tech. and exog. spending shocks} & 
  { [0.01, 2]} & { Normal} & { 0.50} & { 0.25} \\ 
  ${ \mu }_{w}$ & { Wage mark-up shock MA} & { [0.01, 0.99]}
  & { Beta} & { 0.50} & { 0.20} \\ 
  ${ \mu }_{p}$ & { Price mark-up shock MA} & { [0.01, 0.99]}
  & { Beta} & { 0.50} & { 0.20} \\ 
  ${ \alpha }$ & { Share of capital in production} & { %
  [0.01, 1]} & { Normal} & { 0.30} & { 0.05} \\ 
  ${ \psi }$ & { Elast. of capital utilization adjustment cost} & 
  { [0.01, 1]} & { Beta} & { 0.50} & { 0.15} \\ 
  ${ \varphi }$ & { Investment adjustment cost} & { [3,15]}
  & { Normal} & { 4.00} & { 1.50} \\ 
  ${ \sigma }_{c}$ & { Elast. of inertemporal substitution} & 
  { [1, 3]} & { Normal} & { 1.50} & { 0.38} \\ 
  ${ \lambda }$ & { Habit persistence} & { [0.001, 0.99]} & 
  { Beta} & { 0.70} & { 0.10} \\ 
  ${ \phi }_{p}$ & { Fixed costs in production} & { [1, 3]}
  & { Normal} & { 1.25} & { 0.13} \\ 
  ${ \iota }_{w}$ & { Wage indexation} & { [0.01, 0.99]} & 
  { Beta} & { 0.50} & { 0.15} \\ 
  ${ \xi }_{w}$ & { Wage stickiness} & { [0.5, 0.95]} & 
  { Beta} & { 0.50} & { 0.10} \\ 
  ${ \iota }_{p}$ & { Price indexation} & { [0.01, 0.99]} & 
  { Beta} & { 0.50} & { 0.15} \\ 
  ${ \xi }_{p}$ & { Price stickiness} & { [0.1, 0.95]} & 
  { Beta} & { 0.50} & { 0.10} \\ 
  ${ \sigma }_{l}$ & { Labor supply elasticity} & { [1, 10]}
  & { Normal} & { 2.00} & { 0.75} \\ 
  ${ r}_{\pi }$ & { Taylor rule: inflation weight} & { [1, 3]%
  } & { Normal} & { 1.50} & { 0.25} \\ 
  ${ r}_{\Delta y}$ & { Taylor rule: output gap change weight} & 
  { [0.001, 0.5]} & { Normal} & { 0.13} & { 0.05} \\ 
  ${ r}_{y}$ & { Taylor rule: output gap weight} & { [0.001,
  0.5]} & { Normal} & { 0.13} & { 0.05} \\ 
  ${ \rho }$ & { Taylor rule: interest rate smoothing} & { %
  [0.5, 0.975]} & { Beta} & { 0.75} & { 0.10} \\ 
  ${ \rho }_{a}$ & { Productivity shock AR} & { [0.01, 0.99]}
  & { Beta} & { 0.50} & { 0.20} \\ 
  ${ \rho }_{b}$ & { Risk premium shock AR} & { [0.01, 0.99]}
  & { Beta} & { 0.50} & { 0.20} \\ 
  ${ \rho }_{g}$ & { Exogenous spending shock AR } & { %
  [0.01, 0.99]} & { Beta} & { 0.50} & { 0.20} \\ 
  ${ \rho }_{i}$ & { Investment shock AR} & { [0.01, 0.99]}
  & { Beta} & { 0.50} & { 0.20} \\ 
  ${ \rho }_{r}$ & { Monetary policy shock AR} & { [0.01,
  0.99]} & { Beta} & { 0.50} & { 0.20} \\ 
  ${ \rho }_{p}$ & { Price mark-up shock AR} & { [0.01, 0.99]%
  } & { Beta} & { 0.50} & { 0.20} \\ 
  ${ \rho }_{w}$ & { Wage mark-up shock AR} & { [0.001, 0.99]%
  } & { Beta} & { 0.50} & { 0.20} \\ 
  ${ \sigma }_{a}$ & { Productivity shock std. dev.} & { %
  [0.01, 3]} & { IGamma} & { 0.10} & { 2.00} \\ 
  ${ \sigma }_{b}$ & { Risk premium shock std. dev.} & { %
  [0.025, 5]} & { IGamma} & { 0.10} & { 2.00} \\ 
  ${ \sigma }_{g}$ & { Exogenous spending shock std. dev.} & 
  { [0.01, 3]} & { IGamma} & { 0.10} & { 2.00} \\ 
  ${ \sigma }_{i}$ & { Investment shock std. dev.} & { %
  [0.01, 3]} & { IGamma} & { 0.10} & { 2.00} \\ 
  ${ \sigma }_{r}$ & { Monetary policy shock std. dev.} & { %
  [0.01, 3]} & { IGamma} & { 0.10} & { 2.00} \\ 
  ${ \sigma }_{p}$ & { Price mark-up shock std. dev.} & { %
  [0.01, 3]} & { IGamma} & { 0.10} & { 2.00} \\ 
  ${ \sigma }_{w}$ & { Wage mark-up shock std. dev.} & { %
  [0.01, 3]} & { IGamma} & { 0.10} & { 2.00} \\ 
  $\overline{\gamma }$ & { Trend growth: real GDP, Infl., Wages} & 
  { [0.1, 0.8]} & { Normal} & { 0.40} & { 0.10} \\ 
  ${ r}$ & { Discount rate} & { [0.01, 2]} & { Gamma}
  & { 0.25} & { 0.10} \\ 
  $\overline{\pi }$ & { Steady state inflation rate} & { [0.1, 2]}
  & { Gamma} & { 0.62} & { 0.10} \\ 
  $\overline{l}$ & { Steady state hours worked} & { [-10,10]} & 
  { Normal} & { 0.00} & { 2.00} \\ \hline \hline
  \end{tabular}\notes{ \textbf{Legend:} Prior distributions are taken from \citet{smets2007}
  Dynare code.} }
  \end{table}

\begin{table}[H]
  \caption{SW: Estimates, Rejection Rates (Monte Carlo)} \label{tab:SW_MC}
  \centering
  \setlength\tabcolsep{2.5pt}
  \renewcommand{\arraystretch}{0.85} 
  { \small
  \begin{tabular}{lc||cc|cc|cc}
  \hline \hline
  $\theta$ & {\textsc{true}} & \textsc{mean} & \textsc{std} & \textsc{rej}$_c$ & \textsc{rej}$_r$ & \textsc{len}$_c$ & \textsc{len}$_r$ \\ \hline
    ${\rho }_{ga}$ & 0.58 & 0.49 & 0.17 & 0.07 & 0.01 & 0.94 & 2.31 \\
    ${\mu }_{w}$ & 0.90 & 0.75 & 0.11 & 0.07 & 0.02 & 0.37 & 1.05 \\
    ${\mu }_{p}$ & 0.82 & 0.68 & 0.11 & 0.01 & 0.00 & 0.78 & 2.80 \\
    ${\alpha }$ & 0.23 & 0.24 & 0.02 & 0.02 & 0.00 & 0.11 & 0.31 \\
    ${\psi }$ & 0.50 & 0.36 & 0.14 & 0.12 & 0.01 & 0.85 & 2.84 \\
    ${\varphi }$ & 6.15 & 4.80 & 1.09 & 0.11 & 0.01 & 8.03 & 19.52 \\
    ${\sigma }_{c}$ & 1.51 & 1.34 & 0.13 & 0.17 & 0.01 & 0.75 & 2.88 \\
    ${\lambda }$ & 0.71 & 0.69 & 0.07 & 0.03 & 0.00 & 0.32 & 1.37 \\
    ${\phi }_{p}$ & 1.67 & 1.44 & 0.07 & 0.17 & 0.03 & 0.64 & 1.24 \\
    ${\iota }_{w}$ & 0.53 & 0.60 & 0.12 & 0.00 & 0.00 & 1.31 & 2.87 \\
    ${\xi }_{w}$ & 0.78 & 0.68 & 0.09 & 0.10 & 0.01 & 0.44 & 1.50 \\
    ${\iota }_{p}$ & 0.26 & 0.35 & 0.12 & 0.05 & 0.00 & 0.79 & 2.90 \\
    ${\xi }_{p}$ & 0.68 & 0.66 & 0.07 & 0.03 & 0.01 & 0.30 & 0.87 \\
    ${\sigma }_{l}$ & 2.27 & 1.52 & 0.46 & 0.04 & 0.00 & 4.67 & 17.77 \\
    ${r}_{\pi }$ & 2.04 & 1.71 & 0.18 & 0.11 & 0.01 & 1.53 & 2.80 \\
    ${r}_{\Delta y}$ & 0.21 & 0.17 & 0.03 & 0.03 & 0.00 & 0.24 & 0.69 \\
    ${r}_{y}$ & 0.10 & 0.08 & 0.03 & 0.10 & 0.01 & 0.18 & 0.43 \\
    ${\rho }$ & 0.83 & 0.76 & 0.06 & 0.04 & 0.00 & 0.28 & 0.65 \\
    ${\rho }_{a}$ & 0.97 & 0.93 & 0.09 & 0.04 & 0.01 & 0.07 & 0.36 \\
    ${\rho }_{b}$ & 0.28 & 0.37 & 0.14 & 0.12 & 0.01 & 0.43 & 1.10 \\
    ${\rho }_{g}$ & 0.97 & 0.83 & 0.14 & 0.24 & 0.04 & 0.30 & 0.84 \\
    ${\rho }_{i}$ & 0.70 & 0.68 & 0.08 & 0.07 & 0.03 & 0.34 & 0.55 \\
    ${\rho }_{r}$ & 0.17 & 0.38 & 0.17 & 0.18 & 0.04 & 0.83 & 1.33 \\
    ${\rho }_{p}$ & 0.96 & 0.94 & 0.04 & 0.01 & 0.03 & 0.12 & 0.26 \\
    ${\rho }_{w}$ & 0.96 & 0.89 & 0.10 & 0.00 & 0.01 & 0.18 & 0.33 \\
    ${\sigma }_{a}$ & 0.46 & 0.40 & 0.12 & 0.21 & 0.06 & 0.35 & 0.71 \\
    ${\sigma }_{b}$ & 0.23 & 0.17 & 0.04 & 0.53 & 0.07 & 0.11 & 0.29 \\
    ${\sigma }_{g}$ & 0.50 & 0.37 & 0.06 & 0.71 & 0.20 & 0.16 & 0.46 \\
    ${\sigma }_{i}$ & 0.41 & 0.37 & 0.07 & 0.20 & 0.04 & 0.23 & 0.34 \\
    ${\sigma }_{r}$ & 0.22 & 0.15 & 0.04 & 0.48 & 0.13 & 0.14 & 0.27 \\
    ${\sigma }_{p}$ & 0.12 & 0.09 & 0.02 & 0.14 & 0.02 & 0.11 & 0.29 \\
    ${\sigma }_{w}$ & 0.29 & 0.25 & 0.03 & 0.28 & 0.02 & 0.13 & 0.30 \\
    $\overline{\gamma }$ & 0.45 & 0.44 & 0.03 & 0.34 & 0.05 & 0.06 & 0.13 \\
    ${r}$ & 0.12 & 0.22 & 0.07 & 0.06 & 0.01 & 0.43 & 1.30 \\
    $\overline{\pi }$ & 0.69 & 0.67 & 0.18 & 0.17 & 0.13 & 0.62 & 0.69 \\
    $\overline{l}$ & 1.35 & 1.39 & 1.02 & 0.17 & 0.13 & 3.06 & 3.64 \\ \hline \hline
  \end{tabular}
   \notes{ \textbf{Legend:} $n=192$, $k = 4$, $200$ Monte Carlo replications. Rejection rates for specification test (5\% level): 0.02 for all variables, and $0.31, 0.10,0.15, 0.02, 0.02, 0.06$, and $0.04$, respectively}
  }
\end{table}

\begin{table}[H]
  \caption{SW Model: Estimates and Standard Errors} \label{tab:SWest_full}
  \centering
  \setlength\tabcolsep{4.5pt}
   \renewcommand{\arraystretch}{0.85} 
   { \small
  \begin{tabular}{ll|ccc||cc}
  \hline\hline
  \multirow{2}{*}{${\theta }$} & \multirow{2}{*}{Parameter Interpretation} & 
  \multicolumn{3}{c||}{OT Estimate} & \multicolumn{2}{c}{Posterior} \\ 
  &  & {\ \textsc{est}} & {\ \textsc{sd}$_{c}$} & {\ \textsc{sd}$_{r}$} & {\ 
  \textsc{mean}} & \textsc{std} \\ \hline
  ${\rho }_{ga}$ & {\ Corr.: tech. and exog. spending shocks} & 0.47 & 0.27 & 
  0.53 & {\ 0.58} & 0.08 \\ 
  ${\mu }_{w}$ & {\ Wage mark-up shock MA} & 0.88 & 0.13 & 0.20 & {\ 0.90} & 0.04 \\ 
  ${\mu }_{p}$ & {\ Price mark-up shock MA} & 0.78 & 0.25 & 0.86 & {\ 0.81} & 0.08 \\ 
  ${\alpha }$ & {\ Share of capital in production} & 0.24 & 0.04 & 0.05 & {\
  0.23} & 0.02 \\ 
  ${\psi }$ & {\ Elast. of capital utilization adjustment cost} & 0.44 & 0.22
  & 0.78 & {\ 0.49} & 0.10 \\ 
  ${\varphi }$ & {\ Investment adjustment cost} & 3.00 & 1.50 & 3.42 & {\ 6.12}
  & 0.99 \\ 
  ${\sigma }_{c}$ & {\ Elast. of Intertemporal substitution} & 1.01 & 0.12 & 
  0.52 & {\ 1.50} & 0.14 \\ 
  ${\lambda }$ & {\ Habit persistence} & 0.74 & 0.12 & 0.38 & {\ 0.71} & 0.04 \\ 
  ${\phi }_{p}$ & {\ Fixed costs in production} & 1.50 & 0.26 & 0.60 & {\ 1.68}
  & 0.08 \\ 
  ${\iota }_{w}$ & {\ Wage indexation} & 0.85 & 0.30 & 0.95 & {\ 0.56} & 0.13 \\ 
  ${\xi }_{w}$ & {\ Wage stickiness} & 0.84 & 0.12 & 0.04 & {\ 0.77} & 0.05 \\ 
  ${\iota }_{p}$ & {\ Price indexation} & 0.27 & 0.32 & 0.42 & {\ 0.25} &  0.09 \\ 
  ${\xi }_{p}$ & {\ Price stickiness} & 0.80 & 0.06 & 0.12 & {\ 0.69} & 0.05 \\ 
  ${\sigma }_{l}$ & {\ Labor supply elasticity} & 1.00 & 2.38 & 2.92 & {\ 2.25}
  & 0.55 \\ 
  ${r}_{\pi }$ & {\ Taylor rule: inflation weight} & 1.80 & 1.06 & 0.77 & {\
  2.03} & 0.16\\ 
  ${r}_{\Delta y}$ & {\ Taylor rule: output gap change weight} & 0.16 & 0.05 & 
  0.10 & {\ 0.21} & 0.02 \\ 
  ${r}_{y}$ & {\ Taylor rule: output gap weight} & 0.16 & 0.16 & 0.08 & {\ 0.10%
  } & 0.02 \\ 
  ${\rho }$ & {\ Taylor rule: interest rate smoothing} & 0.90 & 0.08 & 0.07 & {%
  \ 0.82} & 0.02 \\ 
  ${\rho }_{a}$ & {\ Productivity shock AR} & 0.94 & 0.03 & 0.22 & {\ 0.97} & 0.01 \\ 
  ${\rho }_{b}$ & {\ Risk premium shock AR} & 0.68 & 0.10 & 0.19 & {\ 0.28} & 0.12 \\ 
  ${\rho }_{g}$ & {\ Exogenous spending shock AR } & 0.86 & 0.09 & 1.22 & {\
  0.97} & 0.01 \\ 
  ${\rho }_{i}$ & {\ Investment shock AR} & 0.47 & 0.15 & 0.13 & {\ 0.70} & 0.06 \\ 
  ${\rho }_{r}$ & {\ Monetary policy shock AR} & 0.47 & 0.28 & 0.71 & {\ 0.17}
  & 0.07 \\ 
  ${\rho }_{p}$ & {\ Price mark-up shock AR} & 0.96 & 0.03 & 0.04 & {\ 0.96} & 0.02 \\ 
  ${\rho }_{w}$ & {\ Wage mark-up shock AR} & 0.92 & 0.11 & 0.21 & {\ 0.96} & 0.02 \\ 
  ${\sigma }_{a}$ & {\ Productivity shock std. dev.} & 0.32 & 0.08 & 0.13 & {\
  0.46} & 0.03 \\ 
  ${\sigma }_{b}$ & {\ Risk premium shock std. dev.} & 0.11 & 0.02 & 0.05 & {\
  0.23} & 0.03 \\ 
  ${\sigma }_{g}$ & {\ Exogenous spending shock std. dev.} & 0.33 & 0.04 & 0.24
  & {\ 0.50} & 0.03 \\ 
  ${\sigma }_{i}$ & {\ Investment shock std. dev.} & 0.33 & 0.07 & 0.16 & {\
  0.41} & 0.04 \\ 
  ${\sigma }_{r}$ & {\ Monetary policy shock std. dev.} & 0.09 & 0.05 & 0.07 & 
  {\ 0.22} & 0.01 \\ 
  ${\sigma }_{p}$ & {\ Price mark-up shock std. dev.} & 0.05 & 0.04 & 0.24 & {%
  \ 0.12} & 0.02 \\ 
  ${\sigma }_{w}$ & {\ Wage mark-up shock std. dev.} & 0.25 & 0.03 & 0.11 & {\
  0.28} & 0.02 \\ 
  $\overline{\gamma }$ & {\ Trend growth: real GDP, Infl., Wages} & 0.46 & 0.01
  & 0.03 & {\ 0.45} & 0.02 \\ 
  ${r}$ & {\ Discount rate} & 0.20 & 0.09 & 0.23 & {\ 0.12} & 0.05 \\ 
  $\overline{\pi }$ & {\ Steady state inflation rate} & 0.80 & 0.26 & 0.25 & {%
  \ 0.68} & 0.10 \\ 
  $\overline{l}$ & {\ Steady state hours worked} & 0.16 & 0.65 & 0.77 & {\ 1.31%
  } & 0.87 \\ \hline\hline
  \end{tabular} }
   \end{table}

\begin{table}[H]
  \caption{SW Model: Estimates, Standard Errors under Stochastic Singularity}\label{tab:SWsing}
  \centering
  \setlength\tabcolsep{4.0pt}
   \renewcommand{\arraystretch}{0.85} 
 { \small
 \begin{tabular}{l|ccc||ccc||ccc}
 \hline\hline
 & \multicolumn{3}{c||}{6 shocks} & \multicolumn{3}{c||}{5 shocks} & 
 \multicolumn{3}{c}{4 shocks} \\ \hline
 & {\textsc{est}} & {\textsc{sd}$_{1}$} & {\textsc{sd}$_{r}$} & {\textsc{est}}
 & {\textsc{sd}$_{1}$} & {\textsc{sd}$_{r}$} & {\textsc{est}} & {\textsc{sd}$%
 _{1}$} & {\textsc{sd}$_{r}$} \\ \hline
 ${\rho }_{ga}$ & 0.45 & 0.24 & 2.70 & 0.47 & 0.18 & 0.25 & 0.50 & 0.12 & 0.17
 \\ 
 ${\mu }_{w}$ & 0.62 & 0.38 & 1.33 & {--} & {--} & {--} & {--} & {--} & {--}
 \\ 
 ${\mu }_{p}$ & 0.98 & 0.06 & 0.91 & 0.81 & 0.15 & 0.26 & {--} & {--} & {--}
 \\ 
 ${\alpha }$ & 0.25 & 0.03 & 0.11 & 0.29 & 0.03 & 0.09 & 0.28 & 0.02 & 0.23
 \\ 
 ${\psi }$ & 0.63 & 0.31 & 0.80 & 0.45 & 0.35 & 0.37 & 0.34 & 0.35 & 0.26 \\ 
 ${\varphi }$ & 3.00 & 1.39 & 7.25 & 3.96 & 4.98 & 7.07 & 3.85 & 3.68 & 7.71
 \\ 
 ${\sigma }_{c}$ & 1.37 & 0.28 & 2.16 & 1.00 & 0.02 & 0.05 & 1.00 & 0.04 & 
 0.04 \\ 
 ${\lambda }$ & 0.36 & 0.10 & 1.27 & 0.99 & 0.01 & 0.01 & 0.99 & 0.02 & 0.01
 \\ 
 ${\phi }_{p}$ & 1.55 & 0.23 & 0.52 & 1.39 & 0.20 & 0.73 & 1.32 & 0.13 & 1.48
 \\ 
 ${\iota }_{w}$ & 0.74 & 0.22 & 1.28 & 0.44 & 0.20 & 1.69 & 0.61 & 
 0.22 & 0.96 \\ 
 ${\xi }_{w}$ & 0.83 & 0.23 & 0.23 & 0.50 & 0.13 & 0.40 & 0.50 & 0.10
 & 0.29 \\ 
 ${\iota }_{p}$ & 0.20 & 0.15 & 1.98 & 0.41 & 0.20 & 0.79 & 0.27 & 
 0.11 & 0.55 \\ 
 ${\xi }_{p}$ & 0.61 & 0.09 & 0.65 & 0.54 & 0.12 & 0.41 & 0.35 & 0.09
 & 0.92 \\ 
 ${\sigma }_{l}$ & 1.00 & 3.27 & 5.48 & 1.00 & 1.61 & 3.42 & 1.00 & 2.34 & 
 2.92 \\ 
 ${r}_{\pi }$ & 1.72 & 1.03 & 0.53 & 1.87 & 0.87 & 0.68 & 1.87 & 0.73 & 1.23
 \\ 
 ${r}_{\Delta y}$ & 0.12 & 0.08 & 0.62 & 0.15 & 0.15 & 0.41 & 0.15 & 0.35 & 
 0.78 \\ 
 ${r}_{y}$ & 0.12 & 0.13 & 0.06 & 0.11 & 0.16 & 0.51 & 0.12 & 0.51 & 1.01 \\ 
 ${\rho }$ & 0.89 & 0.09 & 0.27 & 0.74 & 0.12 & 0.19 & 0.73 & 0.15 & 0.24 \\ 
 ${\rho }_{a}$ & 0.87 & 0.06 & 0.88 & 0.99 & 0.01 & 0.01 & 0.99 & 0.01 & 0.04
 \\ 
 ${\rho }_{b}$ & {--} & {--} & {--} & {--} & {--} & {--} & {--} & {--} & {--}
 \\ 
 ${\rho }_{g}$ & 0.82 & 0.09 & 0.41 & 0.95 & 0.04 & 0.04 & 0.94 & 0.04 & 0.07
 \\ 
 ${\rho }_{i}$ & 0.53 & 0.11 & 0.28 & 0.70 & 0.14 & 0.18 & 0.68 & 0.13 & 0.15
 \\ 
 ${\rho }_{r}$ & 0.61 & 0.16 & 1.56 & 0.95 & 0.04 & 0.05 & 0.95 & 
 0.04 & 0.03 \\ 
 ${\rho }_{p}$ & 0.80 & 0.13 & 2.12 & 0.97 & 0.04 & 0.05 & {--} & {--} & {--}
 \\ 
 ${\rho }_{w}$ & 0.94 & 0.08 & 0.07 & {--} & {--} & {--} & {--} & {--} & {--}
 \\ 
 ${\sigma }_{a}$ & 0.36 & 0.06 & 0.40 & 0.42 & 0.07 & 0.19 & 0.47 & 0.07 & 
 0.32 \\ 
 ${\sigma }_{b}$ & {--} & {--} & {--} & {-- } & {--} & {--} & {--} & {--} & {%
 --} \\ 
 ${\sigma }_{g}$ & 0.35 & 0.03 & 0.08 & 0.41 & 0.05 & 0.08 & 0.40 & 0.05 & 
 0.10 \\ 
 ${\sigma }_{i}$ & 0.38 & 0.07 & 0.24 & 0.38 & 0.07 & 0.10 & 0.41 & 0.06 & 
 0.11 \\ 
 ${\sigma }_{r}$ & 0.07 & 0.03 & 0.47 & 0.06 & 0.05 & 0.05 & 0.06 & 0.04 & 
 0.07 \\ 
 ${\sigma }_{p}$ & 0.23 & 0.03 & 0.22 & 0.13 & 0.02 & 0.13 & {--} & {--} & {--%
 } \\ 
 ${\sigma }_{w}$ & 0.04 & 0.04 & 0.12 & {--} & {--} & {--} & {--} & {--} & {--%
 } \\ 
 $\overline{\gamma }$ & 0.45 & 0.01 & 0.05 & 0.46 & 0.02 & 0.04 & 0.46 & 0.02
 & 0.05 \\ 
 ${r}$ & 0.12 & 0.13 & 0.91 & 0.21 & 0.08 & 0.12 & 0.22 & 0.08 & 0.13 \\ 
 $\overline{\pi }$ & 0.78 & 0.26 & 0.24 & 0.85 & 0.25 & 0.26 & 0.84 & 0.23 & 
 0.25 \\ 
 $\overline{l}$ & 0.21 & 0.89 & 0.90 & 0.62 & 0.91 & 1.03 & 0.58 & 0.85 & 1.39
 \\ \hline\hline
 \end{tabular}%
 \notes{ \textbf{Legend:} The following shocks, and their associated
 parameters, are suppressed in the following order: risk premium
 ($\rho_b,\sigma_b$), wage markup ($\mu_w,\rho_w,\sigma_w$), and price markup
 ($\mu_p,\rho_p,\sigma_p$).} }
  \end{table}

\begin{table}[H]
    \centering
   \caption{SW Model: Properties of Filtered Shock Processes} \label{tab:SWshocks}
  \centering
  \setlength\tabcolsep{0.0pt}
   \renewcommand{\arraystretch}{0.9} 
  { \small  
\begin{tabular}{l|ccccccc}
\hline \hline
\multicolumn{8}{c}{}\\
\multicolumn{8}{c}{(a) Cross Correlation Between Shock Processes} \\
\multicolumn{1}{c}{}& \hphantom{-Investment} & \hphantom{-Investment} & \hphantom{-Investment} & \hphantom{-Investment} & \hphantom{-Investment} & \hphantom{-Investment} & \hphantom{-Investment}\\
\hline
\multicolumn{1}{c}{}& \multicolumn{7}{c}{\textsc{true}} \\ \hline
&  \hphantom{-}TFP & \hphantom{-}Risk & \hphantom{-}Spending & \hphantom{-}Investment & \hphantom{-}Monetary & \hphantom{-}Price & \hphantom{-}Wage \\
TFP                &   \hphantom{-}1.00 &  &  &  &  &  &  \\ 
Risk               &   \hphantom{-}0.00 & \hphantom{-}1.00 &  &  &  &  &  \\ 
Spending           &   \hphantom{-}0.37 & \hphantom{-}0.00 & \hphantom{-}1.00 &  &  &  &  \\ 
Investment         &   \hphantom{-}0.00 & \hphantom{-}0.00 & \hphantom{-}0.00 & \hphantom{-}1.00 &  &  &  \\ 
Monetary           &   \hphantom{-}0.00 & \hphantom{-}0.00 & \hphantom{-}0.00 & \hphantom{-}0.00 & \hphantom{-}1.00 &  &  \\ 
Price              &   \hphantom{-}0.00 & \hphantom{-}0.00 & \hphantom{-}0.00 & \hphantom{-}0.00 & \hphantom{-}0.00 & \hphantom{-}1.00 &  \\ 
Wage               &   \hphantom{-}0.00 & \hphantom{-}0.00 & \hphantom{-}0.00 & \hphantom{-}0.00 & \hphantom{-}0.00 & \hphantom{-}0.00 & \hphantom{-}1.00 \\
\hline
\multicolumn{1}{c}{}& \multicolumn{7}{c}{\textsc{kalman filter}} \\ \hline
&  \hphantom{-}TFP & \hphantom{-}Risk & \hphantom{-}Spending & \hphantom{-}Investment & \hphantom{-}Monetary & \hphantom{-}Price & \hphantom{-}Wage \\
TFP                &   \hphantom{-}1.00 &  &  &  &  &  &  \\ 
Risk               &   -0.48 & \hphantom{-}1.00 &  &  &  &  &  \\ 
Spending           &   \hphantom{-}0.58 & -0.35 & \hphantom{-}1.00 &  &  &  &  \\ 
Investment         &    \hphantom{-}0.16 & -0.11 & \hphantom{-}0.08 & \hphantom{-}1.00 &  &  &  \\ 
Monetary           &    \hphantom{-}0.03 & \hphantom{-}0.28 & -0.00 & \hphantom{-}0.40 & \hphantom{-}1.00 &  &  \\ 
Price              &   -0.04 & -0.08 & -0.17 & \hphantom{-}0.27 & \hphantom{-}0.20 & \hphantom{-}1.00 &  \\ 
Wage               &   \hphantom{-}0.19 & -0.09 & \hphantom{-}0.17 & \hphantom{-}0.14 & \hphantom{-}0.00 & -0.02 & \hphantom{-}1.00 \\
\hline
\multicolumn{1}{c}{}& \multicolumn{7}{c}{\textsc{ot filter}} \\ \hline
&  \hphantom{-}TFP & \hphantom{-}Risk & \hphantom{-}Spending & \hphantom{-}Investment & \hphantom{-}Monetary & \hphantom{-}Price & \hphantom{-}Wage \\
TFP                &   \hphantom{-}1.00 &  &  &  &  &  &  \\ 
Risk               &   \hphantom{-}0.07 & \hphantom{-}1.00 &  &  &  &  &  \\ 
Spending           &   \hphantom{-}0.41 & \hphantom{-}0.05 & \hphantom{-}1.00 &  &  &  &  \\ 
Investment         & -0.12 & \hphantom{-}0.03 & -0.05 & \hphantom{-}1.00 &  &  &  \\ 
Monetary           &   -0.00 & \hphantom{-}0.03 & \hphantom{-}0.03 & -0.01 & \hphantom{-}1.00 &  &  \\ 
Price              &   -0.02 & \hphantom{-}0.04 & -0.06 & -0.00 & -0.05 & \hphantom{-}1.00 &  \\ 
Wage               &   \hphantom{-}0.12 & -0.01 & \hphantom{-}0.08 & -0.01 & -0.00 & -0.01 & \hphantom{-}1.00 \\ 
\hline \hline
\multicolumn{8}{c}{}\\
\multicolumn{8}{c}{(b) First-Order Autocorrelation of Shock Processes} \\  
\multicolumn{8}{c}{}\\ \hline
               & TFP & Risk & Spending & Investment & Monetary & Price & Wage\\ \hline
\textsc{true}  & 0.94  &0.68  &0.86  &0.47  &0.47  & 0.42 &0.04  \\ 
\textsc{kalman filter}   & 0.97  &0.62  &0.97  & 0.83 & 0.09 &-0.04  &0.00   \\ 
\textsc{ot filter}    & 0.91  &0.66  &0.84  &0.45  &0.49  &0.29  &0.05  \\ 
\hline  \hline
\end{tabular}
\notes{\textbf{Legend:} Panel (a) presents pairwise sample and model-implied (\textsc{true}) correlations among the 7 shock processes. The first 5 processes are AR(1) and the remaining 2 are ARMA(1,1). Panel (b) sample  and model-implied (\textsc{true}) first-order autocorrelations of the seven filtered shocks. }}
\end{table}

\newpage
\printbibliography[heading=subbibliography, title={Appendix References}]
\end{refsection}

\end{appendices}
\end{document}